\documentclass[11pt,a4paper]{article}
\pdfoutput=1

\usepackage[applemac]{inputenc}
\usepackage{amsmath, amsthm}
\usepackage{jheppub}
\usepackage[center]{caption}
\usepackage{subcaption}
\usepackage{multirow}
\usepackage{booktabs}
\usepackage{amsmath}
\usepackage{mathtools}
\usepackage{amsfonts}
\usepackage{amssymb}
\usepackage{graphicx}
\usepackage{indentfirst}
\usepackage{slashed}
\usepackage{diagbox}
\usepackage{mathrsfs}
\usepackage{hyperref}
\usepackage{bm}
\usepackage{feynmp}

\usepackage{tikz}
\usetikzlibrary{calc,positioning}

\allowdisplaybreaks

\makeatletter
\newcommand\makebig[2]{%
  \@xp\newcommand\@xp*\csname#1\endcsname{\bBigg@{#2}}%
  \@xp\newcommand\@xp*\csname#1l\endcsname{\@xp\mathopen\csname#1\endcsname}%
  \@xp\newcommand\@xp*\csname#1r\endcsname{\@xp\mathclose\csname#1\endcsname}%
}
\makeatother

\makebig{biggg} {3.0}
\makebig{Biggg} {3.5}
\makebig{bigggg}{4.0}
\makebig{Bigggg}{4.5}

\newcommand{\abs}[1]{\left\lvert#1\right\rvert}

\newcommand{\cutRes}{\mathcal{C}}

\def\beq{\begin{equation}}
\def\eeq{\end{equation}}
\def\bsp#1\esp{\begin{split}#1\end{split}}
\newcommand{\bea}{\begin{eqnarray}}
\newcommand{\eea}{\end{eqnarray}}
\newcommand{\bean}{\begin{eqnarray*}}
\newcommand{\eean}{\end{eqnarray*}}

\def\half{\frac{1}{2}}

\def\abs#1{\left| #1\right|}

\def\eqref#1{(\ref{#1})}

\def\a{{\alpha}}
\def\b{{\beta}}
\def\g{{\gamma}}

\def\eps{\epsilon}
\def\ep{\epsilon}

\def\Label#1{\label{#1}
  \smash{\hbox to0pt{\raise1ex\hbox{\tiny[#1]}\hss}}}

\def\beq{\begin{equation}}
\def\eeq{\end{equation}}
\def\bsp#1\esp{\begin{split}#1\end{split}}

\newcommand{\cC}{\mathcal{C}}

\ifx \tadInsert \undefined
  \def\tadInsert [#1]{
    \raisebox{-4mm}{\includegraphics[keepaspectratio=true, width=.8cm]{./#1}}
  }
\fi

\ifx \tadInsertLow \undefined
  \def\tadInsertLow [#1]{
    \raisebox{-4.1mm}{\includegraphics[keepaspectratio=true, width=.81cm]{./#1}}
  }
\fi

\ifx \bubInsertHigh \undefined
  \def\bubInsertHigh [#1]{
    \raisebox{-2.2mm}{\includegraphics[keepaspectratio=true, width=2cm]{./#1}}
  }
\fi

\ifx \bubInsert \undefined
  \def\bubInsert [#1]{
    \raisebox{-3.6mm}{\includegraphics[keepaspectratio=true, width=2cm]{./#1}}
  }
\fi

\ifx \bubInsertLow \undefined
  \def\bubInsertLow [#1]{
    \raisebox{-4.2mm}{\includegraphics[keepaspectratio=true, width=2cm]{./#1}}
  }
\fi

\preprint{CERN-TH-2021-082
}

\title{The diagrammatic coaction beyond one loop}

\author[a,b,c]{Samuel Abreu,}
\author[d,e,f]{Ruth Britto,}
\author[a]{Claude Duhr,}
\author[c]{Einan Gardi,}
\author[c]{and James Matthew}

\affiliation[a]{Theoretical Physics Department, CERN, Geneva, Switzerland}
\affiliation[b]{Mani L.~Bhaumik Institute for Theoretical Physics,
Department of Physics and Astronomy, UCLA, Los Angeles, CA 90095, USA}
\affiliation[c]{Higgs Centre for Theoretical Physics, 
School of Physics and Astronomy, \\
The University of Edinburgh, Edinburgh EH9 3FD, Scotland, UK}
\affiliation[d]{School of Mathematics, Trinity College, Dublin 2, Ireland}
\affiliation[e]{Hamilton Mathematics Institute, Trinity College, Dublin 2, Ireland}
\affiliation[f]{Institut de Physique Th{\'e}orique, Universit\'e Paris Saclay, 
CEA, CNRS, F-91191 Gif-sur-Yvette cedex, France}

\emailAdd{samuel.abreu@cern.ch}
\emailAdd{brittor@tcd.ie}
\emailAdd{claude.duhr@cern.ch}
\emailAdd{einan.gardi@ed.ac.uk}
\emailAdd{james.matthew01@hotmail.com}

\abstract{
The diagrammatic coaction maps any given Feynman graph into pairs of graphs 
and cut graphs such that, conjecturally, when these graphs are replaced by the 
corresponding Feynman integrals one obtains a coaction on the respective 
functions. The coaction on the functions is constructed by pairing a basis of differential forms,
corresponding to master integrals, with a basis of integration contours, corresponding to 
independent cut integrals.
At one loop, 
a general diagrammatic coaction was established using dimensional regularisation, 
which may be realised in terms of a global coaction on hypergeometric functions, or equivalently, 
order by order in the $\epsilon$ expansion, via a local coaction on multiple polylogarithms. 
The present paper takes the first steps in generalising the diagrammatic coaction beyond one loop.
We first establish general properties that govern the diagrammatic coaction at any loop order.
We then focus on examples of two-loop topologies for which all integrals 
expand into polylogarithms. In each case we determine bases of master integrals
and cuts in terms of hypergeometric functions, 
and then use the global coaction to establish the diagrammatic coaction of all master integrals
in the topology.
The diagrammatic coaction encodes the complete set of discontinuities of Feynman integrals, 
as well as the differential equations they satisfy, providing a general tool to understand 
their physical and mathematical properties.
}

\keywords{Feynman integrals, coaction, multiple polylogarithms.}

\begin{document}
\maketitle





\section{Introduction}\label{sec:intro}

In recent years there has been significant progress in understanding the mathematical properties of dimensionally-regularised Feynman integrals.
Amongst many interesting developments, it has been shown that one-loop Feynman graphs can be endowed with a diagrammatic 
coaction~\cite{Abreu:2017enx,Abreu:2017mtm}.
This coaction maps every graph into a linear combination of pairs of graphs in which subsets of the edges have been either pinched or cut and, crucially, it agrees
with the coaction on the associated master integrals. More precisely,
to every Feynman graph one can associate a Feynman integral, such that the graphical operations of pinching and cutting edges can be directly interpreted as operations on the corresponding Feynman integrals. The diagrammmatic coaction of refs.~\cite{Abreu:2017enx,Abreu:2017mtm} then stands in one-to-one correspondence with a coaction on integrals, and it encodes many of the analytic properties of the integrals such
as their differential equations and discontinuities \cite{Abreu:2014cla,Abreu:2015zaa,Abreu:2017enx,Abreu:2017mtm,Ananthanarayan:2021cch}.
While one-loop Feynman integrals are well understood from this perspective (see also the work of ref.~\cite{Tapuskovic:2019cpr}),
much less is known about the multi-loop case. 
The purpose of this paper is to take the first steps towards generalizing the diagrammatic coaction 
beyond one loop.

Divergences of both short- and long-distance origin are an inherent feature of Feynman integrals appearing in physical applications.
For instance, we are often interested in integrals with massless propagators and massless external momenta, 
in which long-distance divergences are abundant. Dimensional regularisation provides a very convenient
framework to regularise all divergences, and the one-loop diagrammatic coaction was developed within this framework~\cite{Abreu:2017enx,Abreu:2017mtm}.
Feynman integrals computed in $D$ space-time dimensions can be interpreted in (at least) two different ways:
as a Laurent series in the dimensional regulator $\eps=(d_0-D)/2$, 
with $d_0$ a positive integer, or as a function of hypergeometric type 
(cf., e.g.,~refs.~\cite{Bytev:2009kb,Kalmykov:2009tw,delaCruz:2019skx,Klausen:2019hrg,Nasrollahpoursamami:2017shc}).
Consequently, there are different ways in which a putative coaction can act on 
Feynman integrals in dimensional regularisation.

When these Feynman integrals are evaluated as Laurent series in $\epsilon$, the coefficients of the Laurent
series are 
periods\footnote{More precisely, the coefficients are periods when the kinematic variables are evaluated at algebraic numbers. We continue to refer to this type of function as a period.}~\cite{Bogner:2007mn} in the sense of Kontsevich and Zagier~\cite{periods}. 
Following this interpretation one may obtain a coaction by uplifting the periods in the 
Laurent expansion to their motivic analogues. 
The latter are naturally equipped with a (motivic) coaction~\cite{Brown:coaction}, 
which acts on Feynman integrals order by order in the Laurent expansion. 
The most familiar class of periods that appear in Feynman integral calculations 
(besides rational and algebraic functions) are the so-called multiple polylogarithms 
(MPLs)~\cite{2001math......3059G}, which are iterated integrals generalising 
the well-known logarithm and dilogarithm functions. At one loop, MPLs are thought to be sufficient to express any integral. Starting at two loops,
more complicated classes of periods appear. Nevertheless, it is well known that many multi-loop integrals---and indeed many complete on-shell scattering amplitudes in massless theories---may be fully expressed in terms of MPLs, whose coaction is very well 
understood~\cite{2002math......8144G,B:MTMZ,2011arXiv1102.1310B}. For this reason, 
we focus in this paper exclusively on Feynman integrals
whose Laurent expansions only involve MPLs. Following ref.~\cite{brown2019lauricella}, 
we refer to the coaction acting on the coefficients in the Laurent expansion as the \emph{local coaction}.

Considering the second perspective on Feynman integrals mentioned above,
where they are not expanded in the dimensional regulator and evaluate to functions of hypergeometric type,
leads to a second way to define a coaction.
In ref.~\cite{Abreu:2019wzk} we proposed a very compact formula for a coaction on 
large classes of hypergeometric functions which depend on the dimensional regulator 
$\eps$ and whose Laurent expansion in $\eps$ only involves MPLs. 
While a priori the definition of this coaction is completely distinct from the 
(motivic, local) coaction on MPLs, based on explicit 
calculations we conjectured that the two coactions are compatible in the sense that they commute 
 with the Laurent expansion in $\epsilon$. This conjecture was subsequently 
proven in ref.~\cite{brown2019lauricella} for the special case of Lauricella functions.
Following ref.~\cite{brown2019lauricella}, we refer to the coaction on (generalised) hypergeometric 
functions as the \emph{global coaction}.

The aforementioned local and global coactions are constructed knowing which type
of functions Feynman integrals evaluate to. However, they ignore the graphical origin of these Feynman integrals. 
One would like to have a coaction on Feynman integrals 
whose entries are determined by the topological data that defines the 
underlying Feynman graph, without any reference to the special functions that they 
evaluate to. From the physics perspective it is intuitively clear what this topological data is.
As mentioned above, there are two distinct operations that could be applied on any propagator (internal edge)  of a given graph:
 \emph{pinching}---eliminating a propagator and identifying the two vertices it connects;
and \emph{cutting}---putting a propagator on shell, that is, setting the inverse propagator to zero.
Thus, the topological data associated with a given Feynman graph consists of all graphs that may be obtained from it
upon pinching or cutting a subset of its edges.
The coaction conjectured in refs.~\cite{Abreu:2017enx,Abreu:2017mtm} for one-loop graphs,
of which special cases were proven in ref.~\cite{Tapuskovic:2019cpr}, is defined  in terms of this topological data and is an example of such
a coaction. We refer to this coaction as the \emph{diagrammatic coaction}. 

The diagrammatic coaction must be such that when the Feynman graphs are replaced by the associated integrals, obtained by using
the Feynman rules or, equivalently, by directly mapping to Feynman parameter space, one would readily land on a coaction on integrals.
Upon evaluating these integrals to all orders in the dimensional regulator one would recover the global coaction on the corresponding hypergeometric function, and upon evaluating them order-by-order in $\epsilon$ one would recover the local coaction on the Laurent coefficients.

The existence of a purely graphical coaction on Feynman graphs, which is consistent with these requirements,  is highly nontrivial. 
While the one-loop case is well understood 
from refs.~\cite{Abreu:2017enx,Abreu:2017mtm,Tapuskovic:2019cpr}, a complete picture 
of how to generalise the diagrammatic coaction beyond one loop is still lacking.\footnote{A proposal for a diagrammatic coaction beyond one loop was recently put forward in refs.~\cite{Kreimer:2020mwn,Kreimer:2021jni} for finite/renormalised Feynman integrals without the use of dimensional regularisation.}
As anticipated, the purpose of this paper is to take the first steps towards that goal.

Our construction is based on the well-known notion of master integrals. A given graph defines a topology, for which one may define a basis of master integrals. The basis elements all share the same propagators or a subset thereof. As usual, the requirement is that any Feynman integral with the given set of propagators, each raised to any arbitrary integer power, and involving any polynomial numerator depending on internal and external momenta, may be reducible to a linear combination of these basis elements with rational functions as coefficients by means of integration-by-parts relations~\cite{Chetyrkin:1981qh,Tkachov:1981wb}. 
The coaction we seek to construct must apply to any integral of the given topology. Naturally, we consider its application on the corresponding basis of master integrals, from which one may deduce the coaction on any other integral of that topology. This fixes a basis for the space of \emph{integrands} under consideration. In mathematical terms, it identifies the basis elements of the relevant (twisted) cohomology group (cf., e.g., refs.~\cite{Bitoun:2018afx,Mastrolia:2018uzb}). A major difference compared to the one-loop case is that there can be more than one master integral with the same set of propagators. 

Having defined the space of integrands our construction of the coaction is directly guided by the global coaction on the corresponding hypergeometric functions obtained upon evaluating the master integrals. 
Once a basis of integrands has been chosen, we can construct a corresponding (dual) set of \emph{contours} and express each element of the global coaction of these master integrals in terms of the same class of hypergeometric functions~\cite{Abreu:2019wzk}. This set of contours defines a basis for the relevant homology group (see ref.~\cite{Abreu:2019wzk} and references therein regarding the use of intersection theory and twisted cohomology in this context).  

A crucial observation regarding Feynman integrals is that independent integration contours encircling a set of propagator poles directly correspond to independent cut graphs. However, while at one loop there is a unique integration contour that encircles a given set of propagator poles, in the multi-loop case there may be multiple independent contours that all encircle the same set of poles. 
In other words there are multiple independent cuts sharing the same set of on-shell propagators~\cite{Primo:2016ebd,Frellesvig:2017aai,Bosma:2017hrk,Bosma:2017ens}.
This is related to the fact that there are in general multiple master integrals associated with a given topology.

If a diagrammatic coaction as described above indeed exists, it must be possible to identify the elements appearing in the global coaction of each of the master integrals in terms of pinches and cuts of the same set.  
Thus, we identify a clear route to constructing the diagrammatic coaction for a given integral topology: assuming the knowledge of the master integrals and their cuts in terms of hypergeometric functions, one should be able to interpret each element of the global coaction---a priori written in terms of hypergeometric functions---in terms of (cut) graphs. Whenever there are multiple master integrals for a given (sub-) topology, there are also multiple independent cuts sharing the same set of on-shell propagators, making it possible to find a basis of cuts which stands in one-to-one correspondence with the basis of master integrals.
We will show that following this procedure we can indeed establish a diagrammatic interpretation of the coaction for every master integral in a variety of two-loop examples.

The structure of this paper is as follows. In section~\ref{sec:propDiagCoac} we briefly review the main features of the coaction on integrals, as well as the  structure of the diagrammatic coaction at one loop. 
We describe the main properties that the diagrammatic coaction must have, and highlight the differences between the one-loop case and the general multi-loop case. 
Next, in section~\ref{sec:one-loop} we revisit the construction of the diagrammatic coaction at one loop from the perspective of the present paper. To this end we consider the example of a one-loop bubble integral with a single massive propagator, which evaluates in dimensional regularisation to a Gauss hypergeometric function. Using the global coaction of the latter~\cite{Abreu:2019wzk}, we illustrate the procedure described above and interpret each element of the coaction in terms of integrals corresponding to (cut) graphs.
The same general procedure is then applied in section~\ref{sec:first-examples} to a simple two-loop example, namely the sunset integral with one internal massive propagator. We briefly explain there how cuts are computed (with more details in appendix~\ref{CutsSection}) and then show that the diagrammatic interpretation of the global coaction naturally follows. In section~\ref{sec:Two-loop_examples} we present new results for the diagrammatic coaction of several nontrivial two-loop examples, which we use to highlight the novel features that arise beyond one loop. In section~\ref{conclusions} we summarise our findings and give some outlook. We provide some details on the computation of selected cuts in appendix~\ref{CutsSection}, comment on how differential equations
can be used to constrain the form of cut integrals in appendix~\ref{DiffEqCutsSection}, and list the expressions for all the master integrals and the corresponding cuts which appear in our coaction examples in appendix~\ref{Expressions}.

\section{Properties of the diagrammatic coaction}\label{sec:propDiagCoac}

The main purpose of this paper is to provide evidence for the existence of a diagrammatic coaction beyond one loop. Let us begin by discussing the general properties such a coaction is expected to have.
In refs.~\cite{Abreu:2017enx,Abreu:2019wzk} we have conjectured that the three 
coactions mentioned in the previous section (local, global and diagrammatic)
are all manifestations of a general formula for a coaction on integrals, 
which can be written in the compact form:
\begin{equation}\label{eq:coacGen}
	\Delta\int_\gamma\omega=\sum_{ij}c_{ij}\int_{\gamma}\omega_i
	\otimes \int_{\gamma_{j}}\omega\,.
\end{equation}
Here the set $\left\{\gamma_j\right\}$ forms a basis of (equivalence classes of) contours that
generate the homology group associated with the integral, 
and the set  $\left\{\omega_i\right\}$ forms a basis of (equivalence classes of) forms that 
generate the corresponding dual cohomology group. 
It is expected on general grounds that the dimensions of the homology and 
cohomology groups are always equal.
The integral whose coaction is considered, $\int_\gamma\omega$, involves some contour 
$\gamma$ and some integrand $\omega$ belonging to the respective spaces.
The $c_{ij}$ are algebraic functions of $\eps$ and the parameters on which the integrals 
depend;  in a Feynman integral these would be masses and Mandelstam invariants. 
The $c_{ij}$ are uniquely fixed by the choice of bases 
$\left\{\gamma_j\right\}$ and~$\left\{\omega_i\right\}$. 
Bases for which $c_{ij}=\delta_{ij}$
naturally induce a pairing between 
the form $\omega_i$ and the contour~$\gamma_i$, and then the coaction takes the simple form:
\begin{equation}\label{eq:coacGenDual}
	\Delta\int_\gamma\omega=\sum_{i}\int_{\gamma}\omega_i \otimes \int_{\gamma_{i}}\omega\,.
\end{equation}
In what follows we will refer to such bases as \emph{dual bases}, and 
say that $\omega_i$ and $\gamma_i$ are dual to each other.

The matrix 
\begin{equation}
P_{ji}\equiv \int_{\gamma_j}\omega_i 
\end{equation}
is called the \emph{period matrix}, and
the integral $\int_\gamma\omega$ can be written as a linear combination of the entries of
this matrix. Taking the local perspective introduced in the previous section,
dual bases, where the coaction takes the form of eq.~(\ref{eq:coacGenDual}), 
can be identified as those in which the period matrix reduces to the unit 
matrix at leading order in  $\epsilon$, that is $P_{ji}=\delta_{ji}+{\cal O}(\epsilon)$\,.
Having fixed both bases $\left\{\omega_i\right\}$ and $\left\{\gamma_j\right\}$, the
coaction in eq.~\eqref{eq:coacGen} applied to the elements of the period matrix takes
the form of matrix multiplication:
\begin{equation}
\label{coaction_on_P}
\Delta \left[P_{kl}\right] = \sum_{i,j} c_{ij} \,P_{ki}\otimes P_{jl}\,.
\end{equation}
We note that the $c_{ij}$ can be computed as the inverse of a matrix of intersection 
numbers, see ref.~\cite{Abreu:2019wzk}.
The coaction on any integral in the span of the elements of the period matrix can be deduced
from eq.~\eqref{coaction_on_P}.

Equation~\eqref{eq:coacGen} is very general, and it is conjectured to define a coaction 
on large classes of integrals. In particular, it encompasses the well-known (local) coaction on
multiple polylogarithms as well as the (global) coaction on hypergeometric 
functions~\cite{Abreu:2017enx,Abreu:2019wzk}.
Our interest here, however, is not in generic period integrals, but more specifically in Feynman 
integrals. Compared to generic period integrals, these have several important properties that have been
thoroughly studied in the physics literature.
In the rest of this section, we discuss the implications that these properties have on the
form of eq.~(\ref{eq:coacGen}) when applied to Feynman integrals.

\subsection{Left and right entries of the coaction on Feynman integrals}\label{sec:coactionEntries}

Let us start by discussing some general properties the diagrammatic coaction should have.
We argue that the integrals appearing in the left and right entries of the 
coaction in eq.~\eqref{eq:coacGen} can be understood respectively as master integrals related by differential equations, and cut integrals computed as residues and related to discontinuities \cite{Cutkosky:1960sp}.

We start by discussing the left entries, $\int_{\gamma}\omega_i$. 
The contour $\gamma$  is the same contour defining the Feynman integral $\int_{\gamma}\omega$ on which the coaction acts on the left-hand side of eq.~\eqref{eq:coacGen}, i.e.,~it is the usual contour corresponding to unrestricted integration over all loop momenta.
In turn the integrand $\omega_i$ is an element in a restricted set of differential forms, which form a basis of the cohomology group associated to the Feynman integral. 
Given that these are integrated over the usual contour $\gamma$, this basis of 
the cohomology group directly corresponds to what in the physics literature is called a
basis of \emph{master integrals}, cf., e.g., refs.~\cite{Lee:2013hzt,Bitoun:2017nre,Bitoun:2018afx,Mastrolia:2018uzb}. 

With the cohomology basis $\left\{\omega_i\right\}$ fixed, for the examples considered here we can use known mathematical methods to identify a corresponding basis of the homology group, 
that is a basis of contours~$\left\{\gamma_j\right\}$. 
In the context of the global coaction on hypergeometric functions
this process was outlined in ref.~\cite{Abreu:2019wzk}. 
As already stated in section \ref{sec:intro}, in this paper we focus on Feynman integrals
in dimensional regularisation for which the coefficients of the
 Laurent expansion around $\epsilon=0$ are linear combinations of
polylogarithms. It is clear that there is substantial freedom in choosing the 
bases of forms and contours, and it is usually convenient to make a choice such that the
integrals are \emph{pure}, i.e., the Laurent coefficients can be expressed in terms of MPLs of uniform weight with coefficients
that are algebraic numbers~\cite{ArkaniHamed:2010gh}. With such bases, 
the coefficients of the $\epsilon$-expansion are linear combinations
of polylogarithms of a well-defined weight at each order in $\epsilon$; the weight
increases by one with each power of $\epsilon$, and we normalise the integrals
such that the weight of the coefficient of $\epsilon^k$ is $k$.

Equation~\eqref{coaction_on_P} encodes many properties of the integrals under consideration.
In particular, from the construction of the period matrix in terms of complementary twisted homology and cohomology generators, 
it follows that the set of left entries $\left\{P_{ki}\right\}$ in eq.~(\ref{coaction_on_P}), for fixed $k$ and labelled by $i$,
is closed under differentiation. That is, their derivatives can be expressed in terms of the same set, leading to a system of first-order differential equations.
It is also clear that the same system of differential equations is satisfied for any choice of contour $k$,
i.e.~the differential equations are a statement about the cohomology. In particular, they are satisfied by $\{\int_{\gamma}\omega_i\}$, which we identified as the basis of master integrals. We conclude that eq.~(\ref{eq:coacGen}) is consistent with the well-known property of any basis of master integrals, namely that it
is closed under differentiation and satisfies a set of linear first-order differential 
equations~\cite{Kotikov:1991pm,Kotikov:1991hm,Kotikov1991,Gehrmann2000,Henn:2013pwa}.
We recall further that differential operators act in the right entry of the coaction~\cite{Duhr:2012fh,Brown:coaction,brown2019lauricella},
\beq\label{eq:Delta:DEQ}
\Delta\partial_x = (\textrm{id}\otimes \partial_x)\Delta\,.
\eeq
It then follows from eq.~(\ref{coaction_on_P}) and eq.~(\ref{eq:Delta:DEQ}) that
\beq\label{eq:Delta:DEQ_on_P}
\Delta\left[\partial_x P_{kl} \right] = \sum_{i,j} c_{ij}P_{ki}  \otimes \partial_x P_{jl}\,.
\eeq
As an illustration, consider the local coaction acting on a basis
of pure functions. One may then extract the components of eq.~\eqref{eq:Delta:DEQ_on_P} in which the 
right entries have weight zero. 
On the left-hand side the chosen component is the trivial element of the coaction, $\partial_x P_{kl} \otimes 1$. On the right-hand side,
the corresponding components, where the right entries have weight zero, are simply derivatives of logarithms. These non-transcendental factors can be freely moved to the left entries through scalar multiplication. Thus, comparing the left- and right-hand sides,  one derives\footnote{The connection between the coaction and differential equations is explained in more
detail in refs.~\cite{Abreu:2017enx,Abreu:2017mtm}. The interested reader is referred specifically to 
section 9, where the coaction is used to derive the differential equations of generic one-loop integrals.} an explicit differential equation for the $P_{kl}$ in terms of the set $\left\{P_{ki}\right\}$, indexed by $i$. The fact that the set of differential equations remains the same for any contour $k$ simply follows from eq.~(\ref{eq:Delta:DEQ_on_P}) noting that it is determined by the right entries, $c_{ij}\partial_x P_{jl}$, which do not depend on $k$.

Besides differential equations, the coaction also encodes the discontinuities 
of the integrals, i.e., the variation of the integral under analytic continuation 
in one of its external parameters~\cite{Duhr:2012fh,Brown:coaction,brown2019lauricella}.
Discontinuities of Feynman integrals are known to be computable in terms of cut integrals, cf.,~e.g.,~\cite{Cutkosky:1960sp,tHooft:1973wag,Abreu:2014cla,Abreu:2015zaa,Bloch:2015efx,Abreu:2017ptx,Bourjaily:2020wvq,Ananthanarayan:2021cch}, which can be defined through modifications of the integration contour that select some combination of residues. 
With reference to the coaction of the period matrix, eq.~(\ref{coaction_on_P}), the right entries incorporate such contour modifications and can be understood as being drawn from the set of discontinuities of $P_{kl}$.
The same linear relations hold between the discontinuities of $P_{kl}$ and the right entries in its coaction independently of~$l$, i.e.~they are a statement about the homology. It follows that in eq.~(\ref{eq:coacGen}) the discontinuities of $\int_\gamma\omega$ are related in the same way to its cut integrals $\{ \int_{\gamma_{j}}\omega\}$ appearing in the right entries.
Discontinuities act on the left entry of the coaction, according to 
\beq\label{eq:Delta:Disc}
\Delta\textrm{Disc} = (\textrm{Disc}\otimes\textrm{id})\Delta\,.
\eeq
It follows from eq.~(\ref{coaction_on_P}) and eq.~(\ref{eq:Delta:Disc}) that
\begin{equation}
\label{eq:Delta:Disc_on_P}
\Delta\left[ \textrm{Disc}_{x} P_{kl}\right] = \sum_{i,j} c_{ij} \, \left(\textrm{Disc}_{x}  P_{ki}\right)\otimes P_{jl}\,.
\end{equation}
As in the discussion of differential equations above, we can make this discussion more concrete by considering again the local coaction acting on a pure basis.
One may extract the component of this coaction such that the left entry has transcendental weight zero. These weight-zero terms 
are simply rational numbers corresponding to the leading terms in the $\epsilon$-expansion of $\textrm{Disc}_{x}  P_{ki}$.\footnote{We normalise the discontinuity operator to cancel overall factors of $2\pi i$.}
Equation~(\ref{eq:Delta:Disc_on_P}) therefore relates the discontinuity of  $P_{kl}$ with respect to a kinematic variable $x$ to a linear combination (summed over $j$) of the right entries $P_{jl}$  with rational coefficients.
We finally note that it follows from eq.~(\ref{eq:Delta:DEQ_on_P}), 
which holds for any contour~$k$, that the cut integrals are constrained by
the fact that they must satisfy  the \emph{same} differential equations as their uncut 
analogues, see, e.g., refs.~\cite{Anastasiou:2002yz,Primo:2016ebd,Frellesvig:2017aai,Bosma:2017hrk,Bosma:2017ens}. 

These considerations lead us to two important conclusions. First, any putative 
diagrammatic coaction on an $L$-loop integral should admit a representation that only involves integrals with $L$ loops.
Second, it follows from eq.~\eqref{coaction_on_P} (and the way the coaction interacts with differentiation and discontinuity operations) that the coaction on cut integrals will have the same diagrammatic structure as the one for the corresponding uncut ones. This form for the coaction is very restrictive, and it is not at all obvious from the structure of the local or global coactions on motivic periods or hypergeometric functions when considered outside the context of Feynman integrals.

\subsection{Coaction on one-loop integrals}\label{sec:oneLoopCoac}

The properties discussed in the previous section are very general
and independent of the number of loops.
To make them more concrete, we provide below a brief review of the diagrammatic
coaction on one-loop integrals of refs.~\cite{Abreu:2017enx,Abreu:2017mtm}. This will set the scene for the
general construction in section \ref{sec:spanCuts}.

As an illustration, we discuss the diagrammatic coaction 
on the one-loop bubble integral with two massive propagators (for definitions
and details on the computation of the master integrals and cuts, see ref.~\cite{Abreu:2017mtm}):
\begin{eqnarray}
\label{diagConjBubM1M2}
	\Delta
	\left[\bubInsertLow[bub2m12Edges]\right]
	&=& \bubInsertLow[bub2m12Edges] \otimes \bubInsert[bub2m12CutPEdges] 
	+\tadInsert[tad1]\otimes \left( \bubInsert[bub2m12Cut1Edges] + \half\bubInsert[bub2m12CutPEdges] \right) 
	\nonumber \\
	&&+\tadInsert[tad2]\otimes \left(\bubInsert[bub2m12Cut2Edges] + \half\bubInsert[bub2m12CutPEdges]  \right)\,,
\end{eqnarray}
where $e_1$ and $e_2$ denote the two edges of the graph.
In ref.~\cite{Abreu:2017mtm} it was established by explicit computation in $D=2-2\epsilon$ space-time dimensions that, 
when all graphs in eq.~(\ref{diagConjBubM1M2}) are replaced by their Laurent expansion in $\epsilon$, this expression agrees with the local version of the motivic coaction,
which acts on the one-loop bubble integral order by order in $\eps$ (this was explicitly checked
to order $\epsilon^4$). 
The left entries are uncut Feynman integrals:
the bubble integral itself and the two tadpole integrals obtained by pinching one 
of the two edges in the graph. This set of integrals is indeed closed under differentiation. 
The right entries, instead, correspond to the different cuts of the bubble 
integral, where either one or both of the propagators are cut.

To generalise this coaction to one-loop integrals with an arbitrary number of edges, we
recall two properties of one-loop integrals. 
Let $G$ denote a one-loop Feynman graph and $E_G$ the set of internal edges (i.e., propagators).
First, at one loop a set of master integrals 
may be obtained by considering all possible pinches of the graph under consideration
(for generic masses and Mandelstam invariants), i.e.\ we can label
the master integrals by all the non-empty subsets of $E_G$.
For example, for the generic one-loop bubble graph in eq.~\eqref{diagConjBubM1M2}, 
we have three master integrals, with edge sets $\{e_1,e_2\}$, $\{e_1\}$ and $\{e_2\}$.
Second, the geometry underlying one-loop Feynman integrals is such that there
are only singularities at configurations of the loop momentum where either a subset of
the propagators are on-shell, or the loop momentum becomes infinite, see e.g.~ref.~\cite{Abreu:2017ptx}.
It turns out that the contours encircling the singularity at infinity can be written
as linear combinations of those that do not. 
More explicitly, if we denote by $\Gamma_C$ a contour which encircles the poles of the 
propagators corresponding to a subset of edges $C$ (and no other poles), and by
$\Gamma_{\infty C}$ the contour that encircles the propagator poles in $C$ as well as the 
singularity at infinity, we have the relation~\cite{fotiadi,teplitz,Froissart,Froissart2}:
\beq\label{eq:one-loop contours}
\Gamma_{\infty C} = -2x_C \,\Gamma_{C} + 
\sum_{C\subset X \subseteq E_G}(-1)^{\lceil |C|/2\rceil + \lceil |X|/2\rceil}\,
 \Gamma_X\,,
\eeq
where $x_C=1$ if $|C|$ is odd, and $x_C=0$ if $|C|$ is even. We note however that
the set of contours $\{\Gamma_C:C\subseteq E_G\}$, if $C=\emptyset$ is allowed,
is larger than the number of master integrals by exactly one element. 
An extra relation between contours
can be found by specialising eq.~\eqref{eq:one-loop contours} to  
$C=\emptyset$~\cite{Abreu:2017ptx}:
\beq\label{eq:pole_cancellation}
\sum_{e\in E_G} \cC_{e}I_G + \sum_{\substack{e_j,e_k\in E_G\\ j<k}}\cC_{e_je_k}I_G
= -\eps\, I_G\mod i\pi\,,
\eeq
where $I_G$ denotes the Feynman integral associated to the one-loop graph $G$, 
and $\cC_CI_G$ denotes the corresponding integral where the propagators in the set 
$C\subseteq E_G$ are cut.\footnote{Note that while this relation only holds up 
to additive terms proportional to $i\pi$, such terms are immaterial in as far as 
the right entry of the coaction is concerned~\cite{Abreu:2017enx,Abreu:2017mtm}.} 
Using this relation, we can eliminate the contour
$\Gamma_\emptyset$ which corresponds to an uncut integral from the basis of contours.
In conclusion, a basis of contours associated with the one-loop
graph $G$ is given by $\{\Gamma_C:C\subseteq E_G,\,C\neq\emptyset\}$, 
and there are as many independent contours as master integrals.
The fact that the bases of integrands and contours at one loop can be uniquely 
labelled by the propagators of the Feynman integral
is at the core of the simplicity of the diagrammatic coaction on one-loop 
integrals~\cite{Abreu:2017enx,Abreu:2017mtm}.

To write an explicit formula for the one-loop diagrammatic coaction, however, we must
settle on a basis for the integrands and the contours. Regarding the integrands,
we find it convenient  to choose scalar integrals in $D=n_{|E_G|}-2\epsilon$ space-time dimensions~\cite{Abreu:2017mtm},
where $n_{|E_G|}$ is an even number depending on the number of edges $|E_G|$: 
$n_{|E_G|}=|E_G|$ for even~$|E_G|$ and $n_{|E_G|}=|E_G|+1$ for odd $|E_G|$.  
This choice (which was made in particular in eq.~(\ref{diagConjBubM1M2})) has the advantage of having a simple diagrammatic representation: the master integrals are fully 
specified by the associated graph $G$, and furthermore, evaluate to pure functions (once properly normalised, see ref.~\cite{Abreu:2017mtm}).
We denote this choice of the master integrals as 
\begin{equation}
\label{basis_omega_G}
	J_G=\int_{\Gamma_\emptyset}\omega_G\,,
\end{equation}
where $\omega_G$ is the associated integrand and where $\Gamma_\emptyset$ denotes the uncut contour corresponding to the usual unrestricted momentum integration.

Having chosen the basis of integrands, we next turn to the basis of contours.
We recall that the space of contours is spanned by $\{\Gamma_C:C\subseteq E_G,\,C\neq\emptyset\}$.
However, if we were to choose these as our basis we would find that the bases of integrand and contours 
are not dual (in the sense defined in eq.~\eqref{eq:coacGenDual}).
In ref.~\cite{Abreu:2017mtm}, we proposed a choice of basis
of contours which is dual to the basis of integrands of eq.~(\ref{basis_omega_G}). The elements of this basis,
denoted~$\gamma_C$, are defined as
\begin{equation}
\label{gamma_C}
\gamma_C\equiv \Gamma_C+a_C\sum_{e\in E_G\setminus C}\Gamma_{Ce}\,,
\end{equation}
with $a_C=1/2$ for $|C|$ odd and 0 for $|C|$ even, and the contour $\gamma_C$ is dual
to the integrand~$\omega_{G_C}$ which has the propagators in $C$ and no others.
The term proportional to $a_C$ is called a \emph{deformation term}, as it deviates from 
the naive expectation that $\Gamma_C$ would be dual to~$\omega_{G_C}$.

With these choices of bases, the coaction on one-loop Feynman integrals
is given by
\begin{equation}\label{eq:oneLoopMF}
	\Delta\left(\int_{\Gamma_\emptyset}\omega_G \right)=\sum_{\emptyset\neq C\subseteq E_G}
	\int_{\Gamma_\emptyset}\omega_{G_C}\otimes
	\int_{\gamma_C}\omega_G\,,
\end{equation}
where $G_C$ is the graph $G$ with the subset of edges $E_G\setminus C$ pinched.
It satisfies the general properties 
discussed in section 
\ref{sec:coactionEntries} (see refs.~\cite{Abreu:2017enx,Abreu:2017mtm}), and eq.~\eqref{diagConjBubM1M2} is easily obtained 
as a special case. Furthermore,
we stress that the coaction in eq.~\eqref{eq:oneLoopMF} achieves the goals we have set in 
section \ref{sec:intro}. It is a diagrammatic coaction, 
in that it is defined in terms of the topological data of the Feynman graph, and we have argued and provided
evidence~\cite{Abreu:2017enx,Abreu:2017mtm,Abreu:2018nzy,Abreu:2019eyg,Abreu:2019wzk} that 
it reproduces both the local coaction if each integral is replaced by
its Laurent expansion in $\epsilon$,
and the global coaction if each integral is replaced
by the corresponding all-order in $\epsilon$ expression.
As such, it is the template for what we would like to achieve beyond one loop.

We close this brief summary by noting an important property of the contours $\gamma_C$
defined in eq.~\eqref{gamma_C}, which we recall are constructed to be dual
to the forms $\omega_{G_C}$ that only feature the propagators in the set $C$. 
These dual contours only involve contours that encircle the singularities of (at least) all of
the propagators in~$C$. Let us now explain the significance of this property.

Consider the coaction for $\mathcal{C}_YJ_G$, that is for the 
integral corresponding to the diagram $G$ in which all the propagators in $Y$ 
(and only these) are cut. Under the cut conditions, any diagram which does not
feature all the propagators in $Y$ will vanish, and from 
eq.~\eqref{eq:oneLoopMF} it follows that
\begin{equation}\label{eq:oneLoopMFDiff}
	\Delta \int_{\Gamma_{Y}}\omega_G =\sum_{Y\subseteq  C\subseteq E_G}
	\int_{\Gamma_Y}\omega_{G_C}\otimes
	   \int_{\gamma_C}\omega_G\,.
\end{equation}
According to eq.~(\ref{gamma_C}), the contour $\gamma_C$ in the right entry of the coaction necessarily has all the propagators
in $C$ (or more) on shell, and since $Y\subseteq C$, we see that the coaction of  $\mathcal{C}_YJ_G$
is \emph{fully determined} by diagrams where \emph{all} the propagators in $Y$ are on shell.
This matches our physical expectation: all elements of the coaction of a cut integral with the propagators in $Y$ on shell are expected to also have these propagators on shell. In particular, given the discussion below eq.~(\ref{eq:Delta:DEQ_on_P}), it follows
from eq.~\eqref{eq:oneLoopMFDiff} that the differential equation for $\mathcal{C}_YJ_G$
is fully determined by Feynman integrals where all the propagators in $Y$ are cut. This applies to the integrals appearing in that differential equation (which all have precisely the propagators in $Y$ on shell) as well as to the coefficients in the differential equation, which are determined by the derivatives of the right entries in eq.~(\ref{eq:oneLoopMFDiff}) that have all the propagators in $Y$, plus additional ones, on shell.
While the diagrammatic coaction will be more complex beyond one loop, we expect that this property will carry over: it can be written in a form such that if a given propagator features on a left-entry diagram, it will be cut in the corresponding right entry.

\subsection{General formula for a coaction beyond one loop}
\label{sec:spanCuts}

In order to understand how to generalise the coaction from eq.~\eqref{eq:oneLoopMF} beyond one loop,
we start by highlighting a major difference between one-loop Feynman
integrals and $L$-loop ones (for \hbox{$L > 1$}). 
Beyond one loop, there is no direct correspondence between pinches and the set of master 
integrals. More concretely, many topologies feature several master integrals 
that share exactly the same set of propagators but differ for example by numerators 
(or by propagators raised to different integer powers).
As a consequence, a simple representation in terms of graphs as at one loop will 
not be sufficient, and we will need to consider graphs with additional decorations 
to distinguish different master integrals that share the same set of propagators.

In the language of the previous section, let $G$ be an $L$-loop graph. To such
a graph, we can associate several independent integrals
(labelled by an index $k$) of the form
\begin{equation}
\int_{\Gamma_\emptyset}\omega^{(k)}_{G}
\end{equation}
which all have the same set of propagators.
Similarly, for
the graphs $G_C$ obtained by contracting the edges of $G$ that are not in $C$, we
have in general several associated integrals of the form:
\begin{equation}
\label{eq:C-MI}
\int_{\Gamma_\emptyset}\omega^{(k)}_{G_C}\,.
\end{equation}
For some sets $C$ there are no associated master integrals (for example a case where $G_C$ has fewer loops than $G$; then
the corresponding integral vanishes in dimensional regularisation). This leads us
to define the set $M_G$ of \emph{master topologies} of $G$, such that $C \in M_G$ 
if and only if there exists at least one master integral of the form given in 
eq.~(\ref{eq:C-MI}).\footnote{
	To keep our discussion as simple as possible, we always assume that the integrals we select
	as master integrals have the smallest possible number of propagators. For instance, if an $n$-propagator
	 integral $I_n$ is related to an $(n-1)$-propagator integral $I_{n-1}$ by $I_n=r I_{n-1}$ 
	for some rational function $r$, then we would choose $I_{n-1}$ as the master integral 
	rather than $I_n$. This is consistent with building the set $M_G$ by starting with the 
	topologies with the smallest number of propagators. 
}
Determining 
a basis of forms $\omega^{(k)}_{G_C}$ for $M_G$ is closely 
related to the study of integration-by-parts (IBP) relations
(see refs.~\cite{Chetyrkin:1981qh,Tkachov:1981wb,Laporta:2001dd,Lee:2013hzt,Ita:2015tya,Larsen:2015ped,Bitoun:2017nre,Bitoun:2018afx,Mastrolia:2018uzb})
and dimension-shift identities \cite{Bern:1992em,Tarasov:1996br,Lee:2009dh}. 
While a solution to this problem is not known for
an arbitrary $L$-loop Feynman integral, several public IBP 
codes~\cite{vonManteuffel:2012np,Smirnov:2019qkx,Klappert:2020nbg} are able to construct such a basis 
for large classes of graph topologies, and in particular for
all the examples we will consider in subsequent sections.

Just as the independent integrands are not uniquely identified by sets of propagators, in the multi-loop case
it is no longer sufficient to specify a set of cut propagators to define a cut 
integral~\cite{Primo:2016ebd,Frellesvig:2017aai,Bosma:2017hrk,Bosma:2017ens}. 
Independent integration contours associated with a given set of cut propagators
must also be labelled with extra decorations.
While much less is known about dependencies between integration contours beyond one 
loop---in particular, the multi-loop generalisation of the homology relation in eq.~\eqref{eq:one-loop contours} is not 
known---we can leverage our one-loop knowledge to show that some cut integrals are not
independent.

To this end, given a graph $G$, let us consider a contour defining an integral where only the propagators in a set 
$X$ 
are cut (as noted above, there might be several independent contours 
satisfying this condition).
If the diagram obtained from $G$ by pinching all the uncut propagators, $E_G\setminus X$, has the same number of 
loops $L$ as the uncut integral $G$ itself, we refer to this contour as defining a \emph{genuine $L$-loop cut}.
In contrast, non-genuine $L$-loop cuts would be cuts that leave at least one loop uncut.
As we will now show, it is always possible to express non-genuine $L$-loop cuts
in terms of genuine $L$-loop cuts. To understand why this statement should hold, assume that a cut integral contains an uncut one-loop subdiagram. 
We can then use the one-loop relation in eq.~\eqref{eq:pole_cancellation}, 
which crucially is valid for completely arbitrary kinematics, to express the one-loop 
Feynman integral associated to this subdiagram in terms of cut integrals. 
Iterating this procedure, we can eliminate all uncut subdiagrams, replacing them with a linear combination of terms 
which all feature at least one cut propagator in every loop. We thus arrive at a 
representation solely in terms of genuine $L$-loop cuts. 
As a consequence, we only need to consider genuine $L$-loop cuts when constructing
a spanning set of cut integrals.

 For 
each $C\in M_G$, we define the contours 
$\Gamma_C^{(k)}$
as a set (enumerated by $k$) of 
independent 
contours that encircle the poles of the propagators in $C$ and only those.
In view of the argument of the previous paragraph, we require these contours to be genuine $L$-loop cuts. Another constraint on the basis of contours is that its size should be the same as the number of independent integrals, i.e., for a given $C\in M_G$, $k$ takes the same values in the bases $\left\{\Gamma_C^{(k)}\right\}$ and $\left\{\omega_{G_C}^{(k)}\right\}$.

A priori, the basis of contours $\left\{\Gamma_C^{(k)}\right\}$ 
and the basis of forms $\left\{\omega_{G_C}^{(k)}\right\}$ discussed above need not be dual in the sense defined above eq.~\eqref{eq:coacGenDual}. However, given eq.~\eqref{gamma_C} and the discussion in the last paragraph
of section \ref{sec:oneLoopCoac}, we expect that it is possible to choose a
basis of integrands such that the dual contours are given by
\begin{equation}
\label{gammaC_sum_Gamma}
	\gamma_C^{(k)}=\sum_{{\substack{X\in M_G\\ C\subseteq X}}}
	\sum_{i}
\alpha_X^{(k,i)}\Gamma_X^{(i)}\,,
\end{equation}
that is, the dual contour to $\omega^{(k)}_{G_C}$ is a linear combination of contours
that encircle all the poles of the propagators in $C$ or more (but not fewer).
The coefficients $\alpha_X^{(k,i)}$ can in general depend on the same variables as the Feynman 
integral.

It follows from the definitions above and from the general coaction 
formula in eq.~\eqref{eq:coacGenDual} that the coaction on Feynman integrals beyond one loop
takes the form
\begin{equation}\label{eq:coac_L_loop}
	\Delta\left(\int_{\Gamma_\emptyset}\omega_G^{(k)}\right)=
	\sum_{C\in M_G}\sum_i
	\int_{\Gamma_\emptyset}\omega_{G_C}^{(i)}\otimes
	\int_{\gamma_C^{(i)}}\omega_G^{(k)}\,,
\end{equation}
where~$i$ indexes the elements of the basis forms~$\omega_{G_C}^{(i)}$ for a given $C$, as well as their dual contours~$\gamma_C^{(i)}$.
This formula reduces to eq.~\eqref{eq:oneLoopMF} for one-loop integrals: in that case~$M_G$ corresponds to all non-empty subsets of $E_G$ and there is just a single value of $i$ for every~$C$.

We stress that even though eqs.~\eqref{eq:oneLoopMF} and 
\eqref{eq:coac_L_loop} look very similar, the one-loop coaction \eqref{eq:oneLoopMF}
is fully explicit (the bases have been fixed and all integrals have been explicitly defined for a generic mass configuration and any number of legs, see refs.~\cite{Abreu:2017enx,Abreu:2017mtm}), whereas the $L$-loop generalisation
\eqref{eq:coac_L_loop} is not.
In particular, the set of master integrals and their dual contours needs to be identified on a case-by-case basis. Moreover,
the explicit definition of the contours and the calculation of the associated cuts of multi-loop integrals 
is not as well understood as at one loop. In the following sections we will use the connection between the diagrammatic coaction in eq.~\eqref{eq:coac_L_loop} and the global coaction on hypergeometric functions
to make eq.~\eqref{eq:coac_L_loop} fully explicit in a series of two-loop examples. Before doing so, we collect in the remainder of this section properties that the coaction must fulfil, independently of the loop order.

\subsection{Properties of the coaction}

Our pursuit of an explicit  diagrammatic coaction of the form of eq.~\eqref{eq:coac_L_loop} is guided by two additional principles: the consistency with degenerate limits, and the cancellation of spurious poles in $\epsilon$ that may arise in some of the individual terms. We discuss these two principles in turn.

\subsubsection{Degeneracy of external parameters}

An important property of the diagrammatic coaction is that it should be consistent
with taking degenerate values of the external parameters (e.g., setting masses or Mandelstam invariants to zero, or making them equal). Indeed, in physics applications, one is usually interested in degenerate configurations.
Starting with a generic integral and taking a limit to a degenerate kinematic configuration 
does not necessarily commute with the $\eps$-expansion, and the Laurent coefficients may develop logarithmic singularities. 
If the limit is taken prior to the $\eps$-expansion, 
experience from explicit computations shows that the limit is always smooth, and 
the aforementioned logarithmic singularities manifest themselves as additional poles 
in the dimensional regulator. This implies that we should be able to safely take limits 
to degenerate kinematics configurations in the diagrammatic 
coaction.\footnote{We expect this more generally to be a property of the global coaction 
on hypergeometric functions.}

Let us illustrate this point on the example of the one-loop bubble graph.
The bubble integral in eq.~\eqref{diagConjBubM1M2} is an example of a generic 
Feynman integral, i.e., of an integral where all propagators are massive and 
the masses and Mandelstam invariants are all distinct and take generic non-zero values.
Consider now the limit $m_2\to 0$. Then, eq.~\eqref{diagConjBubM1M2} reduces to
\begin{equation}
\label{diagConjBubM1}
	\Delta
	\left[\bubInsertLow[bub1mEdges]\right]
	\!=\! \bubInsertLow[bub1mEdges]\! \otimes\! \bubInsert[bub1mCutPEdges] 
	+\tadInsert[tad1]\otimes \left(\!\bubInsert[bub1mCut1Edges] 
	+ \half\bubInsert[bub1mCutPEdges] \right) \,,
\end{equation}
where the thin line represents a massless propagator.
In this equation, all terms appearing in eq.~\eqref{diagConjBubM1M2} which involve the tadpole with edge 
$e_2$ have disappeared, because scaleless integrals vanish in dimensional regularisation. 
Similarly, all diagrams where a single massless propagator is cut  
vanish~\cite{Abreu:2015zaa,Abreu:2017ptx}. 

We will verify through explicit calculations in a set of examples that the 
diagrammatic coactions we will construct at two loops
also have the property of being consistent with degenerate limits.
In particular, a new feature of some examples we will examine is that
the degenerate configurations correspond to a reduction in the number of master integrals, and we will see what the consequences are for the coaction
formula (see in particular section~\ref{sec:doubleEdged}).


\subsubsection{Cancellation of poles}

Feynman integrals in dimensional regularisation are meromorphic functions of the 
dimensional regulator $\eps$. For generic values of the masses and Mandelstam invariants, 
the poles at $\eps\to 0$ can only be of ultraviolet (UV) origin, 
corresponding to the limit where the components of the loop momenta become infinite. 
If a Feynman integral has enough propagators to balance the power of loop momenta in the numerator, there are 
no singularities in this limit. 
It follows that, for each number of loops $L$,
only a finite number of generic Feynman integrals without numerator are UV-divergent.

Let us see what this implies for the diagrammatic coaction. We have already established
that the left entries of the coaction  of a Feynman integral
includes graphs obtained by removing (pinching) propagators 
of the original Feynman integral, which we here assume to be
finite.  As propagators get removed, the integrals become less well convergent in the UV, and 
after a sufficient number has been removed
they will ultimately have UV poles despite the original integral being finite. 
Self-consistency of the local coaction requires these poles to cancel. This
in turn implies constraints on the combinations of (cut) diagrams that can
appear in the diagrammatic coaction.
Let us recall how this happens at one loop (see refs.~\cite{Abreu:2017enx,Abreu:2017mtm} for more details) by returning to the example in 
eq.~\eqref{diagConjBubM1M2}. The bubble and tadpole integrals are considered
in $D=2-2\eps$ dimensions, where the bubble is finite and the tadpole has 
a UV pole whose residue is independent of the tadpole mass. In our normalisation,
\begin{equation}
	\tadInsert[tad1]=-\frac{1}{\epsilon}+\mathcal{O}(\epsilon^0)\,,
\end{equation}
and similarly for the other tadpole integral. Since all the cuts of the bubble are finite
this is the only source of poles in $\eps$ in the coaction.
Collecting terms appropriately in eq.~\eqref{diagConjBubM1M2}, we find that the 
entries of the coaction that are proportional to $1/\epsilon$ are of the form
\begin{equation}
	-\frac{1}{\epsilon}\otimes
	\left(\bubInsert[bub2m12Cut1Edges]+ \bubInsert[bub2m12Cut2Edges] + 
	\bubInsert[bub2m12CutPEdges]
	\right)\,.
\end{equation}
This is precisely the combination of cuts that appears on the left-hand-side of 
eq.~\eqref{eq:pole_cancellation}. Replacing this combination by the right-hand-side of eq.~\eqref{eq:pole_cancellation} we
immediately note that the $1/\epsilon$ pole cancels and a coaction term $1\otimes I_G$ is recovered.
Beyond one loop, a similar mechanism must be at play: there must be relations between cuts that make the coaction finite despite
of the presence of UV singularities in the left entry.

For non-generic integrals, there can also be poles in $\epsilon$ of
infared origin. These must be studied on a case-by-case basis, and we refer the 
reader to ref.~\cite{Abreu:2017mtm} for examples on how
the diagrammatic coaction and the local coaction are consistent. 
Beyond one loop, however, we note that there is yet another situation to consider. 
It can be that the uncut Feynman integral is finite, 
but cuts have poles in~$\epsilon$. Indeed, the integrations
involved in computing a cut integral are closely related to phase-space integrations,
and if the cut propagators are massless they might lead to the same type 
of singularities that appear in phase-space integrations. 
As for the previous type of singularities we discussed, 
cuts must appear in specific combinations that
guarantee that all poles cancel in the formula for the coaction of a finite integral~\cite{Abreu:2014cla}.


\section{Diagrammatic coactions from hypergeometric functions \label{sec:one-loop}}

At one loop, we know from refs.~\cite{Abreu:2017enx,Abreu:2017mtm} how to construct 
a coaction on Feynman graphs with all the features outlined in the previous section. 
The situation is very different for higher loops. 
As already alluded to previously, there are many obstacles for establishing a
general diagrammatic coaction beyond one loop. 
First, while at one loop it is possible to write down a complete basis of master 
integrals for any integral, the same is not true beyond one loop, 
and one needs to determine 
the basis of master integrals from scratch for every family of integrals.\footnote{
	This should better be done in a consistent manner: if a Feynman graph
	$A$ is obtained by pinching the propagators of another graph $B$, and we have
	already constructed the basis of masters for the integral associated with $A$,
	then this information should be reused in constructing the basis for the master
	integral associated with $B$.
}
Second, the master integrals may not be expressible in terms of multiple polylogarithms. 
Indeed, it is known that starting from two loops 
a wider class of functions is required.  Finally, little is known about how to find 
relations between cuts and build a basis of integration contours at higher loops. 

In order to make progress in our understanding of the new features beyond one loop,
it is therefore important to analyse explicit examples of higher-loop integrals.
In the remainder of this paper, we will do so for some classes of two-loop integrals. As anticipated in the introduction,  
our strategy will be to use the global coaction on hypergeometric 
functions to compute the coaction of the set of master integrals associated with the chosen diagram. 
The entries of this coaction will themselves be hypergeometric functions, which can be identified as elements of the diagrammatic coaction: the left entries with a spanning set of master integrals, and the right entries as a spanning set of cuts of the integral under consideration. 
By separately computing these (cut) integrals in dimensional regularisation, and then expressing the entries in the
aforementioned global coaction in terms of them, we will explicitly construct diagrammatic 
coactions for each case considered. 
Before we apply this method at two loops, we review in this section the global coaction 
on hypergeometric functions, and how it can be used to reproduce, and even prove, the conjectured 
diagrammatic coaction of the one-loop bubble given in eq.~\eqref{diagConjBubM1}.

\subsection{The coaction on hypergeometric functions}
\label{sec:hypgeo_coac}
Let us start by defining the class of hypergeometric integrals on which the coaction of ref.~\cite{Abreu:2019wzk} can be applied. We consider integrands of the form $\omega=\Phi\varphi$, with
\begin{equation}
\label{eq:Phiphi}
\Phi({\bf u}) = \prod_{I} P_I({\bf u})^{a_I \eps}\quad \textrm{and} \quad
\varphi_{n_1\ldots n_K}({\bf u}) = d{\bf u} \prod_{I=1}^K P_I({\bf u})^{n_I}\,,
\end{equation}
where $n_I$ are integers and $P_I$ are polynomials in the kinematic variables~$x_j$ and the integration variables $u_i$ with $d\bold{u}=du_1\wedge\ldots\wedge du_n$. In the following we consider the family of integrals defined by letting the integers $n_I$ vary, 
with the $a_I$ held fixed. The framework to discuss this type of integrals is that of \emph{twisted (co)homology},
where the \emph{twist} is defined by $\Phi$ \cite{AomotoKita}---see also 
refs.~\cite{Mizera:2017rqa,Mastrolia:2018uzb,Frellesvig:2019uqt,Frellesvig:2019kgj}
for applications of \emph{twisted (co)homology} to Feynman integrals.
The distinction between
standard (co)homology and twisted (co)homology will not be crucial for most of
the discussion in this paper. 
If the structure of the polynomials $P_I({\bf u})$ is simple enough, one can construct 
explicit bases for the homology and cohomology groups associated to this integral. 
These bases can be used 
with eq.~\eqref{eq:coacGen} to construct a coaction on these hypergeometric functions. 
For more details about the construction of this coaction we refer to 
ref.~\cite{Abreu:2019wzk}. Here we simply summarise the case of Gauss' hypergeometric 
function $_2F_1$. 

Gauss' hypergeometric function admits the Euler representation
\begin{equation}\label{eq:2f1Simp}
	\,_2F_1\left(\alpha,\beta;\gamma;x\right)
	=\frac{\Gamma(\gamma)}{\Gamma(\alpha)\Gamma(\gamma-\alpha)}
	\int_0^1 u^{\alpha-1} 
	(1-u)^{\gamma-\alpha-1}
	(1-xu)^{-\beta} du\,.
\end{equation}
We will impose the condition that $\alpha=n_\alpha+a\eps$, $\b=n_\b+b\eps$ and
$\g=n_\g+c\eps$ with integer $n_\alpha$, $n_\beta$ and $n_\g$, so that
\beq\bsp
\Phi(u)& = u^{a\eps} 
	(1-u)^{(c-a)\eps}
	(1-xu)^{-b\eps}\,,\\
	 \varphi_{n_\alpha n_\beta n_\gamma}(u)& = 
u^{n_\alpha-1}(1-u)^{n_\gamma-n_\alpha-1}
	(1-xu)^{-n_\beta}\,du\,.
\esp	\eeq
Under this condition, the integral in eq.~\eqref{eq:2f1Simp} defines a meromorphic 
function of $\eps$, and  the Laurent coefficients of the expansion around 
$\epsilon=0$ are linear combinations of MPLs with rational coefficients. 

It is well known that the 
homology and cohomology groups associated with Gauss' hypergeometric function
are two-dimensional. We choose as a basis of the homology group the contours
\begin{equation}
\label{Gamma_basis_2F1}
	\gamma_1=[0,1]\,,\qquad
	\gamma_2=[0,1/x]\,,
\end{equation}
which go between two zeros of the integrand (strictly speaking, between zeros of the polynomials that define the twist). 
As a dual basis for the integrands, we take
\begin{align}
\label{omega_2F1}
\begin{split}
	&\omega_1=(c-a)\epsilon\,\Phi(u)\varphi_{101}(u) = (c-a)\epsilon\,u^{a\eps} (1-u)^{-1+(c-a)\eps}
	(1-xu)^{-b\eps}du\,,\\
	&\omega_2= -b\epsilon x\,\Phi(u)\varphi_{112}(u) = -b\epsilon x\,u^{a\eps} (1-u)^{(c-a)\eps}
	(1-xu)^{-1-b\eps}du\,.
\end{split}
\end{align}
The normalisation is chosen to satisfy the duality condition $c_{ij} = \delta_{ij}$, simplifying the form of the coaction. 
From eq.~\eqref{eq:coacGenDual} we then obtain~\cite{Abreu:2019wzk}:
\begin{align}
\nonumber\Delta\Big({}_2F_1(\a,\b;\g;x)\Big) &=
{}_2F_1(1+a\eps,b\ep;1+c\ep;x) \otimes {}_2F_1(\a,\b;\g;x) \\
\label{eq:coaction2F1}&- \frac{b\ep}{1+c\ep}\,
{}_2F_1(1+a\ep,1+b\ep;2+c\ep;x) \\
\nonumber& ~~\otimes \frac{\Gamma(1-\b)\Gamma(\g)}{\Gamma(1-\b+\a)\Gamma(\g-\a)}
x^{1-\a}{}_2F_1\left(\a,1+\a-\g;1-\b+\a;\frac{1}{x}\right)\,.
\end{align}
Let us make a few comments about this result. First, the particular form of the coaction in eq.~\eqref{eq:coaction2F1}
depends on the choice of bases made for the contours and integrands. Other
choices would lead to equivalent formulas for the coaction, related
to eq.~\eqref{eq:coaction2F1} through standard contiguous and analytic
continuation relations of each of the entries. 
Second, as already noted, whenever $n_{\alpha}, n_\beta, n_\gamma$ are integers, then Gauss' hypergeometric function
can be expanded into a Laurent series involving only MPLs. The expansion of the right-hand side in the global coaction in eq.~\eqref{eq:coaction2F1}
will be consistent with computing the local coaction on MPLs at each order in the expansion. This was conjectured in ref.~\cite{Abreu:2019wzk}, and proven
in ref.~\cite{brown2019lauricella}. Note that the right entries in our coaction, both global and local, always have to be interpreted modulo branch cuts, i.e., in the right entries any two expressions that are related by analytic continuation are considered identical. In the case of MPLs, this corresponds to working modulo $i\pi$, which is in practice how we implement this constraint in this paper. Alternatively, one can work with single-valued versions of hypergeometric functions (and MPLs), which is the approach taken in ref.~\cite{brown2019lauricella}.


\subsection{The diagrammatic coaction on one-loop integrals reloaded}
\label{sec:oneLoopRe}

Let us now discuss how we can relate the global coaction on hypergeometric functions
to the diagrammatic coaction on one-loop integrals, having in mind that we would like 
to use the same strategy to construct a diagrammatic coaction beyond one loop.
We illustrate our method in this section on the example of the one-loop bubble integral with one massive propagator. 

In eq.~\eqref{diagConjBubM1} we presented
the diagrammatic coaction of the one-loop bubble graph with $m_1\neq 0$ and $m_2=0$ in $D=2-2\eps$ space-time dimensions. 
In ref.~\cite{Abreu:2017enx,Abreu:2017mtm} it was checked that when all graphs are 
replaced by the first few orders of their Laurent expansion in $\eps$, then 
eq.~\eqref{diagConjBubM1} reproduces the (local) coaction on MPLs. Based on this 
empirical evidence, we conjectured in refs.~\cite{Abreu:2017enx,Abreu:2017mtm} that this 
should hold true to all orders in the Laurent expansion. We will now show how we can 
recover eq.~\eqref{diagConjBubM1} from the coaction on Gauss' hypergeometric function. 
Since the global and local coactions are proven to be equivalent in the case considered~\cite{brown2019lauricella}, 
this will in effect prove the conjecture of refs.~\cite{Abreu:2017enx,Abreu:2017mtm} for this integral to all orders in $\eps$.

We start from the well-known fact that the one-loop bubble integral in $D=2-2\eps$ dimensions can be evaluated in terms of Gauss' hypergeometric function (cf.,~e.g., ref.~\cite{Anastasiou:1999ui}):
\begin{equation}\bsp\label{bubPM}
J_2(p^2;m_1^2)&\, = \frac{i(p^2-m_1^2)}{2}\,\frac{e^{\gamma_E\eps}}{\pi^{1-\eps}}\int\frac{d^Dk}{(k^2-m_1^2)(k+p)^2}\\
&\,=-\frac{1}{2}
e^{\gamma_E  \epsilon } \Gamma (\epsilon) \left(m_1^2-p^2\right)^{-\epsilon}
	\, _2F_1\left(-\epsilon ,1+\epsilon ;1-\epsilon
   ;\frac{p^2}{p^2-m_1^2}\right)\, .
\esp\end{equation}
Note that we have normalised the integral
so that the coefficients in the Laurent expansion 
are pure functions (see appendix B in ref.~\cite{Abreu:2017mtm} for details).

Using eq.~(\ref{eq:coaction2F1}) and the representation in eq.~\eqref{bubPM}, 
it is straightforward to obtain the global coaction on this integral. 
We first need to specialise eq.~(\ref{eq:coaction2F1}) to the hypergeometric function 
in eq.~(\ref{bubPM}), which gives
\begin{align}
\label{2F1_for_one-mass_bubble}
\begin{split}
	\Delta\Big({}_2F_1(-\epsilon,1+\epsilon;1-\epsilon;x)\Big)&=
	(1-x)^{-\epsilon}\otimes{}_2F_1(-\epsilon,1+\epsilon;1-\epsilon;x)\\
	&-\frac{\epsilon}{1-\epsilon}{}_2F_1(1-\epsilon,1+\epsilon;2-\epsilon;x)
	\otimes \frac{2\Gamma^2(1-\epsilon)}{\Gamma(1-2\epsilon)} x^{1+\epsilon}\,.
\end{split}
\end{align}
Then, to account for the prefactor in front of the ${}_2F_1$ function in eq.~(\ref{bubPM}), we recall 
that $\Delta(f\cdot g)=\Delta(f)\cdot\Delta(g)$ and that \cite{Abreu:2017mtm,brown2019lauricella}
 \begin{equation}
\label{grouplikeFactor}
\Delta\left[x^\epsilon\,{e^{\gamma_E\eps}\Gamma(1+\eps)}\right] 
\,= 
\left[x^\epsilon{e^{\gamma_E\eps}\Gamma(1+\eps)}\right] 
\otimes 
\left[x^\epsilon{e^{\gamma_E\eps}\Gamma(1+\eps)}\right] \,.
\end{equation}
The representation of the coaction obtained by following these steps, however, has no clear
interpretation in terms of Feynman integrals. To reconcile this result with the coaction of 
eq.~\eqref{diagConjBubM1}, we recall the expressions for the cuts of this 
one-loop integral \cite{Abreu:2017ptx},
\begin{equation}\label{bubPMCutM}
	\cutRes_{e_1}
J_2(p^2;m_1^2)=\frac{e^{\gamma_E  \epsilon }}{\Gamma (1-\epsilon )}
	\frac{m_1^2-p^2}{2\,p^2 }\left(-m_1^2\right)^{-\epsilon }\, 
	_2F_1\left(1,1+\epsilon ;1-\epsilon ;\frac{m_1^2}{p^2}\right),
\end{equation}
\begin{equation}\label{bubPMCutP}
	\cutRes_{e_1,e_2}
J_2(p^2;m_1^2)=\frac{ e^{\gamma_E  \epsilon } \Gamma (1-\epsilon ) }{\Gamma (1-2 \epsilon )}
	\left(p^2\right)^{\epsilon } \left(p^2-m_1^2\right)^{-2 \epsilon }\,,
\end{equation}
and that of the tadpole integral of mass $m_1^2$ in $D=2-2\eps$ dimensions 
(and normalised to start as $1+\mathcal{O}(\epsilon)$),
\begin{equation}\label{eq:tad}
	J_1(m_1^2)=-\frac{e^{\gamma_E\eps}\Gamma(1+\eps)}{\eps}(m_1^2)^{-\epsilon}\,.
\end{equation}
One then needs to rewrite the coaction obtained by using eq.~\eqref{2F1_for_one-mass_bubble} 
in terms of these functions. This step might in general be nontrivial. 
For the Gauss hypergeometric function, however, all the analytic continuation and 
contiguous (or integration-by-parts) relations are known. 
Using such relations, we find that eq.~\eqref{2F1_for_one-mass_bubble} can equivalently
be written as
\begin{align}
\label{2F1_for_one-mass_bubble_rewrite}
\begin{split}
	\Delta\Big({}_2F_1(-\epsilon,1+\epsilon;1-\epsilon;x)\Big)&=
{}_2F_1(-\epsilon,1+\epsilon;1-\epsilon;x) 
	\otimes \frac{\Gamma^2(1-\epsilon)}{\Gamma(1-2\epsilon)} x^{\epsilon}
	\\&\hspace{-100pt}+
(1-x)^{-\epsilon}\otimes
\left(\frac{\Gamma^2(1-\epsilon)}{\Gamma(1-2\epsilon)} x^{\epsilon}
-\frac{1}{x}(1-x)^{-\epsilon}
{}_2F_1\left(1,1+\epsilon;1-\epsilon;1-\frac{1}{x}\right)\right)
\,.
\end{split}
\end{align}
Comparing this form with the results in eqs.~\eqref{bubPMCutM}, \eqref{bubPMCutP} and
\eqref{eq:tad}, it becomes clear that we reproduce the diagrammatic
coaction of eq.~\eqref{diagConjBubM1}. We emphasise that the formula for the coaction obtained in this section is not conjectural: indeed, it relies only on the coaction of Gauss' hypergeometric function conjectured in ref.~\cite{Abreu:2019wzk} and proven in ref.~\cite{brown2019lauricella}. We have therefore presented in this section a complete proof of the diagrammatic coaction in eq.~\eqref{diagConjBubM1} 
that was conjectured in ref.~\cite{Abreu:2017mtm}.

Rewriting the coaction of eq.~\eqref{2F1_for_one-mass_bubble} in a form
that made its relation to (cut) Feynman integrals apparent was easy enough in this
case, because we simply needed to show that it reproduced the diagrammatic coaction
we had previously established. Had we not known the form of the diagrammatic
coaction, however, we could have rediscovered eq.~\eqref{diagConjBubM1}
starting from eq.~\eqref{2F1_for_one-mass_bubble}.
To do this, one needs to keep in mind the general properties of the diagrammatic
coaction discussed in section \ref{sec:propDiagCoac}.
Specifically, in this case to obtain a diagrammatic interpretation of eq.~\eqref{diagConjBubM1} 
it would have been sufficient to assume that the bubble and tadpole integrals 
in eqs.~\eqref{bubPM} and \eqref{eq:tad} form a basis for the left entries of the coaction 
and that the cuts in eqs.~\eqref{bubPMCutM} and \eqref{bubPMCutP} form a basis for the right entries. 
This assumption is motivated by the form of eq.~\eqref{eq:coacGen} and the interpretation of its left entries as 
spanning master integrands and its right entries as spanning cuts of the original integral considered.
Going beyond one loop, this exercise provides us with a roadmap for constructing a coaction on 
specific multi-loop Feynman integrals, as we illustrate in a first example in the next section.


\section{The diagrammatical coaction beyond one loop: first example}
\label{sec:first-examples}

\begin{figure}[h]
	\centering
	\resizebox{4cm}{!}{
\begin{tikzpicture}[baseline={([yshift=-.5ex]current bounding box.center)}]
	\coordinate (G1) at (0,0);
	\coordinate (G2) at (1,0);
	\coordinate (H1) at (-1/3,0);
	\coordinate (H2) at (4/3,0);
	\coordinate (I1) at (1/2,1/2);
	\coordinate (I2) at (1/2,1/8);
	\coordinate (I3) at (1/2,-1/2);
	\coordinate (I4) at (1/2,-1/8);
	\coordinate (J1) at (1,1/4);
	\coordinate (J2) at (1,-1/4);
	\coordinate (K1) at (1/2,-1/8);
	\draw (G1) [line width=0.75 mm] -- (G2);
	\draw (G1) to[out=80,in=100] (G2);
	\draw (G1) to[out=-80,in=-100] (G2);
	\draw (G1) [line width=0.75 mm] -- (H1);
	\draw (G2) [line width=0.75 mm]-- (H2);
	\node at (0.65,.275) [above left=0,scale=0.6] {\small$1$};
	\node at (0.65,0) [above left=0,scale=0.6] {\small$3$};
	\node at (0.65,-.32) [above left=0,scale=0.6] {\small$2$};
\end{tikzpicture}
}
	\caption{One-mass sunset.}
	\label{fig:1mss}
\end{figure}
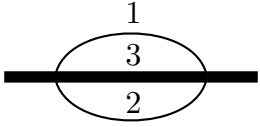

In this section we present our first example of a diagrammatic coaction
beyond one loop. We consider the sunset integral with massive external 
legs, of mass $p^2$, and a single massive propagator of mass $m^2$, see fig.~\ref{fig:1mss}.
We call this integral the \emph{one-mass sunset} to distinguish it 
from similar integrals with a different number of massive propagators,
the zero- and two-mass sunset integrals which will be discussed in section \ref{sec:Sunset}.
The one-mass sunset integral has a salient new feature compared to one-loop integrals:
there are two master integrals that share the same set of propagators. 
It is therefore interesting to see how the diagrammatic coaction applies here.
We note that in other respects this family of integrals is very simple: 
all pinches lead to integrals that vanish in dimensional regularisation. 
This makes it a particularly suitable example to begin with.
To construct the diagrammatic coaction of the one-mass sunset we will follow the approach of
section~\ref{sec:oneLoopRe}, keeping in mind the points highlighted
in section~\ref{sec:propDiagCoac}.

\paragraph{Master integrals.} 
Let us first discuss the left entries of the coaction.
As argued in section \ref{sec:coactionEntries}, these are spanned
by the master integrals associated with the sunset integral under consideration,
that is by a basis of the vector space corresponding to the integrals of the form
\begin{equation}\label{eq:ssetFamily}
S(\nu_1,\nu_2,\nu_3,\nu_4,\nu_5;D;p^2,m^2)=\left(\frac{e^{\gamma_E\epsilon}}{i\pi^{D/2}}\right)^2
\int d^Dk\,d^Dl
\frac{[(k+l)^2]^{-\nu_4}[(l+p)^2]^{-\nu_5}}{[k^2]^{\nu_1}[l^2]^{\nu_2}[(k+l+p)^2-m^2]^{\nu_3}},
\end{equation}
for integer $\nu_i$ with $\nu_4,\nu_5\leq0$ and for $D=n-2\epsilon$, with $n$ even
(the indices related to propagators appear explicitly in fig.~\ref{fig:1mss}).
This space is known to be two-dimensional, and we choose as basis elements
\begin{align}\label{SunsetMastInt1}
&S^{(1)}(p^2,m^2)=\epsilon^2\left({p^2}-{m^2}\right)
S(1,1,1,0,0;2-2\epsilon;p^2,m^2)\\
\nonumber&=(m^2)^{-2\epsilon}\left(1-\frac{p^2}{m^2}\right)e^{2\gamma_E\epsilon}
\Gamma(1+2\epsilon)\Gamma(1-\epsilon)\Gamma(1+\epsilon)
\,{}_2F_1\left(1+2\epsilon,1+\epsilon;1-\epsilon;\frac{p^2}{m^2}\right)\,,
\end{align}
and
\begin{align}\begin{split}\label{SunsetMastInt2}
 S^{(2)}(p^2,m^2)&\,=\,-\epsilon^2S(1,1,1,-1,0;2-2\epsilon;p^2,m^2)\\
&\,=\,(m^2)^{-2\epsilon}e^{2\gamma_E\epsilon}\Gamma(1+2\epsilon)
\Gamma(1-\epsilon)\Gamma(1+\epsilon)
\,{}_2F_1\left(2\epsilon,\epsilon;1-\epsilon;\frac{p^2}{m^2}\right)\,.
\end{split}\end{align}
The normalisation factors in eqs.~\eqref{SunsetMastInt1} and \eqref{SunsetMastInt2}
are chosen so that the functions are pure.
In the rest of this section we suppress the arguments of $S^{(1)}$ and
$S^{(2)}$.

\paragraph{Cut integrals.} 
Having discussed the left entries of the coaction, we now turn our attention 
to the right entries, that is to the cuts of the sunset integral.
The calculation of cut integrals beyond one loop is not yet as well
understood as at one loop, both for conceptual reasons (for instance,
very little is known about the homology group of multi-loop
integrals) and for technical reasons (i.e., the explicit calculation
of multi-loop cut integrals is still a complicated task).
The sunset integral in eq.~\eqref{eq:ssetFamily} presents a major
advantage in this respect: the cuts can be computed iteratively
loop-by-loop, and in each iteration the integrand looks
like the cut of a one-loop integral, up to a small but important 
detail which we will highlight below. Let us see this more explicitly,
and consider an integral where the three propagators of the master integral
$S^{(1)}$ are cut, which we denote by $\mathcal{C}_{1,2,3}S^{(1)}$, where
the subscripts relate to the labels of the $\nu_i$ in eq.~\eqref{eq:ssetFamily}.
By a proper parametrisation of the loop momenta, the cut can be written as 
\begin{equation}\label{OneMassSunsetMaxCutwoNormalization}
	\mathcal{C}_{1,2,3}S^{(1)}\sim\,
	\mathcal{C}_1\int\frac{d^{2-2\epsilon}k}{i\pi^{1-\epsilon}}
	\frac{1}{k^2}\,\,\left(
	\mathcal{C}_{2,3}\int\frac{d^{2-2\epsilon}l}{i\pi^{1-\epsilon}}
	\frac{1}{l^2}\frac{1}{(k+l+p)^2-m^2}\right)\,,
\end{equation}
where we use the symbol $\sim$ because we do not keep track
of overall normalisation factors which would make expressions lengthy.
We refer the reader to appendix \ref{CutsSection} for details.
The expression in parentheses is nothing but the maximal cut of a one-loop bubble integral
with a single massive propagator and a massive external leg
of mass $(k+p)^2$. We have already quoted the result
for the cut of such a bubble integral normalised to its
leading singularity in eq.~\eqref{bubPMCutP}, so we can simply reuse that
expression to get (again, we refer to appendix \ref{CutsSection} for more details):
\begin{equation}\label{eq:tempCut23}
	\mathcal{C}_{1,2,3}S^{(1)}\sim\,
	\mathcal{C}_1\int\frac{d^{2-2\epsilon}k}{i\pi^{1-\epsilon}}
	\frac{1}{k^2}\,\,\Big[(k+p)^2-m^2\Big]^{-1-2\epsilon}
	\Big[(k+p)^2\Big]^{\epsilon}\,.
\end{equation}
The integrand of the remaining integral looks very much like the integrand
of a one-loop one-mass bubble integral. We can thus use one-loop techniques~\cite{Abreu:2014cla,Abreu:2015zaa,Abreu:2017ptx} to compute
its cut, namely we can choose an explicit parametrisation of the loop momentum~(\ref{param})
to impose the remaining cut conditions $k^2=0$.
We can easily integrate over all but one component of the loop momentum $k$, which may be chosen to be the energy component $k_0$,  getting
\begin{equation}\label{eq:tempCut123}
	\mathcal{C}_{1,2,3}S^{(1)}\sim
	\int dk_0\,k_0^{-1-2\epsilon}
	\left(p^2-m^2+2\sqrt{p^2}k_0\right)^{-1-2\epsilon}
	\left(p^2+2\sqrt{p^2}k_0\right)^{2\epsilon}\,.
\end{equation}

At this stage we encounter a major difference between one-loop 
cuts and multi-loop ones. After having imposed all cut
conditions, we have not fully localised the integrand, but we have a one-dimensional integral left to perform.
The integration region over $k_0$ has not been specified in eq.~(\ref{eq:tempCut123}) because it is not determined by the cut conditions.  However, knowing that the space of master integrals is two dimensional, we also expect two independent cuts.  
Further recognising that the integrand is compatible with the general form of eq.~(\ref{eq:Phiphi}), the space of cycles to be considered is determined by the zeros of the polynomials in $k_0$, which are raised to non-integer powers (the factors defining the twist). 
Specifically, we observe that upon choosing such cycles, eq.~(\ref{eq:tempCut123}) lends itself to the 
Euler representation of the Gauss hypergeometric function in eq.~\eqref{eq:2f1Simp} by a simple change of variables.

We conclude that there are two independent cycles, which both encircle the three propagator 
poles of the sunset integral, but differ
with respect to the $k_0$ integration. We choose the basis of cycles such that
\begin{equation}
\label{Gamma12_one-masssunset_def}
	\Gamma^{(1)}_{1,2,3}:\,\,k_0\in\left[-\frac{\sqrt{p^2}}{2},0\right]\,,
	\qquad \quad
	\Gamma^{(2)}_{1,2,3}:\,\,k_0\in\left[\frac{m^2-p^2}{2\sqrt{p^2}},0\right]\,.
\end{equation}
Each of these two integration cycles defines an independent maximal cut of $S^{(1)}$.
Restoring all normalisation factors, we find that the maximal cut associated
with $\Gamma^{(1)}_{1,2,3}$ is
\begin{equation}\label{eq:c1pS1}
	\int_{\Gamma^{(1)}_{1,2,3}}\omega^{(1)}=
	2\epsilon\, e^{2\gamma_E\epsilon}
	\frac{\Gamma(1+\epsilon)\Gamma(1-\epsilon)}
	{\Gamma(1-2\epsilon)}(p^2-m^2)^{-2\epsilon}
	{}_2F_1\left(-2\epsilon,1+2\epsilon;1-\epsilon;\frac{p^2}{p^2-m^2}\right)\,,
\end{equation}
where $\omega^{(i)}$ is the differential form that defines the master integral $S^{(i)}$.
Similarly, the maximal cut associated with $\Gamma^{(2)}_{1,2,3}$ is
\begin{equation}\label{eq:c2pS1}
	\int_{\Gamma^{(2)}_{1,2,3}}\omega^{(1)}=
	4\epsilon\, e^{2\gamma_E\epsilon}\frac{\Gamma^2(1-\epsilon)}
	{\Gamma(1-4\epsilon)}(p^2)^{2\epsilon}(p^2-m^2)^{-4\epsilon}
	{}_2F_1\left(-2\epsilon,-\epsilon;-4\epsilon;1-\frac{m^2}{p^2}\right)\,.
\end{equation}

Let us conclude the discussion of the cuts with some comments.
First, we stress that the fact that there are two independent cut integrals associated with
$S^{(1)}$ is in line with the general discussion of section 
\ref{sec:coactionEntries}: given that there are two master integrals
that share the same set of propagators, there must also be
two independent maximal cuts. Second, we can similarly compute the
two independent cuts corresponding to integrating~$\omega^{(2)}$ 
over~$\Gamma^{(1)}_{1,2,3}$ or~$\Gamma^{(2)}_{1,2,3}$ of eq.~(\ref{Gamma12_one-masssunset_def}).
Finally, we observe that the bases of integrands and contours we have
chosen are not dual in the sense defined in eq.~\eqref{eq:coacGenDual}.
We can however obtain dual bases by adapting the choice of independent contours 
in the definition of the maximal cuts such that 
\begin{equation}
\int_{\gamma^{(j)}_{1,2,3}}\omega^{(i)} = \delta_{ij}+{\cal O}(\epsilon)\,.
\end{equation}
 The new contours
$\gamma^{(1)}_{1,2,3}$ and $\gamma^{(2)}_{1,2,3}$ are related to
those used in eqs.~\eqref{eq:c1pS1} and \eqref{eq:c2pS1} through
\begin{equation}\label{eq:homBasisSunset}
	\gamma^{(1)}_{1,2,3}=\frac{1}{4\epsilon}\Gamma^{(2)}_{1,2,3}\,,\qquad
	\gamma^{(2)}_{1,2,3}=\frac{1}{2\epsilon}
	\left(\Gamma^{(1)}_{1,2,3}-\frac{1}{2}\Gamma^{(2)}_{1,2,3}\right)\,,
\end{equation}
yielding, respectively, 
\begin{align}\begin{split}\label{CutsS1}
\mathcal{C}^{(1)}_{1,2,3}S^{(1)}=\,&e^{2\gamma_E\epsilon}\frac{\Gamma^2(1-\epsilon)}{\Gamma(1-4\epsilon)}(p^2)^{2\epsilon}(p^2-m^2)^{-4\epsilon}{}_2F_1\left(-2\epsilon,-\epsilon;-4\epsilon;1-\frac{m^2}{p^2}\right)\,,\\
\mathcal{C}^{(2)}_{1,2,3}S^{(1)}=\,&e^{2\gamma_E\epsilon}
	\frac{\Gamma(1+\epsilon)\Gamma(1-\epsilon)}
	{\Gamma(1-2\epsilon)}(p^2-m^2)^{-2\epsilon}
	{}_2F_1\left(-2\epsilon,1+2\epsilon;1-\epsilon;\frac{p^2}{p^2-m^2}\right)\\
&-e^{2\gamma_E\epsilon}\frac{\Gamma^2(1-\epsilon)}{\Gamma(1-4\epsilon)}(p^2)^{2\epsilon}(p^2-m^2)^{-4\epsilon}{}_2F_1\left(-2\epsilon,-\epsilon;-4\epsilon;1-\frac{m^2}{p^2}\right),
\end{split}\end{align}
for the first master integrand, and
\begin{align}\begin{split}\label{CutsS2}
\mathcal{C}^{(1)}_{1,2,3}S^{(2)}=\,&e^{2\gamma_E\epsilon}\frac{\epsilon\Gamma^2(1-\epsilon)}{2\,\Gamma(2-4\epsilon)}\frac{(p^2-m^2)^{1-4\epsilon}}{(p^2)^{1-2\epsilon}}{}_2F_1\left(1-2\epsilon,1-\epsilon;2-4\epsilon;1-\frac{m^2}{p^2}\right),\\
\mathcal{C}^{(2)}_{1,2,3}S^{(2)}
=\,&e^{2\gamma_E\epsilon}\frac{\Gamma(1+\epsilon)\Gamma(1-\epsilon)}{\Gamma(1-2\epsilon)}(p^2-m^2)^{-2\epsilon}{}_2F_1\left(1-2\epsilon,2\epsilon;1-\epsilon;\frac{p^2}{p^2-m^2}\right)\\
-&e^{2\gamma_E\epsilon}\frac{\epsilon\Gamma^2(1-\epsilon)}{2\,\Gamma(2-4\epsilon)}
\frac{(p^2-m^2)^{1-4\epsilon}}{(p^2)^{-1+2\epsilon}}{}_2F_1\left(1-2\epsilon,1-\epsilon;2-4\epsilon;1-\frac{m^2}{p^2}\right).
\end{split}\end{align}
for the second.

\paragraph{Global coaction.}

Having determined a complete set of master integrals and dual contours, the coaction is expected to take the simple form of eq.~(\ref{eq:coacGenDual}): 
\begin{equation}\label{eq:coacM1subset}
\Delta \int_{\Gamma_{\emptyset}} \omega^{(i)} =    
\int_{\Gamma_{\emptyset}} \omega^{(1)}
\otimes
\int_{\gamma^{(1)}_{1,2,3}}\omega^{(i)} 
+
\int_{\Gamma_{\emptyset}} \omega^{(2)}
\otimes
\int_{\gamma^{(2)}_{1,2,3}}\omega^{(i)} 
\end{equation}
for $i=1,\,2$. To show this we need to follow the steps taken in
section \ref{sec:oneLoopRe}. That is, we begin by computing the global coaction of each 
master integral in terms of hypergeometric functions, and then express the resulting left entries in terms of the basis of master integrals, and the right entries in terms of their dual basis of contours. 
Considering the coaction on the $S^{(i)}$ given in eqs. (\ref{SunsetMastInt1}) and (\ref{SunsetMastInt2}), 
specialising the general formula of the coaction on the Gauss hypergeometric function in eq.~(\ref{eq:coaction2F1}) to the parameters of these functions, accounting for the overall $\Gamma$-function factors using $\Delta(f\cdot g)=\Delta(f)\cdot\Delta(g)$, and finally employing known analytic continuation and contiguous relations to express the left and right entries using the aforementioned bases, we obtain the expected simple form of the coactions:
\begin{align}\label{sunsetCoaction_S12}
\begin{split}
\Delta S^{(1)}=&\,S^{(1)}\otimes\mathcal{C}_{1,2,3}^{(1)}S^{(1)}+S^{(2)}\otimes
\mathcal{C}_{1,2,3}^{(2)}S^{(1)}\,,\\
\Delta S^{(2)}=&\,S^{(1)}\otimes\mathcal{C}_{1,2,3}^{(1)}S^{(2)}+S^{(2)}\otimes
\mathcal{C}_{1,2,3}^{(2)}S^{(2)}\,,
\end{split}\end{align}
where the $\mathcal{C}_{1,2,3}^{(j)}S^{(i)}$  are given in eqs.~(\ref{CutsS1}) and (\ref{CutsS2}).
Note that the functions $\mathcal{C}_{1,2,3}^{(j)}S^{(i)}$ for fixed $i$ form a basis of the homology group, indexed by $j$, while for fixed $j$, they form a basis for the cohomology group, indexed by $i$. Indeed, the latter are contiguous functions, similarly to their uncut counterparts in eqs.~(\ref {SunsetMastInt1}) and~(\ref{SunsetMastInt2}).

\paragraph{Diagrammatic coaction.} 
The diagrammatic interpretation of the coaction for the two master integrals
of the sunset topology now readily follows: 
eq.~(\ref{sunsetCoaction_S12}) agrees with the general form of the diagrammatic 
coaction in eq.~(\ref{eq:coac_L_loop}). 
As discussed in section \ref{sec:spanCuts}, 
the fact that a master integral is no longer unambiguously identified by 
its propagators, and that a cut integral is no longer unambiguously identified  
by the cut propagators, implies that the diagrams we use to represent them
need to carry additional information.
There are six different quantities we should distinguish: the two master integrals
$S^{(1)}$ and $S^{(2)}$, the two cuts associated with the contour 
$\gamma^{(1)}_{1,2,3}$ denoted $\mathcal{C}^{(1)}_{1,2,3}S^{(1)}$
and $\mathcal{C}^{(1)}_{1,2,3}S^{(2)}$, and 
the two cuts associated with the contour 
$\gamma^{(2)}_{1,2,3}$ denoted $\mathcal{C}^{(2)}_{1,2,3}S^{(1)}$
and $\mathcal{C}^{(2)}_{1,2,3}S^{(2)}$. 
In order to unambiguously identify each of these six quantities
we introduce the following diagrams:
\begin{align}\begin{split}\label{eq:diagramsS1}
	S^{(1)}=
	\begin{tikzpicture}[baseline={([yshift=-.5ex]current bounding box.center)}]
	\coordinate (G1) at (0,0);
	\coordinate (G2) at (1,0);
	\coordinate (H1) at (-1/3,0);
	\coordinate (H2) at (4/3,0);
	\coordinate (I1) at (1/2,1/2);
	\coordinate (I2) at (1/2,1/8);
	\coordinate (I3) at (1/2,-1/2);
	\coordinate (I4) at (1/2,-1/8);
	\coordinate (J1) at (1,1/4);
	\coordinate (J2) at (1,-1/4);
	\coordinate (K1) at (1/2,-1/8);
	\draw (G1) [line width=0.75 mm] -- (G2);
	\draw (G1) to[out=80,in=100] (G2);
	\draw (G1) to[out=-80,in=-100] (G2);
	\draw (G1) [line width=0.75 mm] -- (H1);
	\draw (G2) [line width=0.75 mm]-- (H2);
	\node at (J1) [above=0 mm of J1] {\color{red}\small$(1)$};
	\node at (J2) [below=0 mm of J2] {\vphantom{\small$(1)$}};
	\end{tikzpicture}\,,&\qquad
	S^{(2)}=
	\begin{tikzpicture}[baseline={([yshift=-.5ex]current bounding box.center)}]
	\coordinate (G1) at (0,0);
	\coordinate (G2) at (1,0);
	\coordinate (H1) at (-1/3,0);
	\coordinate (H2) at (4/3,0);
	\coordinate (I1) at (1/2,1/2);
	\coordinate (I2) at (1/2,1/8);
	\coordinate (I3) at (1/2,-1/2);
	\coordinate (I4) at (1/2,-1/8);
	\coordinate (J1) at (1,1/4);
	\coordinate (J2) at (1,-1/4);
	\coordinate (K1) at (1/2,-1/8);
	\draw (G1) [line width=0.75 mm] -- (G2);
	\draw (G1) to[out=80,in=100] (G2);
	\draw (G1) to[out=-80,in=-100] (G2);
	\draw (G1) [line width=0.75 mm] -- (H1);
	\draw (G2) [line width=0.75 mm]-- (H2);
	\node at (J1) [above=0 mm of J1] {\color{blue}\small$(2)$};
	\node at (J2) [below=0 mm of J2] {\vphantom{\small$(2)$}};
	\end{tikzpicture}\,,\\[-5mm]
	\mathcal{C}^{(1)}_{1,2,3}S^{(1)}=
	\begin{tikzpicture}[baseline={([yshift=-.5ex]current bounding box.center)}]
	\coordinate (G1) at (0,0);
	\coordinate (G2) at (1,0);
	\coordinate (H1) at (-1/3,0);
	\coordinate (H2) at (4/3,0);
	\coordinate (I1) at (1/2,1/2);
	\coordinate (I2) at (1/2,1/8);
	\coordinate (I3) at (1/2,-1/2);
	\coordinate (I4) at (1/2,-1/8);
	\coordinate (J1) at (1,1/4);
	\coordinate (J2) at (1,-1/4);
	\coordinate (K1) at (1/2,-1/8);
	\draw (G1) [line width=0.75 mm] -- (G2);
	\draw (G1) to[out=80,in=100] (G2);
	\draw (G1) to[out=-80,in=-100] (G2);
	\draw (G1) [line width=0.75 mm] -- (H1);
	\draw (G2) [line width=0.75 mm]-- (H2);
	\draw (I1) [dashed,color=red,line width=0.5 mm] --(I3);
	\node at (J1) [above=0 mm of J1] {\color{red}\small$(1)$};
	\node at (J2) [below=0 mm of J2] {\vphantom{\small$(1)$}};
	\end{tikzpicture}\,,&\qquad
	\mathcal{C}^{(1)}_{1,2,3}S^{(2)}=
	\begin{tikzpicture}[baseline={([yshift=-.5ex]current bounding box.center)}]
	\coordinate (G1) at (0,0);
	\coordinate (G2) at (1,0);
	\coordinate (H1) at (-1/3,0);
	\coordinate (H2) at (4/3,0);
	\coordinate (I1) at (1/2,1/2);
	\coordinate (I2) at (1/2,1/8);
	\coordinate (I3) at (1/2,-1/2);
	\coordinate (I4) at (1/2,-1/8);
	\coordinate (J1) at (1,1/4);
	\coordinate (J2) at (1,-1/4);
	\coordinate (K1) at (1/2,-1/8);
	\draw (G1) [line width=0.75 mm] -- (G2);
	\draw (G1) to[out=80,in=100] (G2);
	\draw (G1) to[out=-80,in=-100] (G2);
	\draw (G1) [line width=0.75 mm] -- (H1);
	\draw (G2) [line width=0.75 mm]-- (H2);
	\draw (I1) [dashed,color=red,line width=0.5 mm] --(I3);
	\node at (J1) [above=0 mm of J1] {\color{blue}\small$(2)$};
	\node at (J2) [below=0 mm of J2] {\vphantom{\small$(1)$}};
	\end{tikzpicture}\,, \\[-5mm]
	\mathcal{C}^{(2)}_{1,2,3}S^{(1)}=
	\begin{tikzpicture}[baseline={([yshift=-.5ex]current bounding box.center)}]
	\coordinate (G1) at (0,0);
	\coordinate (G2) at (1,0);
	\coordinate (H1) at (-1/3,0);
	\coordinate (H2) at (4/3,0);
	\coordinate (I1) at (1/2,1/2);
	\coordinate (I2) at (1/2,1/8);
	\coordinate (I3) at (1/2,-1/2);
	\coordinate (I4) at (1/2,-1/8);
	\coordinate (J1) at (1,1/4);
	\coordinate (J2) at (1,-1/4);
	\coordinate (K1) at (1/2,-1/8);
	\draw (G1) [line width=0.75 mm] -- (G2);
	\draw (G1) to[out=80,in=100] (G2);
	\draw (G1) to[out=-80,in=-100] (G2);
	\draw (G1) [line width=0.75 mm] -- (H1);
	\draw (G2) [line width=0.75 mm]-- (H2);
	\draw (I1) [dashed,color=blue,line width=0.5 mm] --(I3);
	\node at (J1) [above=0 mm of J1] {\color{red}\small$(1)$};
	\node at (J2) [below=0 mm of J2] {\vphantom{\small$(1)$}};
	\end{tikzpicture}\,,&\qquad
	\mathcal{C}^{(2)}_{1,2,3}S^{(2)}=
	\begin{tikzpicture}[baseline={([yshift=-.5ex]current bounding box.center)}]
	\coordinate (G1) at (0,0);
	\coordinate (G2) at (1,0);
	\coordinate (H1) at (-1/3,0);
	\coordinate (H2) at (4/3,0);
	\coordinate (I1) at (1/2,1/2);
	\coordinate (I2) at (1/2,1/8);
	\coordinate (I3) at (1/2,-1/2);
	\coordinate (I4) at (1/2,-1/8);
	\coordinate (J1) at (1,1/4);
	\coordinate (J2) at (1,-1/4);
	\coordinate (K1) at (1/2,-1/8);
	\draw (G1) [line width=0.75 mm] -- (G2);
	\draw (G1) to[out=80,in=100] (G2);
	\draw (G1) to[out=-80,in=-100] (G2);
	\draw (G1) [line width=0.75 mm] -- (H1);
	\draw (G2) [line width=0.75 mm]-- (H2);
	\draw (I1) [dashed,color=blue,line width=0.5 mm] --(I3);
	\node at (J1) [above=0 mm of J1] {\color{blue}\small$(2)$};
	\node at (J2) [below=0 mm of J2] {\vphantom{\small$(1)$}};
	\end{tikzpicture}\,.
\end{split}\end{align}
In our notation the superscript $(i)$ is associated with the integrand $\omega^{(i)}$ of the respective master integral $S^{(i)}$, and we associated a colour with these indices: here (1) is red and (2) is blue.
These colours are then used to identify distinct cuts, encoding the fact that there is a 
natural association between the master integral $S^{(1)}$ and its dual contour 
$\gamma^{(1)}_{1,2,3}$, and between the master integral $S^{(2)}$ and the dual contour 
$\gamma^{(2)}_{1,2,3}$.

Using these diagrammatic rules, the diagrammatic coaction of the 
one-mass sunset integral is given by
\begin{equation}\label{eq:coactionS1}
\Delta\left[\begin{tikzpicture}[baseline={([yshift=-.5ex]current bounding box.center)}]
\coordinate (G1) at (0,0);
\coordinate (G2) at (1,0);
\coordinate (H1) at (-1/3,0);
\coordinate (H2) at (4/3,0);
\coordinate (I1) at (1/2,1/2);
\coordinate (I2) at (1/2,1/8);
\coordinate (I3) at (1/2,-1/2);
\coordinate (I4) at (1/2,-1/8);
\coordinate (J1) at (1,1/4);
\coordinate (J2) at (1,-1/4);
\coordinate (K1) at (1/2,-1/8);
\draw (G1) [line width=0.75 mm] -- (G2);
\draw (G1) to[out=80,in=100] (G2);
\draw (G1) to[out=-80,in=-100] (G2);
\draw (G1) [line width=0.75 mm] -- (H1);
\draw (G2) [line width=0.75 mm]-- (H2);
\node at (J1) [above=0 mm of J1] {\color{red}\small$(1)$};
\node at (J2) [below=0 mm of J2] {\vphantom{\small$(1)$}};
\end{tikzpicture}\right]=\begin{tikzpicture}[baseline={([yshift=-.5ex]current bounding box.center)}]
\coordinate (G1) at (0,0);
\coordinate (G2) at (1,0);
\coordinate (H1) at (-1/3,0);
\coordinate (H2) at (4/3,0);
\coordinate (I1) at (1/2,1/2);
\coordinate (I2) at (1/2,1/8);
\coordinate (I3) at (1/2,-1/2);
\coordinate (I4) at (1/2,-1/8);
\coordinate (J1) at (1,1/4);
\coordinate (J2) at (1,-1/4);
\coordinate (K1) at (1/2,-1/8);
\draw (G1) [line width=0.75 mm] -- (G2);
\draw (G1) to[out=80,in=100] (G2);
\draw (G1) to[out=-80,in=-100] (G2);
\draw (G1) [line width=0.75 mm] -- (H1);
\draw (G2) [line width=0.75 mm]-- (H2);
\node at (J1) [above=0 mm of J1] {\color{red}\small$(1)$};
\node at (J2) [below=0 mm of J2] {\vphantom{\small$(1)$}};
\end{tikzpicture}\otimes\begin{tikzpicture}[baseline={([yshift=-.5ex]current bounding box.center)}]
\coordinate (G1) at (0,0);
\coordinate (G2) at (1,0);
\coordinate (H1) at (-1/3,0);
\coordinate (H2) at (4/3,0);
\coordinate (I1) at (1/2,1/2);
\coordinate (I2) at (1/2,1/8);
\coordinate (I3) at (1/2,-1/2);
\coordinate (I4) at (1/2,-1/8);
\coordinate (J1) at (1,1/4);
\coordinate (J2) at (1,-1/4);
\coordinate (K1) at (1/2,-1/8);
\draw (G1) [line width=0.75 mm] -- (G2);
\draw (G1) to[out=80,in=100] (G2);
\draw (G1) to[out=-80,in=-100] (G2);
\draw (G1) [line width=0.75 mm] -- (H1);
\draw (G2) [line width=0.75 mm]-- (H2);
\draw (I1) [dashed,color=red,line width=0.5 mm] --(I3);
\node at (J1) [above=0 mm of J1] {\color{red}\small$(1)$};
\node at (J2) [below=0 mm of J2] {\vphantom{\small$(1)$}};
\end{tikzpicture}+\begin{tikzpicture}[baseline={([yshift=-.5ex]current bounding box.center)}]
\coordinate (G1) at (0,0);
\coordinate (G2) at (1,0);
\coordinate (H1) at (-1/3,0);
\coordinate (H2) at (4/3,0);
\coordinate (I1) at (1/2,1/2);
\coordinate (I2) at (1/2,1/8);
\coordinate (I3) at (1/2,-1/2);
\coordinate (I4) at (1/2,-1/8);
\coordinate (J1) at (1,1/4);
\coordinate (J2) at (1,-1/4);
\coordinate (K1) at (1/2,-1/8);
\draw (G1) [line width=0.75 mm] -- (G2);
\draw (G1) to[out=80,in=100] (G2);
\draw (G1) to[out=-80,in=-100] (G2);
\draw (G1) [line width=0.75 mm] -- (H1);
\draw (G2) [line width=0.75 mm]-- (H2);
\node at (J1) [above=0 mm of J1] {\color{blue}\small$(2)$};
\node at (J2) [below=0 mm of J2] {\vphantom{\small$(1)$}};
\end{tikzpicture}\otimes\begin{tikzpicture}[baseline={([yshift=-.5ex]current bounding box.center)}]
\coordinate (G1) at (0,0);
\coordinate (G2) at (1,0);
\coordinate (H1) at (-1/3,0);
\coordinate (H2) at (4/3,0);
\coordinate (I1) at (1/2,1/2);
\coordinate (I2) at (1/2,1/8);
\coordinate (I3) at (1/2,-1/2);
\coordinate (I4) at (1/2,-1/8);
\coordinate (J1) at (1,1/4);
\coordinate (J2) at (1,-1/4);
\coordinate (K1) at (1/2,-1/8);
\draw (G1) [line width=0.75 mm] -- (G2);
\draw (G1) to[out=80,in=100] (G2);
\draw (G1) to[out=-80,in=-100] (G2);
\draw (G1) [line width=0.75 mm] -- (H1);
\draw (G2) [line width=0.75 mm]-- (H2);
\draw (I1) [dashed,color=blue,line width=0.5 mm] --(I3);
\node at (J1) [above=0 mm of J1] {\color{red}\small$(1)$};
\node at (J2) [below=0 mm of J2] {\vphantom{\small$(1)$}};
\end{tikzpicture}\,,
\end{equation}
and
\begin{equation}\label{eq:coactionS2}
\Delta\left[\begin{tikzpicture}[baseline={([yshift=-.5ex]current bounding box.center)}]
\coordinate (G1) at (0,0);
\coordinate (G2) at (1,0);
\coordinate (H1) at (-1/3,0);
\coordinate (H2) at (4/3,0);
\coordinate (I1) at (1/2,1/2);
\coordinate (I2) at (1/2,1/8);
\coordinate (I3) at (1/2,-1/2);
\coordinate (I4) at (1/2,-1/8);
\coordinate (J1) at (1,1/4);
\coordinate (J2) at (1,-1/4);
\coordinate (K1) at (1/2,-1/8);
\draw (G1) [line width=0.75 mm] -- (G2);
\draw (G1) to[out=80,in=100] (G2);
\draw (G1) to[out=-80,in=-100] (G2);
\draw (G1) [line width=0.75 mm] -- (H1);
\draw (G2) [line width=0.75 mm]-- (H2);
\node at (J1) [above=0 mm of J1] {\color{blue}\small$(2)$};
\node at (J2) [below=0 mm of J2] {\vphantom{\small$(1)$}};
\end{tikzpicture}\right]=\begin{tikzpicture}[baseline={([yshift=-.5ex]current bounding box.center)}]
\coordinate (G1) at (0,0);
\coordinate (G2) at (1,0);
\coordinate (H1) at (-1/3,0);
\coordinate (H2) at (4/3,0);
\coordinate (I1) at (1/2,1/2);
\coordinate (I2) at (1/2,1/8);
\coordinate (I3) at (1/2,-1/2);
\coordinate (I4) at (1/2,-1/8);
\coordinate (J1) at (1,1/4);
\coordinate (J2) at (1,-1/4);
\coordinate (K1) at (1/2,-1/8);
\draw (G1) [line width=0.75 mm] -- (G2);
\draw (G1) to[out=80,in=100] (G2);
\draw (G1) to[out=-80,in=-100] (G2);
\draw (G1) [line width=0.75 mm] -- (H1);
\draw (G2) [line width=0.75 mm]-- (H2);
\node at (J1) [above=0 mm of J1] {\color{red}\small$(1)$};
\node at (J2) [below=0 mm of J2] {\vphantom{\small$(1)$}};
\end{tikzpicture}\otimes\begin{tikzpicture}[baseline={([yshift=-.5ex]current bounding box.center)}]
\coordinate (G1) at (0,0);
\coordinate (G2) at (1,0);
\coordinate (H1) at (-1/3,0);
\coordinate (H2) at (4/3,0);
\coordinate (I1) at (1/2,1/2);
\coordinate (I2) at (1/2,1/8);
\coordinate (I3) at (1/2,-1/2);
\coordinate (I4) at (1/2,-1/8);
\coordinate (J1) at (1,1/4);
\coordinate (J2) at (1,-1/4);
\coordinate (K1) at (1/2,-1/8);
\draw (G1) [line width=0.75 mm] -- (G2);
\draw (G1) to[out=80,in=100] (G2);
\draw (G1) to[out=-80,in=-100] (G2);
\draw (G1) [line width=0.75 mm] -- (H1);
\draw (G2) [line width=0.75 mm]-- (H2);
\draw (I1) [dashed,color=red,line width=0.5 mm] --(I3);
\node at (J1) [above=0 mm of J1] {\color{blue}\small$(2)$};
\node at (J2) [below=0 mm of J2] {\vphantom{\small$(1)$}};
\end{tikzpicture}+\begin{tikzpicture}[baseline={([yshift=-.5ex]current bounding box.center)}]
\coordinate (G1) at (0,0);
\coordinate (G2) at (1,0);
\coordinate (H1) at (-1/3,0);
\coordinate (H2) at (4/3,0);
\coordinate (I1) at (1/2,1/2);
\coordinate (I2) at (1/2,1/8);
\coordinate (I3) at (1/2,-1/2);
\coordinate (I4) at (1/2,-1/8);
\coordinate (J1) at (1,1/4);
\coordinate (J2) at (1,-1/4);
\coordinate (K1) at (1/2,-1/8);
\draw (G1) [line width=0.75 mm] -- (G2);
\draw (G1) to[out=80,in=100] (G2);
\draw (G1) to[out=-80,in=-100] (G2);
\draw (G1) [line width=0.75 mm] -- (H1);
\draw (G2) [line width=0.75 mm]-- (H2);
\node at (J1) [above=0 mm of J1] {\color{blue}\small$(2)$};
\node at (J2) [below=0 mm of J2] {\vphantom{\small$(1)$}};
\end{tikzpicture}\otimes\begin{tikzpicture}[baseline={([yshift=-.5ex]current bounding box.center)}]
\coordinate (G1) at (0,0);
\coordinate (G2) at (1,0);
\coordinate (H1) at (-1/3,0);
\coordinate (H2) at (4/3,0);
\coordinate (I1) at (1/2,1/2);
\coordinate (I2) at (1/2,1/8);
\coordinate (I3) at (1/2,-1/2);
\coordinate (I4) at (1/2,-1/8);
\coordinate (J1) at (1,1/4);
\coordinate (J2) at (1,-1/4);
\coordinate (K1) at (1/2,-1/8);
\draw (G1) [line width=0.75 mm] -- (G2);
\draw (G1) to[out=80,in=100] (G2);
\draw (G1) to[out=-80,in=-100] (G2);
\draw (G1) [line width=0.75 mm] -- (H1);
\draw (G2) [line width=0.75 mm]-- (H2);
\draw (I1) [dashed,color=blue,line width=0.5 mm] --(I3);
\node at (J1) [above=0 mm of J1] {\color{blue}\small$(2)$};
\node at (J2) [below=0 mm of J2] {\vphantom{\small$(1)$}};
\end{tikzpicture}\,.
\end{equation}
We emphasise that this diagrammatic coaction holds as a function of $\epsilon$, and to all orders in the Laurent expansion, and it is not conjectural. Indeed, it was obtained by starting from the representation of the sunset integrals and their cuts in terms of Gauss' hypergeometric function. As explained above, the diagrammatic coaction in eqs.~\eqref{eq:coactionS1} and~\eqref{eq:coactionS2} follows from the global version of the coaction on Gauss' hypergeometric function known from refs.~\cite{Abreu:2019wzk,brown2019lauricella}.

\paragraph{Relations between cuts, discontinuities and uncut integrals.}
We see that only the cuts corresponding to the contours 
$\gamma_{1,2,3}^{(i)}$, $i=1,2$, enter the diagrammatic coaction in eqs.~\eqref{eq:coactionS1}
 and~\eqref{eq:coactionS2}.
Of course, there are additional cuts we could consider, e.g., cut integrals where not all propagators are put on shell. Since there are two master integrals, and since the number of independent contours must equal the number of master integrals, these additional cut integrals must be linear combinations of the maximal cuts $\cC_{1,2,3}^{(i)}S^{(j)}$. Indeed, we find for example that
\begin{equation}\label{eq:1pcutAsCut}
	\cC_3S^{(i)} =  \mathcal{C}^{(1)}_{1,2,3}S^{(i)}-\mathcal{C}^{(2)}_{1,2,3}S^{(i)} \mod i\pi\,\,,
\end{equation}
where $\cC_3S^{(i)}$ is the integral where the (massive) propagator 3 is cut. 
We also find that the uncut Feynman integral can be written as a linear combination of its maximal cuts, 
similar to relation~\eqref{eq:pole_cancellation} for one-loop integrals:
\begin{equation}\label{eq:unctuAsCut}
	S^{(i)}=\mathcal{C}^{(1)}_{1,2,3}S^{(i)}
	+\mathcal{C}^{(2)}_{1,2,3}S^{(i)} \mod i\pi\,.
\end{equation}

It is well known that cuts of Feynman integrals are closely related to their
discontinuities \cite{Cutkosky:1960sp}. While at one loop it was trivial
to identify the functions capturing the discontinuities associated with propagator
masses or external channels in the diagrammatic coaction \cite{Abreu:2017enx,Abreu:2017mtm}, 
the situation is more complicated at two loops. First, given the discussion in 
section \ref{sec:spanCuts}, beyond one loop the diagrammatic coaction will 
never include one-propagator cuts that compute discontinuities associated with propagator
masses \cite{Abreu:2015zaa}; these are non-genuine $L$-loop cuts and are therefore not part of our basis.
Second, beyond one loop the discontinuities associated
with external channels are usually given by a linear combination of cuts,
some of which leave one or more of the subloops uncut, see e.g.~ref.~\cite{Abreu:2014cla}. Nevertheless,
given that the cuts that appear in the diagrammatic coaction form a basis for all
cuts and that discontinuities are expressible as linear combinations
of cut integrals, discontinuities can be written as linear combinations of our basis
of cuts. In the context of the sunset integral we have discussed in this section, we find
\begin{align}\begin{split}
	\textrm{Disc}_{m^2}S^{(i)}&\sim
	2\epsilon\left(\mathcal{C}^{(1)}_{1,2,3}S^{(i)}-\mathcal{C}^{(2)}_{1,2,3}S^{(i)}\right)\,,\\
	\textrm{Disc}_{p^2}S^{(i)}&\sim
	-4 \epsilon\,\mathcal{C}^{(1)}_{1,2,3}S^{(i)}\,,
\end{split}\end{align}
where the symbol $\sim$ is used because we have not defined
the operator $\textrm{Disc}$, and different definitions might vary by some 
overall normalisation.

\paragraph{Diagrammatic coaction of cut integrals.}
Let us conclude this section by commenting on the coaction on the cuts of the sunset integral.
We consider the generic contour $\Gamma(a,b)$,
corresponding to a linear combination of the two generators of the
homology group defined in eq.~\eqref{eq:homBasisSunset}:
\begin{equation}
	\Gamma(a,b)=
	a\,\gamma_{1,2,3}^{(1)}+b\,\gamma_{1,2,3}^{(2)}.
\end{equation}
Diagrammatically, we write
\begin{equation}
	\begin{tikzpicture}[baseline={([yshift=-.5ex]current bounding box.center)}]
	\coordinate (G1) at (0,0);
	\coordinate (G2) at (1,0);
	\coordinate (H1) at (-1/3,0);
	\coordinate (H2) at (4/3,0);
	\coordinate (I1) at (1/2,1/2);
	\coordinate (I2) at (1/2,1/8);
	\coordinate (I3) at (1/2,-1/2);
	\coordinate (I4) at (1/2,-1/8);
	\coordinate (J1) at (1,1/4);
	\coordinate (J2) at (1,-1/4);
	\coordinate (K1) at (1/2,-1/8);
	\draw (G1) [line width=0.75 mm] -- (G2);
	\draw (G1) to[out=80,in=100] (G2);
	\draw (G1) to[out=-80,in=-100] (G2);
	\draw (G1) [line width=0.75 mm] -- (H1);
	\draw (G2) [line width=0.75 mm]-- (H2);
	\draw (I1) [dashed,color=violet,line width=0.5 mm] --(I3);
	\node at (J1) [above=0 mm of J1] {\color{red}\small$(1)$};
	\node at (J2) [below=0 mm of J2] {\vphantom{\small$(1)$}};
	\end{tikzpicture}
	=
	a\,
	\begin{tikzpicture}[baseline={([yshift=-.5ex]current bounding box.center)}]
	\coordinate (G1) at (0,0);
	\coordinate (G2) at (1,0);
	\coordinate (H1) at (-1/3,0);
	\coordinate (H2) at (4/3,0);
	\coordinate (I1) at (1/2,1/2);
	\coordinate (I2) at (1/2,1/8);
	\coordinate (I3) at (1/2,-1/2);
	\coordinate (I4) at (1/2,-1/8);
	\coordinate (J1) at (1,1/4);
	\coordinate (J2) at (1,-1/4);
	\coordinate (K1) at (1/2,-1/8);
	\draw (G1) [line width=0.75 mm] -- (G2);
	\draw (G1) to[out=80,in=100] (G2);
	\draw (G1) to[out=-80,in=-100] (G2);
	\draw (G1) [line width=0.75 mm] -- (H1);
	\draw (G2) [line width=0.75 mm]-- (H2);
	\draw (I1) [dashed,color=red,line width=0.5 mm] --(I3);
	\node at (J1) [above=0 mm of J1] {\color{red}\small$(1)$};
	\node at (J2) [below=0 mm of J2] {\vphantom{\small$(1)$}};
	\end{tikzpicture}
	+b\,
	\begin{tikzpicture}[baseline={([yshift=-.5ex]current bounding box.center)}]
	\coordinate (G1) at (0,0);
	\coordinate (G2) at (1,0);
	\coordinate (H1) at (-1/3,0);
	\coordinate (H2) at (4/3,0);
	\coordinate (I1) at (1/2,1/2);
	\coordinate (I2) at (1/2,1/8);
	\coordinate (I3) at (1/2,-1/2);
	\coordinate (I4) at (1/2,-1/8);
	\coordinate (J1) at (1,1/4);
	\coordinate (J2) at (1,-1/4);
	\coordinate (K1) at (1/2,-1/8);
	\draw (G1) [line width=0.75 mm] -- (G2);
	\draw (G1) to[out=80,in=100] (G2);
	\draw (G1) to[out=-80,in=-100] (G2);
	\draw (G1) [line width=0.75 mm] -- (H1);
	\draw (G2) [line width=0.75 mm]-- (H2);
	\draw (I1) [dashed,color=blue,line width=0.5 mm] --(I3);
	\node at (J1) [above=0 mm of J1] {\color{red}\small$(1)$};
	\node at (J2) [below=0 mm of J2] {\vphantom{\small$(1)$}};
	\end{tikzpicture}\,.
\end{equation}
A similar relation, with the same coefficients $a$ and $b$, holds for the second master integrand. 
Keeping in mind that the change of integration contour only affects the
left entries of the coaction, see eq.~\eqref{eq:coacGen},
the coaction for this contour
follows directly from that of eq.~\eqref{eq:coactionS1}, 
\begin{equation}\label{eq:genCountour1ms}
\Delta\left[\begin{tikzpicture}[baseline={([yshift=-.5ex]current bounding box.center)}]
	\coordinate (G1) at (0,0);
	\coordinate (G2) at (1,0);
	\coordinate (H1) at (-1/3,0);
	\coordinate (H2) at (4/3,0);
	\coordinate (I1) at (1/2,1/2);
	\coordinate (I2) at (1/2,1/8);
	\coordinate (I3) at (1/2,-1/2);
	\coordinate (I4) at (1/2,-1/8);
	\coordinate (J1) at (1,1/4);
	\coordinate (J2) at (1,-1/4);
	\coordinate (K1) at (1/2,-1/8);
	\draw (G1) [line width=0.75 mm] -- (G2);
	\draw (G1) to[out=80,in=100] (G2);
	\draw (G1) to[out=-80,in=-100] (G2);
	\draw (G1) [line width=0.75 mm] -- (H1);
	\draw (G2) [line width=0.75 mm]-- (H2);
	\draw (I1) [dashed,color=violet,line width=0.5 mm] --(I3);
	\node at (J1) [above=0 mm of J1] {\color{red}\small$(1)$};
	\node at (J2) [below=0 mm of J2] {\vphantom{\small$(1)$}};
	\end{tikzpicture}\right]=\begin{tikzpicture}[baseline={([yshift=-.5ex]current bounding box.center)}]
	\coordinate (G1) at (0,0);
	\coordinate (G2) at (1,0);
	\coordinate (H1) at (-1/3,0);
	\coordinate (H2) at (4/3,0);
	\coordinate (I1) at (1/2,1/2);
	\coordinate (I2) at (1/2,1/8);
	\coordinate (I3) at (1/2,-1/2);
	\coordinate (I4) at (1/2,-1/8);
	\coordinate (J1) at (1,1/4);
	\coordinate (J2) at (1,-1/4);
	\coordinate (K1) at (1/2,-1/8);
	\draw (G1) [line width=0.75 mm] -- (G2);
	\draw (G1) to[out=80,in=100] (G2);
	\draw (G1) to[out=-80,in=-100] (G2);
	\draw (G1) [line width=0.75 mm] -- (H1);
	\draw (G2) [line width=0.75 mm]-- (H2);
	\draw (I1) [dashed,color=violet,line width=0.5 mm] --(I3);
	\node at (J1) [above=0 mm of J1] {\color{red}\small$(1)$};
	\node at (J2) [below=0 mm of J2] {\vphantom{\small$(1)$}};
	\end{tikzpicture}\otimes\begin{tikzpicture}[baseline={([yshift=-.5ex]current bounding box.center)}]
\coordinate (G1) at (0,0);
\coordinate (G2) at (1,0);
\coordinate (H1) at (-1/3,0);
\coordinate (H2) at (4/3,0);
\coordinate (I1) at (1/2,1/2);
\coordinate (I2) at (1/2,1/8);
\coordinate (I3) at (1/2,-1/2);
\coordinate (I4) at (1/2,-1/8);
\coordinate (J1) at (1,1/4);
\coordinate (J2) at (1,-1/4);
\coordinate (K1) at (1/2,-1/8);
\draw (G1) [line width=0.75 mm] -- (G2);
\draw (G1) to[out=80,in=100] (G2);
\draw (G1) to[out=-80,in=-100] (G2);
\draw (G1) [line width=0.75 mm] -- (H1);
\draw (G2) [line width=0.75 mm]-- (H2);
\draw (I1) [dashed,color=red,line width=0.5 mm] --(I3);
\node at (J1) [above=0 mm of J1] {\color{red}\small$(1)$};
\node at (J2) [below=0 mm of J2] {\vphantom{\small$(1)$}};
\end{tikzpicture}+\begin{tikzpicture}[baseline={([yshift=-.5ex]current bounding box.center)}]
	\coordinate (G1) at (0,0);
	\coordinate (G2) at (1,0);
	\coordinate (H1) at (-1/3,0);
	\coordinate (H2) at (4/3,0);
	\coordinate (I1) at (1/2,1/2);
	\coordinate (I2) at (1/2,1/8);
	\coordinate (I3) at (1/2,-1/2);
	\coordinate (I4) at (1/2,-1/8);
	\coordinate (J1) at (1,1/4);
	\coordinate (J2) at (1,-1/4);
	\coordinate (K1) at (1/2,-1/8);
	\draw (G1) [line width=0.75 mm] -- (G2);
	\draw (G1) to[out=80,in=100] (G2);
	\draw (G1) to[out=-80,in=-100] (G2);
	\draw (G1) [line width=0.75 mm] -- (H1);
	\draw (G2) [line width=0.75 mm]-- (H2);
	\draw (I1) [dashed,color=violet,line width=0.5 mm] --(I3);
	\node at (J1) [above=0 mm of J1] {\color{blue}\small$(2)$};
	\node at (J2) [below=0 mm of J2] {\vphantom{\small$(1)$}};
	\end{tikzpicture}\otimes\begin{tikzpicture}[baseline={([yshift=-.5ex]current bounding box.center)}]
\coordinate (G1) at (0,0);
\coordinate (G2) at (1,0);
\coordinate (H1) at (-1/3,0);
\coordinate (H2) at (4/3,0);
\coordinate (I1) at (1/2,1/2);
\coordinate (I2) at (1/2,1/8);
\coordinate (I3) at (1/2,-1/2);
\coordinate (I4) at (1/2,-1/8);
\coordinate (J1) at (1,1/4);
\coordinate (J2) at (1,-1/4);
\coordinate (K1) at (1/2,-1/8);
\draw (G1) [line width=0.75 mm] -- (G2);
\draw (G1) to[out=80,in=100] (G2);
\draw (G1) to[out=-80,in=-100] (G2);
\draw (G1) [line width=0.75 mm] -- (H1);
\draw (G2) [line width=0.75 mm]-- (H2);
\draw (I1) [dashed,color=blue,line width=0.5 mm] --(I3);
\node at (J1) [above=0 mm of J1] {\color{red}\small$(1)$};
\node at (J2) [below=0 mm of J2] {\vphantom{\small$(1)$}};
\end{tikzpicture}\,,
\end{equation}
and similarly for the second master integrand. 

We note that the diagrammatic coaction of eq.~~\eqref{eq:genCountour1ms}
has exactly the same structure as that in eq.~\eqref{eq:coactionS1}, 
featuring the complete set of master integrals. More generally, the expectation is that the coaction of cut Feynman integrals 
would be simpler, as only a subset of the master integrals survives under any particular cut---those that feature all propagators that are being cut (cf.~eq.~(\ref{eq:oneLoopMFDiff}) for the one-loop case).
In this respect then the one-mass sunset topology is very special: all of its pinches are zero (i.e.~there are no subtopologies), and the associated homology group is 
spanned by the two maximal cuts. As such, any contour can be written in terms of the maximal cuts,
see e.g.~eqs.~\eqref{eq:1pcutAsCut} and \eqref{eq:unctuAsCut}, and they do not set to zero any of the master integrals in the topology. The coaction of any cut of the one-mass sunset will thus always have
the same structure as eqs.~\eqref{eq:coactionS1} and \eqref{eq:genCountour1ms}.


\section{Coactions of further two-loop Feynman integrals\label{sec:Two-loop_examples}}

In this section we present further examples of diagrammatic coactions at two loops.
While we are still far from having a complete picture as we do at one loop, these
examples are aimed at showing that a diagrammatic coaction exists (at least) for a wide
variety of two-loop examples. We will consider up to four-point diagrams with up to 
five propagators, which will allow us to have examples with up to six master integrals.
In all cases the diagrammatic coactions will be obtained with the approach outlined
in the previous section, i.e., we will start from the global coaction of the uncut
integral (in all cases studied here, these integrals evaluate to hypergeometric 
functions considered in ref.~\cite{Abreu:2019wzk}
and their global coaction is known) and match that coaction to a sum over tensor 
products of master integrals and cuts. 
We will also discuss how the diagrammatic coaction is consistent with taking massless limits.

As was made clear in section \ref{sec:spanCuts}, the diagrammatic coaction beyond
one loop requires extra decorations on Feynman graphs corresponding to the index
$i$ on the right-hand side of eq.~\eqref{eq:coac_L_loop}. Throughout this section
we use the same notation used in eqs.~\eqref{eq:coactionS1} and \eqref{eq:coactionS2}.
That is, we encode this information in an explicit superscript where a given index value and a corresponding colour uniquely specify the master integrand; the colour then specifies the dual contour associated with each master integrand, and it is used in drawing the 
dashed lines cutting through the relevant subset of propagators on the right entry. 
When this degeneracy is not present we revert to the notation we used
at one loop, where all cut propagators are denoted by red dashed lines.

All the examples listed in this section are by construction consistent with the global coaction,
which was conjectured to be consistent with the local coaction on MPLs in 
ref.~\cite{Abreu:2019wzk}. In all cases, we have verified explicitly that the local coaction of 
every master integral is indeed consistent with the $\epsilon$ expansion of the entries of 
its diagrammatic coaction up to order $\epsilon^4$, i.e., up to weight four.

\subsection{Double tadpole}\label{sec:doubTad}
Let us first consider an example of a two-loop integral that is simply the product of two
one-loop integrals, namely the product of two tadpoles. We have already quoted the expression
for the one-loop tadpole of mass $m^2$ in eq.~\eqref{eq:tad}. The double tadpole with masses
$m_1^2$ and $m_2^2$ (evaluated in $2-2\epsilon$ dimensions, and normalised to start
as $1+\mathcal{O}(\epsilon)$) is then
\begin{equation}\label{eq:tadsq}
	J(m_1^2,m_2^2)=e^{2\epsilon\gamma_E}\Gamma^{2}(1+\epsilon)(m_1^2)^{-\epsilon}(m_2^2)^{-\epsilon}\,.
\end{equation}
Its maximal cut is also given by the product of the one-loop expressions \cite{Abreu:2017mtm}:
\begin{equation}\label{eq:tadsqCut}
	\mathcal{C}_{1,2}J(m_1^2,m_2^2)=\frac{e^{2\epsilon\gamma_E}}{\Gamma^{2}(1-\epsilon)}
	(-m_1^2)^{-\epsilon}(-m_2^2)^{-\epsilon}\,.
\end{equation}
We note that, modulo $i\pi$, the maximal cut and the uncut integral are in fact equal
(assuming that they are both normalised to start as $1+\mathcal{O}(\epsilon)$).
More precisely, if we consider the expansion in $\epsilon$ of eqs.~\eqref{eq:tadsq} and
\eqref{eq:tadsqCut} and set to zero all powers of $\pi$, we find that the two
expressions agree.\footnote{\label{fn:trivial}
It is clear that this must happen from the
perspective of their global coaction, see e.g.~section~2.2 of ref.~\cite{Abreu:2019wzk}.
} The global coaction of the double-tadpole is therefore very simple, and the diagrammatic coaction is,
\begin{align}
\Delta\left[\begin{tikzpicture}[baseline={([yshift=-.5ex]current bounding box.center)}]
\coordinate (G1) at (0,0);
\coordinate (G2) at (0,4/5);
\coordinate (G3) at (0,-4/5);
\coordinate (H1) at (-1/2,0);
\coordinate (H2) at (1/2,0);
\draw (G1) [line width=0.75 mm]to[out=120,in=180] (G2);
\draw (G2) [line width=0.75 mm]to[out=0,in=60] (G1);
\draw (G1) [line width=0.75 mm]to[out=-120,in=180] (G3);
\draw (G3) [line width=0.75 mm]to[out=0,in=-60] (G1);
\draw (G1) [line width=0.75 mm] -- (H1);
\draw (G1) [line width=0.75 mm]-- (H2);
\end{tikzpicture}\right]=\begin{tikzpicture}[baseline={([yshift=-.5ex]current bounding box.center)}]
\coordinate (G1) at (0,0);
\coordinate (G2) at (0,4/5);
\coordinate (G3) at (0,-4/5);
\coordinate (H1) at (-1/2,0);
\coordinate (H2) at (1/2,0);
\draw (G1) [line width=0.75 mm]to[out=120,in=180] (G2);
\draw (G2) [line width=0.75 mm]to[out=0,in=60] (G1);
\draw (G1) [line width=0.75 mm]to[out=-120,in=180] (G3);
\draw (G3) [line width=0.75 mm]to[out=0,in=-60] (G1);
\draw (G1) [line width=0.75 mm] -- (H1);
\draw (G1) [line width=0.75 mm]-- (H2);
\end{tikzpicture}\otimes\begin{tikzpicture}[baseline={([yshift=-.5ex]current bounding box.center)}]
\coordinate (G1) at (0,0);
\coordinate (G2) at (0,4/5);
\coordinate (G3) at (0,-4/5);
\coordinate (H1) at (-1/2,0);
\coordinate (H2) at (1/2,0);
\coordinate (I1) at (0,2.5/5);
\coordinate (I2) at (0,5.5/5);
\coordinate (I3) at (0,-2.5/5);
\coordinate (I4) at (0,-5.5/5);
\draw (G1) [line width=0.75 mm]to[out=120,in=180] (G2);
\draw (G2) [line width=0.75 mm]to[out=0,in=60] (G1);
\draw (G1) [line width=0.75 mm]to[out=-120,in=180] (G3);
\draw (G3) [line width=0.75 mm]to[out=0,in=-60] (G1);
\draw (G1) [line width=0.75 mm] -- (H1);
\draw (G1) [line width=0.75 mm]-- (H2);
\draw (I1) [dashed,color=red,line width=0.5 mm] --(I2);
\draw (I3) [dashed,color=red,line width=0.5 mm] --(I4);
\end{tikzpicture}\,.
\end{align}
This agrees with the way the coaction acts on products of functions, 
$\Delta(f\cdot g)=\Delta(f)\cdot\Delta(g)$.
We note that the same approach can be used to compute the coaction
of any multi-loop integral that is just a product of one-loop integrals. It is clear
that there is a diagrammatic coaction for those cases, which follows in a straightforward
way from the diagrammatic coaction at one loop.

 \subsection{Sunset}
\label{sec:Sunset}

We next consider genuine two-loop diagrams
similar to those studied in section~\ref{sec:first-examples}.
We will investigate the case where all propagators are massless, a trivial case but one
which appears as a master integral corresponding to a subtopology of examples we will encounter later, and the case where two propagators are massive. Both cases evaluate to 
hypergeometric-type integrals which have been considered in ref.~\cite{Abreu:2019wzk}, 
which means that their global coaction is known and we can follow the approach
of sections~\ref{sec:oneLoopRe} and \ref{sec:first-examples} to construct their
diagrammatic coaction.

\subsubsection{Massless propagators}\label{sec:sunsetMassless}

We begin with the coaction on the sunset with no internal masses, associated
with integrals of the form
\begin{equation}\label{eq:ssetFamily0}
S(\nu_1,\nu_2,\nu_3,\nu_4,\nu_5;D;p^2)=\left(\frac{e^{\gamma_E\epsilon}}{i\pi^{D/2}}\right)^2
\int d^Dk\,d^Dl
\frac{[(k+l)^2]^{-\nu_4}[(l+p)^2]^{-\nu_5}}{[k^2]^{\nu_1}[l^2]^{\nu_2}[(k+l+p)^2]^{\nu_3}},
\end{equation}
for integer $\nu_i$ with $\nu_4,\nu_5\leq0$ and for $D=n-2\epsilon$, with $n$ even.
It is well known that there is a single
master integral associated with this topology. We choose the scalar integral
evaluated in $D=2-2\epsilon$ dimensions, normalised to start as $1+\mathcal{O}(\epsilon)$.
More concretely, we take
\begin{align}\begin{split}\label{SunsetZeroMassMaster}
S(p^2)=&-\frac{p^2}{3}\epsilon^2 S(1,1,1,0,0;2-2\epsilon;p^2)=(-p^2)^{-2\epsilon}e^{2\gamma_E\epsilon}
\frac{\Gamma^3(1-\epsilon)\Gamma(1+2\epsilon)}{\Gamma(1-3\epsilon)}\,.
\end{split}\end{align}
The cut integral can be computed in a straightforward loop-by-loop approach,
as was done in the example of section~\ref{sec:first-examples}.
As for the double tadpole, if the maximal cut is normalised in the same way, we find
that it is equal to the uncut integral modulo $i\pi$ (see footnote \ref{fn:trivial}).
It is then easy to see that the massless sunset satisfies the diagrammatic coaction
\begin{align}\label{eq:massless_sunset}
\Delta\left[\begin{tikzpicture}[baseline={([yshift=-.5ex]current bounding box.center)}]
\coordinate (G1) at (0,0);
\coordinate (G2) at (1,0);
\coordinate (H1) at (-1/3,0);
\coordinate (H2) at (4/3,0);
\coordinate (I1) at (1/2,1/2);
\coordinate (I2) at (1/2,1/8);
\coordinate (I3) at (1/2,-1/2);
\coordinate (I4) at (1/2,-1/8);
\coordinate (J1) at (1,1/4);
\coordinate (J2) at (1,-1/4);
\coordinate (K1) at (1/2,-1/8);
\draw (G1) -- (G2);
\draw (G1) to[out=80,in=100] (G2);
\draw (G1) to[out=-80,in=-100] (G2);
\draw (G1) [line width=0.75 mm] -- (H1);
\draw (G2) [line width=0.75 mm]-- (H2);
\end{tikzpicture}\right]=\begin{tikzpicture}[baseline={([yshift=-.5ex]current bounding box.center)}]
\coordinate (G1) at (0,0);
\coordinate (G2) at (1,0);
\coordinate (H1) at (-1/3,0);
\coordinate (H2) at (4/3,0);
\coordinate (I1) at (1/2,1/2);
\coordinate (I2) at (1/2,1/8);
\coordinate (I3) at (1/2,-1/2);
\coordinate (I4) at (1/2,-1/8);
\coordinate (J1) at (1,1/4);
\coordinate (J2) at (1,-1/4);
\coordinate (K1) at (1/2,-1/8);
\draw (G1) -- (G2);
\draw (G1) to[out=80,in=100] (G2);
\draw (G1) to[out=-80,in=-100] (G2);
\draw (G1) [line width=0.75 mm] -- (H1);
\draw (G2) [line width=0.75 mm]-- (H2);
\end{tikzpicture}\otimes\begin{tikzpicture}[baseline={([yshift=-.5ex]current bounding box.center)}]
\coordinate (G1) at (0,0);
\coordinate (G2) at (1,0);
\coordinate (H1) at (-1/3,0);
\coordinate (H2) at (4/3,0);
\coordinate (I1) at (1/2,1/2);
\coordinate (I2) at (1/2,1/8);
\coordinate (I3) at (1/2,-1/2);
\coordinate (I4) at (1/2,-1/8);
\coordinate (J1) at (1,1/4);
\coordinate (J2) at (1,-1/4);
\coordinate (K1) at (1/2,-1/8);
\draw (G1) -- (G2);
\draw (G1) to[out=80,in=100] (G2);
\draw (G1) to[out=-80,in=-100] (G2);
\draw (G1) [line width=0.75 mm] -- (H1);
\draw (G2) [line width=0.75 mm]-- (H2);
\draw (I1) [dashed,color=red,line width=0.5 mm] --(I3);
\end{tikzpicture}\,.
\end{align}

\subsubsection{Two massive propagators}\label{sec:sunsetTwoMass}
We now consider the more general sunset integral where there are two non-vanishing
(and non-equal) internal masses. This topology is defined by the set of integrals of the form
\begin{align}\begin{split}\label{2mSunsetIntFamily}
&S(\nu_1,\nu_2,\nu_3,\nu_4,\nu_5;D;p^2,m_1^2,m_2^2)=\\
&\hspace*{40pt}=\left(\frac{e^{\gamma_E\epsilon}}{i\pi^{D/2}}\right)^2
\int d^Dk\int d^Dl\,
\frac{[m_2^2-(k+p)^2]^{-\nu_4}[m_1^2-(l+p)^2]^{-\nu_5}}
{[k^2-m_1^2]^{\nu_1}[l^2-m_2^2]^{\nu_2}[(k+l+p)^2]^{\nu_3}}\,,
\end{split}\end{align}
for integer $\nu_i$ with $\nu_4,\nu_5\leq0$ and for $D=n-2\epsilon$, with $n$ even.
There are four master integrals
in this topology. Out of these four, there are three master integrals with
three propagators, which we choose to be
\begin{align}\begin{split}\label{eq:2massBasis}
	S^{(1)}(p^2,m_1^2,m_2^2)=-\epsilon^2
	e^{2\gamma_E\epsilon}\int\frac{d^{2-2\epsilon}k}{i\pi^{1-\epsilon}}
	\int\frac{d^{2-2\epsilon}l}{i\pi^{1-\epsilon}}
	\frac{\sqrt{\lambda\left(p^2,m_1^2,m_2^2\right)}}{(k^2-m_1^2)(l^2-m_2^2)(k+l+p)^2}\,,\\
	S^{(2)}(p^2,m_1^2,m_2^2)=\epsilon^2
	e^{2\gamma_E\epsilon}\int\frac{d^{2-2\epsilon}k}{i\pi^{1-\epsilon}}
	\int\frac{d^{2-2\epsilon}l}{i\pi^{1-\epsilon}}
	\frac{m_2^2-(k+p)^2}{(k^2-m_1^2)(l^2-m_2^2)(k+l+p)^2}\,,\\
	S^{(3)}(p^2,m_1^2,m_2^2)=\epsilon^2
	e^{2\gamma_E\epsilon}\int\frac{d^{2-2\epsilon}k}{i\pi^{1-\epsilon}}
	\int\frac{d^{2-2\epsilon}l}{i\pi^{1-\epsilon}}
	\frac{m_1^2-(l+p)^2}{(k^2-m_1^2)(l^2-m_2^2)(k+l+p)^2}\,,
\end{split}\end{align}
where $\lambda(a,b,c)=a^2+b^2+c^2-2ab-2ac-2bc$ is the usual K\"{a}llen function.
The normalisation is chosen so that the master integrals are pure functions, 
and the weight of the MPLs in the coefficients at order $\epsilon^k$ is $k$.
The fourth master integral is the product of two tadpoles considered in section~\ref{sec:doubTad}.
As usual, for simplicity we will drop the dependence of the $S^{(i)}$ on the kinematic variables
whenever there is no ambiguity.
We give explicit expressions for the three integrals of eq.~\eqref{eq:2massBasis} in 
 appendix~\ref{sec:appTwoMassExp}. These can be written in terms of Appell $F_4$ functions
(see e.g. ref.~\cite{handbook}), which is the most complicated type of hypergeometric
functions considered in ref.~\cite{Abreu:2019wzk}. The global coaction
of the two-mass sunset integral can be directly obtained with the results presented
there.

As in the case of the one-mass sunset integral of eq.~\eqref{eq:tempCut123}, 
the maximal-cut conditions do not completely determine the result of the maximal cut. 
Consistently with the number of master integrals of the top topology (those in  eq.~(\ref{eq:2massBasis})), 
there are three independent maximal cuts, and we denote the corresponding contours by $\Gamma_{1,2,3}^{(i)}$, 
where $i$ ranges from $1$ to $3$.
We compute these cuts following a loop-by-loop approach, similar to the one used in section~\ref{sec:first-examples} for the one-mass sunset case, making use of the one-loop techniques of 
refs.~\cite{Abreu:2014cla,Abreu:2015zaa,Abreu:2017ptx}. To this end we introduce an explicit 
parametrisation of the loop momenta in terms of spherical coordinates, 
which we use to impose the cut conditions. We refer the reader to appendices~\ref{CutsSection}
and \ref{DiffEqCutsSection} for more details on the calculation, and a discussion of two
different representations for these cuts, either in terms of Appell $F_1$ functions or
in terms of Appell $F_4$ functions. Both representations are useful as they highlight different properties
of the cut integrals. 
We collect explicit expressions for the maximal cuts in appendix~\ref{sec:appTwoMassExp}.

Finally, there is another independent contour $\Gamma_{1,2}$ 
which only encircles the poles of the two massive propagators. This is the same contour  
defining the maximal cut of the double tadpole $J$ in eq.~\eqref{eq:tadsqCut}.
Details on this calculation can be found in appendix \ref{CutsSection}, and explicit
results for the cuts associated with this contour are listed in 
appendix~\ref{sec:appTwoMassExp}.

Starting from the global coaction on the Appell $F_4$ functions \cite{Abreu:2019wzk}, we can
obtain the coaction on each of the three three-propagator master integrals
of the sunset topology with two massive propagators.\footnote{We choose to apply the coaction in a symmetric basis of Appell $F_4$ functions, depending on the kinematic variables through the ratios $\frac{m_1^2}{p^2}$ and $\frac{m_2^2}{p^2}$. To this end one needs to first apply the analytic continuation relation (\ref{AnalContRelations}) to the expressions quoted in eqs.~(\ref{STwoMasses}), (\ref{STwoMasses2}) and (\ref{STwoMasses3}). The right entries in the coaction can then be expressed in terms of the cuts quoted in appendix~\ref{sec:appTwoMassExp}.} We find
\begin{align}\begin{split}\label{DeltaJ}
\Delta S^{(1)}=&\,J\otimes\left(\mathcal{C}_{1,2}S^{(1)}+\mathcal{C}_{1,2,3}^{(1)}S^{(1)}+
\mathcal{C}_{1,2,3}^{(2)}S^{(1)}+\mathcal{C}_{1,2,3}^{(3)}S^{(1)}\right)\\
&+S^{(1)}\otimes\mathcal{C}_{1,2,3}^{(1)}S^{(1)}+S^{(2)}\otimes
\mathcal{C}_{1,2,3}^{(2)}S^{(1)}+S^{(3)}\otimes\mathcal{C}_{1,2,3}^{(3)}S^{(1)}\,,\\
\Delta S^{(2)}=&\,J\otimes\left(\mathcal{C}_{1,2}S^{(2)}+\mathcal{C}_{1,2,3}^{(1)}S^{(2)}+
\mathcal{C}_{1,2,3}^{(2)}S^{(2)}+\mathcal{C}_{1,2,3}^{(3)}S^{(2)}\right)\\
&+S^{(1)}\otimes\mathcal{C}_{1,2,3}^{(1)}S^{(2)}+S^{(2)}\otimes
\mathcal{C}_{1,2,3}^{(2)}S^{(2)}+S^{(3)}\otimes\mathcal{C}_{1,2,3}^{(3)}S^{(2)}\,,\\
\Delta S^{(3)}=&\,J\otimes\left(\mathcal{C}_{1,2}S^{(3)}+\mathcal{C}_{1,2,3}^{(1)}S^{(3)}+
\mathcal{C}_{1,2,3}^{(2)}S^{(3)}+\mathcal{C}_{1,2,3}^{(3)}S^{(3)}\right)\\
&+S^{(1)}\otimes\mathcal{C}_{1,2,3}^{(1)}S^{(3)}+S^{(2)}\otimes
\mathcal{C}_{1,2,3}^{(2)}S^{(3)}+S^{(3)}\otimes\mathcal{C}_{1,2,3}^{(3)}S^{(3)}\,,
\end{split}\end{align}
where $J$ is the two-loop tadpole defined in eq.~\eqref{eq:tadsq}.
Generalizing the diagrammatic notation of eq.~\eqref{eq:diagramsS1} in a straightforward
way, we obtain the diagrammatic coactions for the two-mass sunset: 
\begin{align}\begin{split}
\label{eq:Sunset2Mass1}
&\Delta\left[
\,.
\end{split}
\end{align}
We stress that, as for previous examples, these coactions are 
simply obtained as a diagrammatic representation of the global
coaction on Appell $F_4$ functions. If the conjectured formula
for the $F_4$ coaction in ref.~\cite{Abreu:2019wzk} is proven to all orders in $\epsilon$, then 
so will eqs.~\eqref{eq:Sunset2Mass1}, \eqref{eq:Sunset2Mass2} and \eqref{eq:Sunset2Mass3}.
The complexity of the functions involved in
these Feynman integrals makes this a particularly nontrivial example, and we find that 
all the properties highlighted in section~\ref{sec:propDiagCoac} are satisfied.

Let us make some comments about the coactions in eqs.~\eqref{eq:Sunset2Mass1}, \eqref{eq:Sunset2Mass2} and \eqref{eq:Sunset2Mass3}. First, we see that the terms in the coaction that have a double tadpole integral in the left entry have cut integrals where two or three propagators are cut in the right entry. The appearance of so-called deformation terms in the right entry where more propagators are cut than those present in the master integral in the left entry is not surprising: this is consistent with eqs.~(\ref{gammaC_sum_Gamma}) and~(\ref{eq:coac_L_loop}) and is familiar from the one-loop case, see e.g.~eq.~\eqref{diagConjBubM1M2}, where these terms can be traced back to relations among the homology generators (cf.~eq.~\eqref{eq:one-loop contours}). 
Second, we have checked that upon sending one or both masses to zero, eqs.~\eqref{eq:Sunset2Mass1}, \eqref{eq:Sunset2Mass2} and \eqref{eq:Sunset2Mass3} reduce to the corresponding diagrammatic coactions for the one- or zero-mass sunset integrals in eqs.~\eqref{eq:coactionS1}, \eqref{eq:coactionS2} and~\eqref{eq:massless_sunset}. We note that for this reduction to be possible, some of the terms in the coaction need to be rearranged, because the number of master integrals changes as masses are sent to zero. We will illustrate this rearrangement in detail in the next section on the example of the double-edged triangle integral.

To close the discussion of the two-mass sunset integral, we comment on 
the coaction of its cuts. As an example, we consider a cut integral
given by a linear combination of the three maximal cuts
(but which does not involve the two-propagator cut), represented
diagrammatically as
\begin{equation}
	\begin{tikzpicture}[baseline={([yshift=-.5ex]current bounding box.center)}]
	\coordinate (G1) at (0,0);
	\coordinate (G2) at (1,0);
	\coordinate (H1) at (-1/3,0);
	\coordinate (H2) at (4/3,0);
	\coordinate (I1) at (1/2,1/2);
	\coordinate (I2) at (1/2,1/8);
	\coordinate (I3) at (1/2,-1/2);
	\coordinate (I4) at (1/2,-1/8);
	\coordinate (J1) at (1,1/4);
	\coordinate (J2) at (1,-1/4);
	\coordinate (K1) at (1/2,-1/8);
	\draw (G1) -- (G2);
	\draw (G1) [line width=0.75 mm]to[out=80,in=100] (G2);
	\draw (G1) [line width=0.75 mm]to[out=-80,in=-100] (G2);
	\draw (G1) [line width=0.75 mm] -- (H1);
	\draw (G2) [line width=0.75 mm]-- (H2);
	\draw (I1) [dashed,color=violet,line width=0.5 mm] --(I3);
	\node at (J1) [above=0 mm of J1] {\color{orange}\small$(1)$};
	\node at (J2) [below=0 mm of J2] {\vphantom{\small$(3)$}};
	\end{tikzpicture}
	=a\,
	\begin{tikzpicture}[baseline={([yshift=-.5ex]current bounding box.center)}]
	\coordinate (G1) at (0,0);
	\coordinate (G2) at (1,0);
	\coordinate (H1) at (-1/3,0);
	\coordinate (H2) at (4/3,0);
	\coordinate (I1) at (1/2,1/2);
	\coordinate (I2) at (1/2,1/8);
	\coordinate (I3) at (1/2,-1/2);
	\coordinate (I4) at (1/2,-1/8);
	\coordinate (J1) at (1,1/4);
	\coordinate (J2) at (1,-1/4);
	\coordinate (K1) at (1/2,-1/8);
	\draw (G1) -- (G2);
	\draw (G1) [line width=0.75 mm]to[out=80,in=100] (G2);
	\draw (G1) [line width=0.75 mm]to[out=-80,in=-100] (G2);
	\draw (G1) [line width=0.75 mm] -- (H1);
	\draw (G2) [line width=0.75 mm]-- (H2);
	\draw (I1) [dashed,color=orange,line width=0.5 mm] --(I3);
	\node at (J1) [above=0 mm of J1] {\color{orange}\small$(1)$};
	\node at (J2) [below=0 mm of J2] {\vphantom{\small$(3)$}};
	\end{tikzpicture}
	+b\,
	\begin{tikzpicture}[baseline={([yshift=-.5ex]current bounding box.center)}]
	\coordinate (G1) at (0,0);
	\coordinate (G2) at (1,0);
	\coordinate (H1) at (-1/3,0);
	\coordinate (H2) at (4/3,0);
	\coordinate (I1) at (1/2,1/2);
	\coordinate (I2) at (1/2,1/8);
	\coordinate (I3) at (1/2,-1/2);
	\coordinate (I4) at (1/2,-1/8);
	\coordinate (J1) at (1,1/4);
	\coordinate (J2) at (1,-1/4);
	\coordinate (K1) at (1/2,-1/8);
	\draw (G1) -- (G2);
	\draw (G1) [line width=0.75 mm]to[out=80,in=100] (G2);
	\draw (G1) [line width=0.75 mm]to[out=-80,in=-100] (G2);
	\draw (G1) [line width=0.75 mm] -- (H1);
	\draw (G2) [line width=0.75 mm]-- (H2);
	\draw (I1) [dashed,color=blue,line width=0.5 mm] --(I3);
	\node at (J1) [above=0 mm of J1] {\color{orange}\small$(1)$};
	\node at (J2) [below=0 mm of J2] {\vphantom{\small$(3)$}};
	\end{tikzpicture}
	+c\,
	\begin{tikzpicture}[baseline={([yshift=-.5ex]current bounding box.center)}]
	\coordinate (G1) at (0,0);
	\coordinate (G2) at (1,0);
	\coordinate (H1) at (-1/3,0);
	\coordinate (H2) at (4/3,0);
	\coordinate (I1) at (1/2,1/2);
	\coordinate (I2) at (1/2,1/8);
	\coordinate (I3) at (1/2,-1/2);
	\coordinate (I4) at (1/2,-1/8);
	\coordinate (J1) at (1,1/4);
	\coordinate (J2) at (1,-1/4);
	\coordinate (K1) at (1/2,-1/8);
	\draw (G1) -- (G2);
	\draw (G1) [line width=0.75 mm]to[out=80,in=100] (G2);
	\draw (G1) [line width=0.75 mm]to[out=-80,in=-100] (G2);
	\draw (G1) [line width=0.75 mm] -- (H1);
	\draw (G2) [line width=0.75 mm]-- (H2);
	\draw (I1) [dashed,color=black!60!green,line width=0.5 mm] --(I3);
	\node at (J1) [above=0 mm of J1] {\color{orange}\small$(1)$};
	\node at (J2) [below=0 mm of J2] {\vphantom{\small$(3)$}};
	\end{tikzpicture}\,,
\end{equation}
and similarly for the other two master integrands. To construct the coaction on this generic
maximal cut, we first note that it sets to zero the double tadpole since this subtopology
does not feature one of the cut propagators. Then, we recall that a change of integration
contour only affects the left entries of the coaction, see eq.~\eqref{eq:coacGen}, and
it then follows from eq.~\eqref{eq:Sunset2Mass1} that
\begin{align}\begin{split}
\Delta\left[\begin{tikzpicture}[baseline={([yshift=-.5ex]current bounding box.center)}]
\coordinate (G1) at (0,0);
\coordinate (G2) at (1,0);
\coordinate (H1) at (-1/3,0);
\coordinate (H2) at (4/3,0);
\coordinate (I1) at (1/2,1/2);
\coordinate (I2) at (1/2,1/8);
\coordinate (I3) at (1/2,-1/2);
\coordinate (I4) at (1/2,-1/8);
\coordinate (J1) at (1,1/4);
\coordinate (J2) at (1,-1/4);
\coordinate (K1) at (1/2,-1/8);
\draw (G1) -- (G2);
\draw (G1) [line width=0.75 mm]to[out=80,in=100] (G2);
\draw (G1) [line width=0.75 mm]to[out=-80,in=-100] (G2);
\draw (G1) [line width=0.75 mm] -- (H1);
\draw (G2) [line width=0.75 mm]-- (H2);
\draw (I1) [dashed,color=violet,line width=0.5 mm] --(I3);
\node at (J1) [above=0 mm of J1] {\color{orange}\small$(1)$};
\node at (J2) [below=0 mm of J2] {\vphantom{\small$(1)$}};
\end{tikzpicture}\right]
&\,=\begin{tikzpicture}[baseline={([yshift=-.5ex]current bounding box.center)}]
\coordinate (G1) at (0,0);
\coordinate (G2) at (1,0);
\coordinate (H1) at (-1/3,0);
\coordinate (H2) at (4/3,0);
\coordinate (I1) at (1/2,1/2);
\coordinate (I2) at (1/2,1/8);
\coordinate (I3) at (1/2,-1/2);
\coordinate (I4) at (1/2,-1/8);
\coordinate (J1) at (1,1/4);
\coordinate (J2) at (1,-1/4);
\coordinate (K1) at (1/2,-1/8);
\draw (G1) -- (G2);
\draw (G1) [line width=0.75 mm]to[out=80,in=100] (G2);
\draw (G1) [line width=0.75 mm]to[out=-80,in=-100] (G2);
\draw (G1) [line width=0.75 mm] -- (H1);
\draw (G2) [line width=0.75 mm]-- (H2);
\draw (I1) [dashed,color=violet,line width=0.5 mm] --(I3);
\node at (J1) [above=0 mm of J1] {\color{orange}\small$(1)$};
\node at (J2) [below=0 mm of J2] {\vphantom{\small$(1)$}};
\end{tikzpicture}\otimes
\begin{tikzpicture}[baseline={([yshift=-.5ex]current bounding box.center)}]
\coordinate (G1) at (0,0);
\coordinate (G2) at (1,0);
\coordinate (H1) at (-1/3,0);
\coordinate (H2) at (4/3,0);
\coordinate (I1) at (1/2,1/2);
\coordinate (I2) at (1/2,1/8);
\coordinate (I3) at (1/2,-1/2);
\coordinate (I4) at (1/2,-1/8);
\coordinate (J1) at (1,1/4);
\coordinate (J2) at (1,-1/4);
\coordinate (K1) at (1/2,-1/8);
\draw (G1) -- (G2);
\draw (G1) [line width=0.75 mm]to[out=80,in=100] (G2);
\draw (G1) [line width=0.75 mm]to[out=-80,in=-100] (G2);
\draw (G1) [line width=0.75 mm] -- (H1);
\draw (G2) [line width=0.75 mm]-- (H2);
\draw (I1) [dashed,color=orange,line width=0.5 mm] --(I3);
\node at (J1) [above=0 mm of J1] {\color{orange}\small$(1)$};
\node at (J2) [below=0 mm of J2] {\vphantom{\small$(1)$}};
\end{tikzpicture}
+\begin{tikzpicture}[baseline={([yshift=-.5ex]current bounding box.center)}]
\coordinate (G1) at (0,0);
\coordinate (G2) at (1,0);
\coordinate (H1) at (-1/3,0);
\coordinate (H2) at (4/3,0);
\coordinate (I1) at (1/2,1/2);
\coordinate (I2) at (1/2,1/8);
\coordinate (I3) at (1/2,-1/2);
\coordinate (I4) at (1/2,-1/8);
\coordinate (J1) at (1,1/4);
\coordinate (J2) at (1,-1/4);
\coordinate (K1) at (1/2,-1/8);
\draw (G1) -- (G2);
\draw (G1) [line width=0.75 mm]to[out=80,in=100] (G2);
\draw (G1) [line width=0.75 mm]to[out=-80,in=-100] (G2);
\draw (G1) [line width=0.75 mm] -- (H1);
\draw (G2) [line width=0.75 mm]-- (H2);
\draw (I1) [dashed,color=violet,line width=0.5 mm] --(I3);
\node at (J1) [above=0 mm of J1] {\color{blue}\small$(2)$};
\node at (J2) [below=0 mm of J2] {\vphantom{\small$(2)$}};
\end{tikzpicture}\otimes
\begin{tikzpicture}[baseline={([yshift=-.5ex]current bounding box.center)}]
\coordinate (G1) at (0,0);
\coordinate (G2) at (1,0);
\coordinate (H1) at (-1/3,0);
\coordinate (H2) at (4/3,0);
\coordinate (I1) at (1/2,1/2);
\coordinate (I2) at (1/2,1/8);
\coordinate (I3) at (1/2,-1/2);
\coordinate (I4) at (1/2,-1/8);
\coordinate (J1) at (1,1/4);
\coordinate (J2) at (1,-1/4);
\coordinate (K1) at (1/2,-1/8);
\draw (G1) -- (G2);
\draw (G1) [line width=0.75 mm]to[out=80,in=100] (G2);
\draw (G1) [line width=0.75 mm]to[out=-80,in=-100] (G2);
\draw (G1) [line width=0.75 mm] -- (H1);
\draw (G2) [line width=0.75 mm]-- (H2);
\draw (I1) [dashed,color=blue,line width=0.5 mm] --(I3);
\node at (J1) [above=0 mm of J1] {\color{orange}\small$(1)$};
\node at (J2) [below=0 mm of J2] {\vphantom{\small$(1)$}};
\end{tikzpicture}\\
&\,+\begin{tikzpicture}[baseline={([yshift=-.5ex]current bounding box.center)}]
\coordinate (G1) at (0,0);
\coordinate (G2) at (1,0);
\coordinate (H1) at (-1/3,0);
\coordinate (H2) at (4/3,0);
\coordinate (I1) at (1/2,1/2);
\coordinate (I2) at (1/2,1/8);
\coordinate (I3) at (1/2,-1/2);
\coordinate (I4) at (1/2,-1/8);
\coordinate (J1) at (1,1/4);
\coordinate (J2) at (1,-1/4);
\coordinate (K1) at (1/2,-1/8);
\draw (G1) -- (G2);
\draw (G1) [line width=0.75 mm]to[out=80,in=100] (G2);
\draw (G1) [line width=0.75 mm]to[out=-80,in=-100] (G2);
\draw (G1) [line width=0.75 mm] -- (H1);
\draw (G2) [line width=0.75 mm]-- (H2);
\draw (I1) [dashed,color=violet,line width=0.5 mm] --(I3);
\node at (J1) [above=0 mm of J1] {\color{black!60!green}\small$(3)$};
\node at (J2) [below=0 mm of J2] {\vphantom{\small$(3)$}};
\end{tikzpicture}\otimes
\begin{tikzpicture}[baseline={([yshift=-.5ex]current bounding box.center)}]
\coordinate (G1) at (0,0);
\coordinate (G2) at (1,0);
\coordinate (H1) at (-1/3,0);
\coordinate (H2) at (4/3,0);
\coordinate (I1) at (1/2,1/2);
\coordinate (I2) at (1/2,1/8);
\coordinate (I3) at (1/2,-1/2);
\coordinate (I4) at (1/2,-1/8);
\coordinate (J1) at (1,1/4);
\coordinate (J2) at (1,-1/4);
\coordinate (K1) at (1/2,-1/8);
\draw (G1) -- (G2);
\draw (G1) [line width=0.75 mm]to[out=80,in=100] (G2);
\draw (G1) [line width=0.75 mm]to[out=-80,in=-100] (G2);
\draw (G1) [line width=0.75 mm] -- (H1);
\draw (G2) [line width=0.75 mm]-- (H2);
\draw (I1) [dashed,color=black!60!green,line width=0.5 mm] --(I3);
\node at (J1) [above=0 mm of J1] {\color{orange}\small$(1)$};
\node at (J2) [below=0 mm of J2] {\vphantom{\small$(1)$}};
\end{tikzpicture}\,.
\end{split}
\end{align}
A similar coaction can be written for the other two integrands corresponding to the master
integrals $S^{(2)}$ and $S^{(3)}$. Because a generic maximal cut sets  the
double tadpole to zero, this coaction is simpler than that of the uncut integral. This is consistent
with the fact that the space of maximal cuts is three-dimensional, while the uncut topology
has four master integrals. This simplicity is also manifest in the system of differential
equations satisfied by the maximal cuts compared to that of the uncut integrals, 
as discussed in appendix \ref{DiffEqCutsSection}.

\subsection{Double-Edged Triangle}\label{sec:doubleEdged}

\begin{figure}[ht]
	\centering
	\resizebox{4cm}{!}{
	\begin{tikzpicture}
	\coordinate (G1) at (0,0);
	\coordinate (G2) at (1,0.57735026919);
	\coordinate (G3) at (1,-0.57735026919);
	\coordinate (H1) at (-1/3,0);
	\coordinate [above right = 1/3 of G2](H2);
	\coordinate [below right = 1/3 of G3](H3);
	\coordinate (I1) at (1/2,0.28867513459);
	\coordinate [above left=1/4 and 1/8 of I1](I11);
	\coordinate [below right = 1/4 and 1/8 of I1](I12);
	\coordinate (I2) at (1/2,-0.28867513459);
	\coordinate [below left = 1/4 and 1/8 of I2](I21);
	\coordinate [above right = 1/4 and 1/8 of I2](I22);
	\coordinate (I3) at (1,0);
	\coordinate (I31) at (3/4,0);
	\coordinate (I32) at (5/4,0);
	\coordinate [above left=1/4 of G2] (J1);
	\draw (G1) -- (G2);
	\draw (G1) -- (G3);
	\draw (G2) to[out=-110,in=110] (G3);
	\draw (G2) to[out=-70,in=70] (G3);
	\draw (G1) [line width=0.75 mm] -- (H1);
	\draw (G2) [line width=0.75 mm] -- (H2);
	\draw (G3) [line width=0.75 mm] -- (H3);
	\node at (H1) [left=0,scale=0.7] {\small$p_3$};
	\node at (H2) [above right=0,scale=0.7] {\small$p_2$};
	\node at (H3) [below right=0,scale=0.7] {\small$p_1$};
	\node at (0.5,0.57735026919/2) [above left=0,scale=0.7] {\small$3$};
	\node at (0.5,-0.57735026919/2) [below left=0,scale=0.7] {\small$4$};
	\node at (0.7,0) [scale=0.7] {\small$1$};
	\node at (1.3,0) [scale=0.7] {\small$2$};
	\end{tikzpicture}
	}
	\caption{Double-edged triangle.}
	\label{fig:doubleEdgeTriangle}
\end{figure}
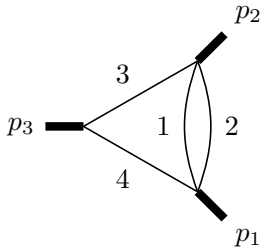

The next example we  consider is the double-edged triangle of
fig.~\ref{fig:doubleEdgeTriangle}. We will always take all propagators
massless, but will consider all possible configurations of massless and
massive external legs, which will allow us to illustrate how the diagrammatic
coaction behaves in limits where masses are set to zero.

The graph of fig.~\ref{fig:doubleEdgeTriangle} defines a family
of master integrals corresponding to 
\begin{align}\begin{split}\label{eq:doubleEdgedGen}
P(\nu_1,\nu_2,\nu_3,\nu_4,&\,\nu_5,\nu_6,\nu_7;D;p_1^2,p_2^2,p_3^2)\\
=&\left(\frac{e^{\gamma_E\epsilon}}{i\pi^{D/2}}\right)^2
\int d^{D}l\int d^{D}k
\frac{[(k+p_3)^2]^{-\nu_5}[(k+p_2)^2]^{-\nu_6}[(l+p_2)^2]^{-\nu_7}}
{[k^2]^{\nu_1}[(k+l+p_2)^2]^{\nu_2}[l^2]^{\nu_3}[(l-p_3)^2]^{\nu_4}}
\end{split}\end{align}
for integer $\nu_i$ with $\nu_5,\nu_6,\nu_7\leq0$ and for $D=n-2\epsilon$, with $n$ even.
In the most complicated case we will consider all external legs are massive and the
space of functions defined by eq.~\eqref{eq:doubleEdgedGen} is spanned by four master
integrals. Two are the sunset integrals corresponding to pinching propagators
3 or 4 in fig.~\ref{fig:doubleEdgeTriangle}. This type of integral
was already discussed in section~\ref{sec:sunsetMassless}. 
The other two master integrals are new and
have not yet been discussed, and we choose them to be
\begin{align}\begin{split}\label{eq:doubleEdged3m}
P^{(1)}(p_1^2,p_2^2,p_3^2)=&-\epsilon^3
\sqrt{\lambda\left(p_1^2,p_2^2,p_3^2\right)}
P(2,1,1,1,0,0,0;4-2\epsilon;p_1^2,p_2^2,p_3^2)\,,\\
P^{(2)}(p_1^2,p_2^2,p_3^2)=&-\epsilon^2p_3^2P(1,1,1,1,0,0,-1;2-2\epsilon;p_1^2,p_2^2,p_3^2)\,,
\end{split}\end{align}
where $\lambda(a,b,c)$ is again the K\"{a}llen function. These integrals evaluate to Appell $F_4$ functions,
and explicit expressions can be found in appendix \ref{sec:doubleEdged3s}.
Their global coaction can be obtained from the results of ref.~\cite{Abreu:2019wzk}.
As is by now well established, to each of the four master integrals corresponds
an independent integration contour: one corresponding to cutting propagators (1,2,3), one
to cutting propagators (1,2,4), and two corresponding to cutting all four propagators.
A basis of independent cut integrals was computed with the same approach 
we used for previous examples. Subsequently we identify the specific cuts dual to the chosen 
master integrals above, by forming linear combinations of the cuts in this basis and imposing 
that eq.~\eqref{eq:coacGenDual} is satisfied. 
We give expressions for each of the four cuts of $P^{(1)}(p_1^2,p_2^2,p_3^2)$ and
$P^{(2)}(p_1^2,p_2^2,p_3^2)$ in appendix \ref{sec:doubleEdged3s}.
Matching the cut integrals to the global coactions, we find the diagrammatic
coactions for the master integrals in eq.~\eqref{eq:doubleEdged3m}:
\begin{align}\nonumber
&\Delta\left[
.
\label{eq:diagCoactP2}
\end{align}

In these diagrammatic coactions we used similar conventions as in previous examples,
that is we distinguish different master integrals and their associated cuts
by using different colours.
These coactions satisfy all the properties of section~\ref{sec:propDiagCoac}, and we will
use them to illustrate the consistency of the coaction in various massless limits. 

\paragraph{Two external masses.} Let us first 
discuss the case where one of the three external legs becomes massless. Given the symmetry of 
the diagram under exchange of~$p_1$ and~$p_2$, there are only two limits to consider:
$p_2^2\to0$ and $p_3^2\to0$. In the limit $p_2^2\to0$, the dimension of the basis of the 
space of functions defined in eq.~\eqref{eq:doubleEdgedGen} reduces from four to two,
because the sunset integral with propagators $1,2,3$ are scaleless and vanish in dimensional regularisation,
 and the two master integrals in 
eq.~\eqref{eq:doubleEdged3m} become linearly dependent. More explicitly,
\begin{equation}\label{eq:relInLimit}
	P^{(2)}(p_1^2,0,p_3^2)= - 6\,P^{(1)}(p_1^2,0,p_3^2)-3\,S(p_1^2)\,,
\end{equation}
with the zero-mass subset $S(p_1^2)$ defined in eq.~\eqref{SunsetZeroMassMaster}.
In order to obtain the diagrammatic coaction for $P^{(1)}(p_1^2,0,p_3^2)$ we 
take the $p_2^2\to0$ limit  in eq.~\eqref{eq:diagCoactP1} and then use the relation
(\ref{eq:relInLimit}) to express the last term on the right hand side of (\ref{eq:diagCoactP1})  in terms of $P^{(1)}$ and $S(p_1^2)$. Further defining
\begin{equation}\label{eq:maxcutLimit}
	\begin{tikzpicture}[baseline={([yshift=-.5ex]current bounding box.center)}]
	\coordinate (G1) at (0,0);
	\coordinate (G2) at (1,0.57735026919);
	\coordinate (G3) at (1,-0.57735026919);
	\coordinate (H1) at (-1/3,0);
	\coordinate [above right = 1/3 of G2](H2);
	\coordinate [below right = 1/3 of G3](H3);
	\coordinate (I1) at (1/2,0.28867513459);
	\coordinate [above left=1/4 and 1/8 of I1](I11);
	\coordinate [below right = 1/4 and 1/8 of I1](I12);
	\coordinate (I2) at (1/2,-0.28867513459);
	\coordinate [below left = 1/4 and 1/8 of I2](I21);
	\coordinate [above right = 1/4 and 1/8 of I2](I22);
	\coordinate (I3) at (1,0);
	\coordinate (I31) at (3/4,0);
	\coordinate (I32) at (5/4,0);
	\coordinate [above left=1/4 of G2] (J1);
	\draw (G1) -- (G2);
	\draw (G1) -- (G3);
	\draw (G2) to[out=-110,in=110] (G3);
	\draw (G2) to[out=-70,in=70] (G3);
	\draw (G1) [line width=0.75 mm] -- (H1);
	\draw (G2) -- (H2);
	\draw (G3) [line width=0.75 mm]-- (H3);
	\draw (I11) [dashed,color=violet,line width=0.5 mm] -- (I12);
	\draw (I21) [dashed,color=violet,line width=0.5 mm] -- (I22);
	\draw (I31) [dashed,color=violet,line width=0.5 mm] -- (I3);
	\draw (I32) [dashed,color=violet,line width=0.5 mm] -- (I3);
	\node at (H1) [left=0,scale=0.7] {$p_3$};
	\node at (H2) [above right=0,scale=0.7] {$p_2$};
	\node at (H3) [below right=0,scale=0.7] {$p_1$};
	\node at (0.3,0.65/2) [scale=0.7] {\small$3$};
	\node at (0.3,-0.65/2) [scale=0.7] {\small$4$};
	\node at (0.75,0.2) [scale=0.7] {\small$1$};
	\node at (1.25,0.2) [scale=0.7] {\small$2$};
	\end{tikzpicture}
	=\left(
	\begin{tikzpicture}[baseline={([yshift=-.5ex]current bounding box.center)}]
	\coordinate (G1) at (0,0);
	\coordinate (G2) at (1,0.57735026919);
	\coordinate (G3) at (1,-0.57735026919);
	\coordinate (H1) at (-1/3,0);
	\coordinate [above right = 1/3 of G2](H2);
	\coordinate [below right = 1/3 of G3](H3);
	\coordinate (I1) at (1/2,0.28867513459);
	\coordinate [above left=1/4 and 1/8 of I1](I11);
	\coordinate [below right = 1/4 and 1/8 of I1](I12);
	\coordinate (I2) at (1/2,-0.28867513459);
	\coordinate [below left = 1/4 and 1/8 of I2](I21);
	\coordinate [above right = 1/4 and 1/8 of I2](I22);
	\coordinate (I3) at (1,0);
	\coordinate (I31) at (3/4,0);
	\coordinate (I32) at (5/4,0);
	\coordinate [above left=1/3 of G2] (J1);
	\draw (G1) -- (G2);
	\draw (G1) -- (G3);
	\draw (G2) to[out=-110,in=110] (G3);
	\draw (G2) to[out=-70,in=70] (G3);
	\draw (G1) [line width=0.75 mm] -- (H1);
	\draw (G2) [line width=0.75 mm] -- (H2);
	\draw (G3) [line width=0.75 mm] -- (H3);
	\draw (I11) [dashed,color=black!60!green,line width=0.5 mm] -- (I12);
	\draw (I21) [dashed,color=black!60!green,line width=0.5 mm] -- (I22);
	\draw (I31) [dashed,color=black!60!green,line width=0.5 mm] -- (I3);
	\draw (I32) [dashed,color=black!60!green,line width=0.5 mm] -- (I3);
	\node at (H1) [left=0,scale=0.7] {$p_3$};
	\node at (H2) [above right=0,scale=0.7] {$p_2$};
	\node at (H3) [below right=0,scale=0.7] {$p_1$};
	\node at (0.3,0.65/2) [scale=0.7] {\small$3$};
	\node at (0.3,-0.65/2) [scale=0.7] {\small$4$};
	\node at (0.75,0.2) [scale=0.7] {\small$1$};
	\node at (1.25,0.2) [scale=0.7] {\small$2$};
	\node at (J1) {\color{black!60!green}{\small$(1)$}};
	\end{tikzpicture}
	-6
	\begin{tikzpicture}[baseline={([yshift=-.5ex]current bounding box.center)}]
	\coordinate (G1) at (0,0);
	\coordinate (G2) at (1,0.57735026919);
	\coordinate (G3) at (1,-0.57735026919);
	\coordinate (H1) at (-1/3,0);
	\coordinate [above right = 1/3 of G2](H2);
	\coordinate [below right = 1/3 of G3](H3);
	\coordinate (I1) at (1/2,0.28867513459);
	\coordinate [above left=1/4 and 1/8 of I1](I11);
	\coordinate [below right = 1/4 and 1/8 of I1](I12);
	\coordinate (I2) at (1/2,-0.28867513459);
	\coordinate [below left = 1/4 and 1/8 of I2](I21);
	\coordinate [above right = 1/4 and 1/8 of I2](I22);
	\coordinate (I3) at (1,0);
	\coordinate (I31) at (3/4,0);
	\coordinate (I32) at (5/4,0);
	\coordinate [above left=1/3 of G2] (J1);
	\draw (G1) -- (G2);
	\draw (G1) -- (G3);
	\draw (G2) to[out=-110,in=110] (G3);
	\draw (G2) to[out=-70,in=70] (G3);
	\draw (G1) [line width=0.75 mm] -- (H1);
	\draw (G2) [line width=0.75 mm] -- (H2);
	\draw (G3) [line width=0.75 mm] -- (H3);
	\draw (I11) [dashed,color=blue,line width=0.5 mm] -- (I12);
	\draw (I21) [dashed,color=blue,line width=0.5 mm] -- (I22);
	\draw (I31) [dashed,color=blue,line width=0.5 mm] -- (I3);
	\draw (I32) [dashed,color=blue,line width=0.5 mm] -- (I3);
	\node at (H1) [left=0,scale=0.7] {$p_3$};
	\node at (H2) [above right=0,scale=0.7] {$p_2$};
	\node at (H3) [below right=0,scale=0.7] {$p_1$};
	\node at (0.3,0.65/2) [scale=0.7] {\small$3$};
	\node at (0.3,-0.65/2) [scale=0.7] {\small$4$};
	\node at (0.75,0.2) [scale=0.7] {\small$1$};
	\node at (1.25,0.2) [scale=0.7] {\small$2$};
	\node at (J1) {\color{black!60!green}{\small$(1)$}};
	\end{tikzpicture}
	\right)\Bigggg\vert_{p_2^2=0}\,,
\end{equation}
and
\begin{equation}\label{eq:2cutLimit}
	\begin{tikzpicture}[baseline={([yshift=-.5ex]current bounding box.center)}]
	\coordinate (G1) at (0,0);
	\coordinate (G2) at (1,0.57735026919);
	\coordinate (G3) at (1,-0.57735026919);
	\coordinate (H1) at (-1/3,0);
	\coordinate [above right = 1/3 of G2](H2);
	\coordinate [below right = 1/3 of G3](H3);
	\coordinate (I1) at (1/2,0.28867513459);
	\coordinate [above left=1/4 and 1/8 of I1](I11);
	\coordinate [below right = 1/4 and 1/8 of I1](I12);
	\coordinate (I2) at (1/2,-0.28867513459);
	\coordinate [below left = 1/4 and 1/8 of I2](I21);
	\coordinate [above right = 1/4 and 1/8 of I2](I22);
	\coordinate (I3) at (1,0);
	\coordinate (I31) at (3/4,0);
	\coordinate (I32) at (5/4,0);
	\coordinate [above left=1/4 of G2] (J1);
	\draw (G1) -- (G2);
	\draw (G1) -- (G3);
	\draw (G2) to[out=-110,in=110] (G3);
	\draw (G2) to[out=-70,in=70] (G3);
	\draw (G1) [line width=0.75 mm] -- (H1);
	\draw (G2) -- (H2);
	\draw (G3) [line width=0.75 mm]-- (H3);
	\draw (I21) [dashed,color=violet,line width=0.5 mm] -- (I22);
	\draw (I31) [dashed,color=violet,line width=0.5 mm] -- (I3);
	\draw (I32) [dashed,color=violet,line width=0.5 mm] -- (I3);
	\node at (H1) [left=0,scale=0.7] {$p_3$};
	\node at (H2) [above right=0,scale=0.7] {$p_2$};
	\node at (H3) [below right=0,scale=0.7] {$p_1$};
	\node at (0.3,0.65/2) [scale=0.7] {\small$3$};
	\node at (0.3,-0.65/2) [scale=0.7] {\small$4$};
	\node at (0.75,0.2) [scale=0.7] {\small$1$};
	\node at (1.25,0.2) [scale=0.7] {\small$2$};
	\end{tikzpicture}
	=\left(
	\begin{tikzpicture}[baseline={([yshift=-.5ex]current bounding box.center)}]
	\coordinate (G1) at (0,0);
	\coordinate (G2) at (1,0.57735026919);
	\coordinate (G3) at (1,-0.57735026919);
	\coordinate (H1) at (-1/3,0);
	\coordinate [above right = 1/3 of G2](H2);
	\coordinate [below right = 1/3 of G3](H3);
	\coordinate (I1) at (1/2,0.28867513459);
	\coordinate [above left=1/4 and 1/8 of I1](I11);
	\coordinate [below right = 1/4 and 1/8 of I1](I12);
	\coordinate (I2) at (1/2,-0.28867513459);
	\coordinate [below left = 1/4 and 1/8 of I2](I21);
	\coordinate [above right = 1/4 and 1/8 of I2](I22);
	\coordinate (I3) at (1,0);
	\coordinate (I31) at (3/4,0);
	\coordinate (I32) at (5/4,0);
	\coordinate [above left=1/3 of G2] (J1);
	\draw (G1) -- (G2);
	\draw (G1) -- (G3);
	\draw (G2) to[out=-110,in=110] (G3);
	\draw (G2) to[out=-70,in=70] (G3);
	\draw (G1) [line width=0.75 mm] -- (H1);
	\draw (G2) [line width=0.75 mm] -- (H2);
	\draw (G3) [line width=0.75 mm] -- (H3);
	\draw (I21) [dashed,color=red,line width=0.5 mm] -- (I22);
	\draw (I31) [dashed,color=red,line width=0.5 mm] -- (I3);
	\draw (I32) [dashed,color=red,line width=0.5 mm] -- (I3);
	\node at (H1) [left=0,scale=0.7] {$p_3$};
	\node at (H2) [above right=0,scale=0.7] {$p_2$};
	\node at (H3) [below right=0,scale=0.7] {$p_1$};
	\node at (0.3,0.65/2) [scale=0.7] {\small$3$};
	\node at (0.3,-0.65/2) [scale=0.7] {\small$4$};
	\node at (0.75,0.2) [scale=0.7] {\small$1$};
	\node at (1.25,0.2) [scale=0.7] {\small$2$};
	\node at (J1) {\color{black!60!green}{\small$(1)$}};
	\end{tikzpicture}
	\!\!\!-2
	\begin{tikzpicture}[baseline={([yshift=-.5ex]current bounding box.center)}]
	\coordinate (G1) at (0,0);
	\coordinate (G2) at (1,0.57735026919);
	\coordinate (G3) at (1,-0.57735026919);
	\coordinate (H1) at (-1/3,0);
	\coordinate [above right = 1/3 of G2](H2);
	\coordinate [below right = 1/3 of G3](H3);
	\coordinate (I1) at (1/2,0.28867513459);
	\coordinate [above left=1/4 and 1/8 of I1](I11);
	\coordinate [below right = 1/4 and 1/8 of I1](I12);
	\coordinate (I2) at (1/2,-0.28867513459);
	\coordinate [below left = 1/4 and 1/8 of I2](I21);
	\coordinate [above right = 1/4 and 1/8 of I2](I22);
	\coordinate (I3) at (1,0);
	\coordinate (I31) at (3/4,0);
	\coordinate (I32) at (5/4,0);
	\coordinate [above left=1/3 of G2] (J1);
	\draw (G1) -- (G2);
	\draw (G1) -- (G3);
	\draw (G2) to[out=-110,in=110] (G3);
	\draw (G2) to[out=-70,in=70] (G3);
	\draw (G1) [line width=0.75 mm] -- (H1);
	\draw (G2) [line width=0.75 mm] -- (H2);
	\draw (G3) [line width=0.75 mm] -- (H3);
	\draw (I11) [dashed,color=blue,line width=0.5 mm] -- (I12);
	\draw (I21) [dashed,color=blue,line width=0.5 mm] -- (I22);
	\draw (I31) [dashed,color=blue,line width=0.5 mm] -- (I3);
	\draw (I32) [dashed,color=blue,line width=0.5 mm] -- (I3);
	\node at (H1) [left=0,scale=0.7] {$p_3$};
	\node at (H2) [above right=0,scale=0.7] {$p_2$};
	\node at (H3) [below right=0,scale=0.7] {$p_1$};
	\node at (0.3,0.65/2) [scale=0.7] {\small$3$};
	\node at (0.3,-0.65/2) [scale=0.7] {\small$4$};
	\node at (0.75,0.2) [scale=0.7] {\small$1$};
	\node at (1.25,0.2) [scale=0.7] {\small$2$};
	\node at (J1) {\color{black!60!green}{\small$(1)$}};
	\end{tikzpicture}
	\right)\Bigggg\vert_{p_2^2=0}\,,
\end{equation}
we obtain the diagrammatic coaction of $P^{(1)}(p_1^2,0,p_3^2)$:
\begin{align}\label{eq;diagCoaction2Asy}
&\Delta\left[\begin{tikzpicture}[baseline={([yshift=-.5ex]current bounding box.center)}]
\coordinate (G1) at (0,0);
\coordinate (G2) at (1,0.57735026919);
\coordinate (G3) at (1,-0.57735026919);
\coordinate (H1) at (-1/3,0);
\coordinate [above right = 1/3 of G2](H2);
\coordinate [below right = 1/3 of G3](H3);
\coordinate (I1) at (1/2,0.28867513459);
\coordinate [above left=1/4 and 1/8 of I1](I11);
\coordinate [below right = 1/4 and 1/8 of I1](I12);
\coordinate (I2) at (1/2,-0.28867513459);
\coordinate [below left = 1/4 and 1/8 of I2](I21);
\coordinate [above right = 1/4 and 1/8 of I2](I22);
\coordinate (I3) at (1,0);
\coordinate (I31) at (3/4,0);
\coordinate (I32) at (5/4,0);
\coordinate [above left=1/4 of G2] (J1);
\draw (G1) -- (G2);
\draw (G1) -- (G3);
\draw (G2) to[out=-110,in=110] (G3);
\draw (G2) to[out=-70,in=70] (G3);
\draw (G1) [line width=0.75 mm] -- (H1);
\draw (G2) -- (H2);
\draw (G3) [line width=0.75 mm]-- (H3);
\node at (H1) [left=0,scale=0.7] {$p_3$};
\node at (H2) [above right=0,scale=0.7] {$p_2$};
\node at (H3) [below right=0,scale=0.7] {$p_1$};
\node at (1/2,0.57735026919/2) [above left=0,scale=0.7] {\small$3$};
\node at (1/2,-0.57735026919/2) [below left=0,scale=0.7] {\small$4$};
\node at (0.7,0) [scale=0.7] {\small$1$};
\node at (1.3,0) [scale=0.7] {\small$2$};
\end{tikzpicture}\right]
=\begin{tikzpicture}[baseline={([yshift=-.5ex]current bounding box.center)}]
\coordinate (G1) at (0,0);
\coordinate (G2) at (1,0.57735026919);
\coordinate (G3) at (1,-0.57735026919);
\coordinate (H1) at (-1/3,0);
\coordinate [above right = 1/3 of G2](H2);
\coordinate [below right = 1/3 of G3](H3);
\coordinate (I1) at (1/2,0.28867513459);
\coordinate [above left=1/4 and 1/8 of I1](I11);
\coordinate [below right = 1/4 and 1/8 of I1](I12);
\coordinate (I2) at (1/2,-0.28867513459);
\coordinate [below left = 1/4 and 1/8 of I2](I21);
\coordinate [above right = 1/4 and 1/8 of I2](I22);
\coordinate (I3) at (1,0);
\coordinate (I31) at (3/4,0);
\coordinate (I32) at (5/4,0);
\coordinate [above left=1/4 of G2] (J1);
\draw (G1) -- (G2);
\draw (G1) -- (G3);
\draw (G2) to[out=-110,in=110] (G3);
\draw (G2) to[out=-70,in=70] (G3);
\draw (G1) [line width=0.75 mm] -- (H1);
\draw (G2) -- (H2);
\draw (G3) [line width=0.75 mm]-- (H3);
\node at (H1) [left=0,scale=0.7] {$p_3$};
\node at (H2) [above right=0,scale=0.7] {$p_2$};
\node at (H3) [below right=0,scale=0.7] {$p_1$};
\node at (1/2,0.57735026919/2) [above left=0,scale=0.7] {\small$3$};
\node at (1/2,-0.57735026919/2) [below left=0,scale=0.7] {\small$4$};
\node at (0.7,0) [scale=0.7] {\small$1$};
\node at (1.3,0) [scale=0.7] {\small$2$};
\end{tikzpicture}\!\!
\otimes\begin{tikzpicture}[baseline={([yshift=-.5ex]current bounding box.center)}]
\coordinate (G1) at (0,0);
\coordinate (G2) at (1,0.57735026919);
\coordinate (G3) at (1,-0.57735026919);
\coordinate (H1) at (-1/3,0);
\coordinate [above right = 1/3 of G2](H2);
\coordinate [below right = 1/3 of G3](H3);
\coordinate (I1) at (1/2,0.28867513459);
\coordinate [above left=1/4 and 1/8 of I1](I11);
\coordinate [below right = 1/4 and 1/8 of I1](I12);
\coordinate (I2) at (1/2,-0.28867513459);
\coordinate [below left = 1/4 and 1/8 of I2](I21);
\coordinate [above right = 1/4 and 1/8 of I2](I22);
\coordinate (I3) at (1,0);
\coordinate (I31) at (3/4,0);
\coordinate (I32) at (5/4,0);
\coordinate [above left=1/4 of G2] (J1);
\draw (G1) -- (G2);
\draw (G1) -- (G3);
\draw (G2) to[out=-110,in=110] (G3);
\draw (G2) to[out=-70,in=70] (G3);
\draw (G1) [line width=0.75 mm] -- (H1);
\draw (G2) -- (H2);
\draw (G3) [line width=0.75 mm]-- (H3);
\draw (I11) [dashed,color=violet,line width=0.5 mm] -- (I12);
\draw (I21) [dashed,color=violet,line width=0.5 mm] -- (I22);
\draw (I31) [dashed,color=violet,line width=0.5 mm] -- (I3);
\draw (I32) [dashed,color=violet,line width=0.5 mm] -- (I3);
\node at (H1) [left=0,scale=0.7] {$p_3$};
\node at (H2) [above right=0,scale=0.7] {$p_2$};
\node at (H3) [below right=0,scale=0.7] {$p_1$};
\node at (0.3,0.65/2) [scale=0.7] {\small$3$};
\node at (0.3,-0.65/2) [scale=0.7] {\small$4$};
\node at (0.75,0.2) [scale=0.7] {\small$1$};
\node at (1.25,0.2) [scale=0.7] {\small$2$};
\end{tikzpicture}\!\!+\begin{tikzpicture}[baseline={([yshift=-.5ex]current bounding box.center)}]
\coordinate (G1) at (0,0);
\coordinate (G2) at (1,0);
\coordinate (H1) at (-1/3,0);
\coordinate (H2) at (4/3,0);
\coordinate (I1) at (1/2,1/2);
\coordinate (I2) at (1/2,1/8);
\coordinate (I3) at (1/2,-1/2);
\coordinate (I4) at (1/2,-1/8);
\coordinate (J1) at (1,1/4);
\coordinate (J2) at (1,-1/4);
\coordinate (K1) at (1/2,-1/8);
\draw (G1) -- (G2);
\draw (G1) to[out=80,in=100] (G2);
\draw (G1) to[out=-80,in=-100] (G2);
\draw (G1) [line width=0.75 mm] -- (H1);
\draw (G2) [line width=0.75 mm]-- (H2);
\node at (K1) [above=0.7 mm of K1,scale=0.7] {\small$1$};
\node at (K1) [above=3.7 mm of K1,scale=0.7] {\small$4$};
\node at (K1) [below=1.3 mm of K1,scale=0.7] {\small$2$};
\node at (H1) [left=0,scale=0.7] {$p_1$};
\node at (H2) [right=0,scale=0.7] {$p_1$};
\end{tikzpicture}\otimes\begin{tikzpicture}[baseline={([yshift=-.5ex]current bounding box.center)}]
\coordinate (G1) at (0,0);
\coordinate (G2) at (1,0.57735026919);
\coordinate (G3) at (1,-0.57735026919);
\coordinate (H1) at (-1/3,0);
\coordinate [above right = 1/3 of G2](H2);
\coordinate [below right = 1/3 of G3](H3);
\coordinate (I1) at (1/2,0.28867513459);
\coordinate [above left=1/4 and 1/8 of I1](I11);
\coordinate [below right = 1/4 and 1/8 of I1](I12);
\coordinate (I2) at (1/2,-0.28867513459);
\coordinate [below left = 1/4 and 1/8 of I2](I21);
\coordinate [above right = 1/4 and 1/8 of I2](I22);
\coordinate (I3) at (1,0);
\coordinate (I31) at (3/4,0);
\coordinate (I32) at (5/4,0);
\coordinate [above left=1/4 of G2] (J1);
\draw (G1) -- (G2);
\draw (G1) -- (G3);
\draw (G2) to[out=-110,in=110] (G3);
\draw (G2) to[out=-70,in=70] (G3);
\draw (G1) [line width=0.75 mm] -- (H1);
\draw (G2) -- (H2);
\draw (G3) [line width=0.75 mm]-- (H3);
\draw (I21) [dashed,color=violet,line width=0.5 mm] -- (I22);
\draw (I31) [dashed,color=violet,line width=0.5 mm] -- (I3);
\draw (I32) [dashed,color=violet,line width=0.5 mm] -- (I3);
\node at (H1) [left=0,scale=0.7] {$p_3$};
\node at (H2) [above right=0,scale=0.7] {$p_2$};
\node at (H3) [below right=0,scale=0.7] {$p_1$};
\node at (0.3,0.65/2) [scale=0.7] {\small$3$};
\node at (0.3,-0.65/2) [scale=0.7] {\small$4$};
\node at (0.75,0.2) [scale=0.7] {\small$1$};
\node at (1.25,0.2) [scale=0.7] {\small$2$};
\end{tikzpicture}\!\!.
\end{align}
While we obtained this diagrammatic coaction by taking the $p_2^2\to0$ limit of
the coaction of $P^{(1)}(p_1^2,p_2^2,p_3^2)$, we could also have computed it directly
following the same steps we have taken in previous cases. We have collected 
all relevant expressions in appendix \ref{sec:doubleEdged2sAsy}, and one can easily
check that the diagrammatic coaction obtained by sending to zero some of the scales agrees with the global coaction obtained with these expressions.

Through this example we see that under massless limits, dual cut contours defined in the massless case may be related in a nontrivial way to dual cuts in the massive case, when the number of master integrals with a given set of propagators changes upon taking the limit. We note nevertheless that the same definition of the dual contours of eqs.~\eqref{eq:maxcutLimit} and \eqref{eq:2cutLimit}
could also have been obtained starting with the diagrammatic
coaction for $P^{(2)}$ given in eq.~\eqref{eq:diagCoactP2} and then using the relation in eq.~\eqref{eq:relInLimit} on both the left- and right-hand sides (and for both the left and right entries of the coaction) showing the internal consistency of our results.

Let us next consider the limit where $p_3^2\to0$. Once again we find that the
space of integrals defined by eq.~\eqref{eq:doubleEdgedGen} is spanned by
two master integrals which, importantly, can be chosen to be the sunset integrals
with propagators (1,2,3) and (1,2,4). In other words, in this limit the 
double-edged triangle is reducible to integrals with fewer propagators.
We find:
\begin{equation}
	P^{(1)}(p_1^2,p_2^2,0)=\frac{1}{2}\left(S(p_2^2)-S(p_1^2)\right)\,,\qquad
	P^{(2)}(p_1^2,p_2^2,0)=0\,.
\end{equation}
Consistently, we also find that
\begin{equation}\label{eq:maxCutLimit}
	\mathcal{C}_{1,2,3,4}^{(i)}P^{(j)}(p_1^2,p_2^2,0)=0\,.
\end{equation}
It then follows that the $p_3^2\to0$ limit of eq.~\eqref{eq:diagCoactP1}
gives
\begin{align}\label{CoactionP2}
\Delta\left[\begin{tikzpicture}[baseline={([yshift=-.5ex]current bounding box.center)}]
\coordinate (G1) at (0,0);
\coordinate (G2) at (1,0.57735026919);
\coordinate (G3) at (1,-0.57735026919);
\coordinate (H1) at (-1/3,0);
\coordinate [above right = 1/3 of G2](H2);
\coordinate [below right = 1/3 of G3](H3);
\coordinate (I1) at (1/2,0.28867513459);
\coordinate [above left=1/4 and 1/8 of I1](I11);
\coordinate [below right = 1/4 and 1/8 of I1](I12);
\coordinate (I2) at (1/2,-0.28867513459);
\coordinate [below left = 1/4 and 1/8 of I2](I21);
\coordinate [above right = 1/4 and 1/8 of I2](I22);
\coordinate (I3) at (1,0);
\coordinate (I31) at (3/4,0);
\coordinate (I32) at (5/4,0);
\coordinate [above left=1/4 of G2] (J1);
\draw (G1) -- (G2);
\draw (G1) -- (G3);
\draw (G2) to[out=-110,in=110] (G3);
\draw (G2) to[out=-70,in=70] (G3);
\draw (G1)  -- (H1);
\draw (G2) [line width=0.75 mm]-- (H2);
\draw (G3) [line width=0.75 mm]-- (H3);
\node at (H2) [below =0 mm of H2] {\vphantom{\small$2$}};
\node at (H2) [above =0 mm of H3] {\vphantom{\small$2$}};
\node at (H1) [left=0,scale=0.7] {$p_3$};
\node at (H2) [above right=0,scale=0.7] {$p_2$};
\node at (H3) [below right=0,scale=0.7] {$p_1$};
\node at (1/2,0.57735026919/2) [above left=0,scale=0.7] {\small$3$};
\node at (1/2,-0.57735026919/2) [below left=0,scale=0.7] {\small$4$};
\node at (0.7,0) [scale=0.7] {\small$1$};
\node at (1.3,0) [scale=0.7] {\small$2$};
\end{tikzpicture}\right]=&\begin{tikzpicture}[baseline={([yshift=-.5ex]current bounding box.center)}]
\coordinate (G1) at (0,0);
\coordinate (G2) at (1,0);
\coordinate (H1) at (-1/3,0);
\coordinate (H2) at (4/3,0);
\coordinate (I1) at (1/2,1/2);
\coordinate (I2) at (1/2,1/8);
\coordinate (I3) at (1/2,-1/2);
\coordinate (I4) at (1/2,-1/8);
\coordinate (J1) at (1,1/4);
\coordinate (J2) at (1,-1/4);
\coordinate (K1) at (1/2,-1/8);
\draw (G1) -- (G2);
\draw (G1) to[out=80,in=100] (G2);
\draw (G1) to[out=-80,in=-100] (G2);
\draw (G1) [line width=0.75 mm] -- (H1);
\draw (G2) [line width=0.75 mm]-- (H2);
\node at (K1) [above=0.7 mm of K1,scale=0.7] {\small$1$};
\node at (K1) [above=3.7 mm of K1,scale=0.7] {\small$3$};
\node at (K1) [below=1.3 mm of K1,scale=0.7] {\small$2$};
\node at (H1) [left=0,scale=0.7] {$p_2$};
\node at (H2) [right=0,scale=0.7] {$p_2$};
\end{tikzpicture}\otimes\begin{tikzpicture}[baseline={([yshift=-.5ex]current bounding box.center)}]
\coordinate (G1) at (0,0);
\coordinate (G2) at (1,0.57735026919);
\coordinate (G3) at (1,-0.57735026919);
\coordinate (H1) at (-1/3,0);
\coordinate [above right = 1/3 of G2](H2);
\coordinate [below right = 1/3 of G3](H3);
\coordinate (I1) at (1/2,0.28867513459);
\coordinate [above left=1/4 and 1/8 of I1](I11);
\coordinate [below right = 1/4 and 1/8 of I1](I12);
\coordinate (I2) at (1/2,-0.28867513459);
\coordinate [below left = 1/4 and 1/8 of I2](I21);
\coordinate [above right = 1/4 and 1/8 of I2](I22);
\coordinate (I3) at (1,0);
\coordinate (I31) at (3/4,0);
\coordinate (I32) at (5/4,0);
\coordinate [above left=1/4 of G2] (J1);
\draw (G1) -- (G2);
\draw (G1) -- (G3);
\draw (G2) to[out=-110,in=110] (G3);
\draw (G2) to[out=-70,in=70] (G3);
\draw (G1)  -- (H1);
\draw (G2) [line width=0.75 mm]-- (H2);
\draw (G3) [line width=0.75 mm]-- (H3);
\draw (I11) [dashed,color=red,line width=0.5 mm] -- (I12);
\draw (I31) [dashed,color=red,line width=0.5 mm] -- (I3);
\draw (I32) [dashed,color=red,line width=0.5 mm] -- (I3);
\node at (H1) [left=0,scale=0.7] {$p_3$};
\node at (H2) [above right=0,scale=0.7] {$p_2$};
\node at (H3) [below right=0,scale=0.7] {$p_1$};
\node at (0.3,0.65/2) [scale=0.7] {\small$3$};
\node at (0.3,-0.65/2) [scale=0.7] {\small$4$};
\node at (0.75,0.2) [scale=0.7] {\small$1$};
\node at (1.25,0.2) [scale=0.7] {\small$2$};
\end{tikzpicture}\!\!\!
+\begin{tikzpicture}[baseline={([yshift=-.5ex]current bounding box.center)}]
\coordinate (G1) at (0,0);
\coordinate (G2) at (1,0);
\coordinate (H1) at (-1/3,0);
\coordinate (H2) at (4/3,0);
\coordinate (I1) at (1/2,1/2);
\coordinate (I2) at (1/2,1/8);
\coordinate (I3) at (1/2,-1/2);
\coordinate (I4) at (1/2,-1/8);
\coordinate (J1) at (1,1/4);
\coordinate (J2) at (1,-1/4);
\coordinate (K1) at (1/2,-1/8);
\draw (G1) -- (G2);
\draw (G1) to[out=80,in=100] (G2);
\draw (G1) to[out=-80,in=-100] (G2);
\draw (G1) [line width=0.75 mm] -- (H1);
\draw (G2) [line width=0.75 mm]-- (H2);
\node at (K1) [above=0.7 mm of K1,scale=0.7] {\small$1$};
\node at (K1) [above=3.7 mm of K1,scale=0.7] {\small$4$};
\node at (K1) [below=1.3 mm of K1,scale=0.7] {\small$2$};
\node at (H1) [left=0,scale=0.7] {$p_1$};
\node at (H2) [right=0,scale=0.7] {$p_1$};
\end{tikzpicture}\otimes\begin{tikzpicture}[baseline={([yshift=-.5ex]current bounding box.center)}]
\coordinate (G1) at (0,0);
\coordinate (G2) at (1,0.57735026919);
\coordinate (G3) at (1,-0.57735026919);
\coordinate (H1) at (-1/3,0);
\coordinate [above right = 1/3 of G2](H2);
\coordinate [below right = 1/3 of G3](H3);
\coordinate (I1) at (1/2,0.28867513459);
\coordinate [above left=1/4 and 1/8 of I1](I11);
\coordinate [below right = 1/4 and 1/8 of I1](I12);
\coordinate (I2) at (1/2,-0.28867513459);
\coordinate [below left = 1/4 and 1/8 of I2](I21);
\coordinate [above right = 1/4 and 1/8 of I2](I22);
\coordinate (I3) at (1,0);
\coordinate (I31) at (3/4,0);
\coordinate (I32) at (5/4,0);
\coordinate [above left=1/4 of G2] (J1);
\draw (G1) -- (G2);
\draw (G1) -- (G3);
\draw (G2) to[out=-110,in=110] (G3);
\draw (G2) to[out=-70,in=70] (G3);
\draw (G1)  -- (H1);
\draw (G2) [line width=0.75 mm]-- (H2);
\draw (G3) [line width=0.75 mm]-- (H3);
\draw (I21) [dashed,color=red,line width=0.5 mm] -- (I22);
\draw (I31) [dashed,color=red,line width=0.5 mm] -- (I3);
\draw (I32) [dashed,color=red,line width=0.5 mm] -- (I3);
\node at (H1) [left=0,scale=0.7] {$p_3$};
\node at (H2) [above right=0,scale=0.7] {$p_2$};
\node at (H3) [below right=0,scale=0.7] {$p_1$};
\node at (0.3,0.65/2) [scale=0.7] {\small$3$};
\node at (0.3,-0.65/2) [scale=0.7] {\small$4$};
\node at (0.75,0.2) [scale=0.7] {\small$1$};
\node at (1.25,0.2) [scale=0.7] {\small$2$};
\end{tikzpicture}\!\!.
\end{align}
It is straightforward to check that this diagrammatic coaction agrees with the global coaction
one obtains using the explicit expressions for the different contributions
that are listed in appendix \ref{sec:doubleEdged2sSy}.
We note that because this example
is a reducible master integral (i.e., it can be written as a linear
combination of integrals with fewer propagators), it does not appear in any of the left
entries of the diagrammatic coaction. This is of course a choice, but it guarantees consistency with
the massless limit of eq.~\eqref{eq:diagCoactP1}.
More generally, this serves as a template for what 
would happen if one were to compute the coaction of a reducible Feynman integral: consistently 
with the properties listed in section~\ref{sec:propDiagCoac}, 
the integral would not appear in the left entries of the coaction tensor and its maximal cuts would
vanish,  but its non-maximal cuts would appear in the right entries.

\paragraph{One external mass.} To complete the discussion of the double-edged triangle
we consider the two independent one-mass configurations, which can both be easily
obtained from the two-mass configurations. We start with the case where $p_1^2=p^2_2=0$,
and taking $p_1^2\to0$ in eq.~\eqref{eq;diagCoaction2Asy} we directly obtain
\begin{align}\label{eq:diagCoaction1mSy}
\Delta\left[\begin{tikzpicture}[baseline={([yshift=-.5ex]current bounding box.center)}]
\coordinate (G1) at (0,0);
\coordinate (G2) at (1,0.57735026919);
\coordinate (G3) at (1,-0.57735026919);
\coordinate (H1) at (-1/3,0);
\coordinate [above right = 1/3 of G2](H2);
\coordinate [below right = 1/3 of G3](H3);
\coordinate (I1) at (1/2,0.28867513459);
\coordinate [above left=1/4 and 1/8 of I1](I11);
\coordinate [below right = 1/4 and 1/8 of I1](I12);
\coordinate (I2) at (1/2,-0.28867513459);
\coordinate [below left = 1/4 and 1/8 of I2](I21);
\coordinate [above right = 1/4 and 1/8 of I2](I22);
\coordinate (I3) at (1,0);
\coordinate (I31) at (3/4,0);
\coordinate (I32) at (5/4,0);
\coordinate [above left=1/4 of G2] (J1);
\draw (G1) -- (G2);
\draw (G1) -- (G3);
\draw (G2) to[out=-110,in=110] (G3);
\draw (G2) to[out=-70,in=70] (G3);
\draw (G1) [line width=0.75 mm] -- (H1);
\draw (G2) -- (H2);
\draw (G3) -- (H3);
\node at (H2) [below =0 mm of H2] {\vphantom{\small$2$}};
\node at (H2) [above =0 mm of H3] {\vphantom{\small$2$}};
\node at (H1) [left=0,scale=0.7] {$p_3$};
\node at (H2) [above right=0,scale=0.7] {$p_2$};
\node at (H3) [below right=0,scale=0.7] {$p_1$};
\node at (1/2,0.57735026919/2) [above left=0,scale=0.7] {\small$3$};
\node at (1/2,-0.57735026919/2) [below left=0,scale=0.7] {\small$4$};
\node at (0.7,0) [scale=0.7] {\small$1$};
\node at (1.3,0) [scale=0.7] {\small$2$};
\end{tikzpicture}\right]=\begin{tikzpicture}[baseline={([yshift=-.5ex]current bounding box.center)}]
\coordinate (G1) at (0,0);
\coordinate (G2) at (1,0.57735026919);
\coordinate (G3) at (1,-0.57735026919);
\coordinate (H1) at (-1/3,0);
\coordinate [above right = 1/3 of G2](H2);
\coordinate [below right = 1/3 of G3](H3);
\coordinate (I1) at (1/2,0.28867513459);
\coordinate [above left=1/4 and 1/8 of I1](I11);
\coordinate [below right = 1/4 and 1/8 of I1](I12);
\coordinate (I2) at (1/2,-0.28867513459);
\coordinate [below left = 1/4 and 1/8 of I2](I21);
\coordinate [above right = 1/4 and 1/8 of I2](I22);
\coordinate (I3) at (1,0);
\coordinate (I31) at (3/4,0);
\coordinate (I32) at (5/4,0);
\coordinate [above left=1/4 of G2] (J1);
\draw (G1) -- (G2);
\draw (G1) -- (G3);
\draw (G2) to[out=-110,in=110] (G3);
\draw (G2) to[out=-70,in=70] (G3);
\draw (G1) [line width=0.75 mm] -- (H1);
\draw (G2) -- (H2);
\draw (G3) -- (H3);
\node at (H1) [left=0,scale=0.7] {$p_3$};
\node at (H2) [above right=0,scale=0.7] {$p_2$};
\node at (H3) [below right=0,scale=0.7] {$p_1$};
\node at (1/2,0.57735026919/2) [above left=0,scale=0.7] {\small$3$};
\node at (1/2,-0.57735026919/2) [below left=0,scale=0.7] {\small$4$};
\node at (0.7,0) [scale=0.7] {\small$1$};
\node at (1.3,0) [scale=0.7] {\small$2$};
\end{tikzpicture}\otimes\begin{tikzpicture}[baseline={([yshift=-.5ex]current bounding box.center)}]
\coordinate (G1) at (0,0);
\coordinate (G2) at (1,0.57735026919);
\coordinate (G3) at (1,-0.57735026919);
\coordinate (H1) at (-1/3,0);
\coordinate [above right = 1/3 of G2](H2);
\coordinate [below right = 1/3 of G3](H3);
\coordinate (I1) at (1/2,0.28867513459);
\coordinate [above left=1/4 and 1/8 of I1](I11);
\coordinate [below right = 1/4 and 1/8 of I1](I12);
\coordinate (I2) at (1/2,-0.28867513459);
\coordinate [below left = 1/4 and 1/8 of I2](I21);
\coordinate [above right = 1/4 and 1/8 of I2](I22);
\coordinate (I3) at (1,0);
\coordinate (I31) at (3/4,0);
\coordinate (I32) at (5/4,0);
\coordinate [above left=1/4 of G2] (J1);
\draw (G1) -- (G2);
\draw (G1) -- (G3);
\draw (G2) to[out=-110,in=110] (G3);
\draw (G2) to[out=-70,in=70] (G3);
\draw (G1) [line width=0.75 mm] -- (H1);
\draw (G2) -- (H2);
\draw (G3) -- (H3);
\draw (I11) [dashed,color=violet,line width=0.5 mm] -- (I12);
\draw (I21) [dashed,color=violet,line width=0.5 mm] -- (I22);
\draw (I31) [dashed,color=violet,line width=0.5 mm] -- (I3);
\draw (I32) [dashed,color=violet,line width=0.5 mm] -- (I3);
\node at (H1) [left=0,scale=0.7] {$p_3$};
\node at (H2) [above right=0,scale=0.7] {$p_2$};
\node at (H3) [below right=0,scale=0.7] {$p_1$};
\node at (0.3,0.65/2) [scale=0.7] {\small$3$};
\node at (0.3,-0.65/2) [scale=0.7] {\small$4$};
\node at (0.75,0.2) [scale=0.7] {\small$1$};
\node at (1.25,0.2) [scale=0.7] {\small$2$};
\end{tikzpicture}.
\end{align}
For the other one-mass configuration with $p_2^2=p_3^2=0$, we start from
eq.~\eqref{CoactionP2} and take $p^2_2\to0$ to obtain
\begin{align}\label{CoactionP1}
\Delta\left[\begin{tikzpicture}[baseline={([yshift=-.5ex]current bounding box.center)}]
\coordinate (G1) at (0,0);
\coordinate (G2) at (1,0.57735026919);
\coordinate (G3) at (1,-0.57735026919);
\coordinate (H1) at (-1/3,0);
\coordinate [above right = 1/3 of G2](H2);
\coordinate [below right = 1/3 of G3](H3);
\coordinate (I1) at (1/2,0.28867513459);
\coordinate [above left=1/4 and 1/8 of I1](I11);
\coordinate [below right = 1/4 and 1/8 of I1](I12);
\coordinate (I2) at (1/2,-0.28867513459);
\coordinate [below left = 1/4 and 1/8 of I2](I21);
\coordinate [above right = 1/4 and 1/8 of I2](I22);
\coordinate (I3) at (1,0);
\coordinate (I31) at (3/4,0);
\coordinate (I32) at (5/4,0);
\coordinate [above left=1/4 of G2] (J1);
\draw (G1) -- (G2);
\draw (G1) -- (G3);
\draw (G2) to[out=-110,in=110] (G3);
\draw (G2) to[out=-70,in=70] (G3);
\draw (G1)  -- (H1);
\draw (G2) -- (H2);
\draw (G3) [line width=0.75 mm]-- (H3);
\node at (H2) [below =0 mm of H2] {\vphantom{\small$2$}};
\node at (H2) [above =0 mm of H3] {\vphantom{\small$2$}};
\node at (H1) [left=0,scale=0.7] {$p_3$};
\node at (H2) [above right=0,scale=0.7] {$p_2$};
\node at (H3) [below right=0,scale=0.7] {$p_1$};
\node at (1/2,0.57735026919/2) [above left=0,scale=0.7] {\small$3$};
\node at (1/2,-0.57735026919/2) [below left=0,scale=0.7] {\small$4$};
\node at (0.7,0) [scale=0.7] {\small$1$};
\node at (1.3,0) [scale=0.7] {\small$2$};
\end{tikzpicture}\right]=\begin{tikzpicture}[baseline={([yshift=-.5ex]current bounding box.center)}]
\coordinate (G1) at (0,0);
\coordinate (G2) at (1,0);
\coordinate (H1) at (-1/3,0);
\coordinate (H2) at (4/3,0);
\coordinate (I1) at (1/2,1/2);
\coordinate (I2) at (1/2,1/8);
\coordinate (I3) at (1/2,-1/2);
\coordinate (I4) at (1/2,-1/8);
\coordinate (J1) at (1,1/4);
\coordinate (J2) at (1,-1/4);
\coordinate (K1) at (1/2,-1/8);
\draw (G1) -- (G2);
\draw (G1) to[out=80,in=100] (G2);
\draw (G1) to[out=-80,in=-100] (G2);
\draw (G1) [line width=0.75 mm] -- (H1);
\draw (G2) [line width=0.75 mm]-- (H2);
\node at (K1) [above=0.7 mm of K1,scale=0.7] {\small$1$};
\node at (K1) [above=3.7 mm of K1,scale=0.7] {\small$4$};
\node at (K1) [below=1.3 mm of K1,scale=0.7] {\small$2$};
\node at (H1) [left=0,scale=0.7] {$p_{1}$};
\node at (H2) [right=0,scale=0.7] {$p_1$};
\end{tikzpicture}\otimes\begin{tikzpicture}[baseline={([yshift=-.5ex]current bounding box.center)}]
\coordinate (G1) at (0,0);
\coordinate (G2) at (1,0.57735026919);
\coordinate (G3) at (1,-0.57735026919);
\coordinate (H1) at (-1/3,0);
\coordinate [above right = 1/3 of G2](H2);
\coordinate [below right = 1/3 of G3](H3);
\coordinate (I1) at (1/2,0.28867513459);
\coordinate [above left=1/4 and 1/8 of I1](I11);
\coordinate [below right = 1/4 and 1/8 of I1](I12);
\coordinate (I2) at (1/2,-0.28867513459);
\coordinate [below left = 1/4 and 1/8 of I2](I21);
\coordinate [above right = 1/4 and 1/8 of I2](I22);
\coordinate (I3) at (1,0);
\coordinate (I31) at (3/4,0);
\coordinate (I32) at (5/4,0);
\coordinate [above left=1/4 of G2] (J1);
\draw (G1) -- (G2);
\draw (G1) -- (G3);
\draw (G2) to[out=-110,in=110] (G3);
\draw (G2) to[out=-70,in=70] (G3);
\draw (G1) -- (H1);
\draw (G2) -- (H2);
\draw (G3) [line width=0.75 mm] -- (H3);
\draw (I21) [dashed,color=red,line width=0.5 mm] -- (I22);
\draw (I31) [dashed,color=red,line width=0.5 mm] -- (I3);
\draw (I32) [dashed,color=red,line width=0.5 mm] -- (I3);
\node at (H1) [left=0,scale=0.7] {$p_3$};
\node at (H2) [above right=0,scale=0.7] {$p_2$};
\node at (H3) [below right=0,scale=0.7] {$p_1$};
\node at (0.3,0.65/2) [scale=0.7] {\small$3$};
\node at (0.3,-0.65/2) [scale=0.7] {\small$4$};
\node at (0.75,0.2) [scale=0.7] {\small$1$};
\node at (1.25,0.2) [scale=0.7] {\small$2$};
\end{tikzpicture}.
\end{align}
We note that we could also have started from eq.~\eqref{eq;diagCoaction2Asy}
and set $p_3^2=0$. Given eqs.~\eqref{eq:2cutLimit} and \eqref{eq:maxCutLimit},
it is clear that we would obtain the same diagrammatic coaction.

The diagrammatic coactions in eqs.~\eqref{eq:diagCoaction1mSy} and \eqref{CoactionP1}
were obtained from massless limits of other diagrammatic coactions, but they
can be easily seen to be consistent with the associated global coactions using 
the results in appendices \ref{sec:doubleEdged1sSy} and \ref{sec:doubleEdged1sAsy}.

\subsection{Adjacent Triangles\label{sec:adjacentTriangles}}

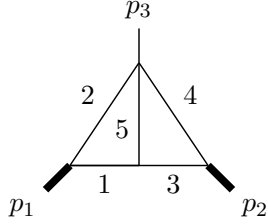
\begin{figure}[htb]
	\centering
	\resizebox{4cm}{!}{
	\begin{tikzpicture}[baseline={([yshift=-.5ex]current bounding box.center)}]
	\coordinate (G1) at (0,0);
	\coordinate (G2) at (-2/3,0);
	\coordinate (G3) at (2/3,0);
	\coordinate (G4) at (0,1);
	\coordinate [below left=1/3 of G2] (H1);
	\coordinate [below right=1/3 of G3] (H2);
	\coordinate [above=1/3 of G4] (H3);
	\coordinate (I1) at (1/3,0);
	\coordinate (I2) at (-1/3,0);
	\coordinate (I3) at (1/3,1/2);
	\coordinate (I4) at (-1/3,1/2);
	\coordinate (I5) at (0,1/3);
	\coordinate (I11) at (1/3,-8/40);
	\coordinate (I12) at (1/3,8/40);
	\coordinate (I21) at (-1/3,-8/40);
	\coordinate (I22) at (-1/3,8/40);
	\coordinate (I31) at (1/3+12/80,1/2+4/40);
	\coordinate (I32) at (1/3-12/80,1/2-4/40);
	\coordinate (I41) at (-1/3-12/80,1/2+4/40);
	\coordinate (I42) at (-1/3+12/80,1/2-4/40);
	\coordinate (I51) at (-10/40,1/3);
	\coordinate (I52) at (10/40,1/3);
	\draw (G1) -- (G2);
	\draw (G2) -- (G3);
	\draw (G3) -- (G4);
	\draw (G4) -- (G1);
	\draw (G2) -- (G4);
	\draw (G2) [line width=0.75 mm]-- (H1);
	\draw (G3) [line width=0.75 mm]-- (H2);
	\draw (G4) -- (H3);
	\node at (H1) [below left=0,scale=0.7] {$\small{p_1}$};
	\node at (H2) [below right=0,scale=0.7] {$\small{p_2}$};
	\node at (H3) [above=0,scale=0.7] {$\small{p_3}$};
	\node at (-1/3,0) [below=0,scale=0.7] {$\small{1}$};
	\node at (0,0.35) [left=0,scale=0.7] {$\small{5}$};
	\node at (-1/3,1/2) [above left=0,scale=0.7] {$\small{2}$};
	\node at (1/3,0) [below=0,scale=0.7] {$\small{3}$};
	\node at (1/3,1/2) [above right=0,scale=0.7] {$\small{4}$};
	\end{tikzpicture}
	}
	\caption{Adjacent triangles with two massive external legs.}
	\label{fig:adjTriangles}
\end{figure}

As a next example we consider the topology defined by the diagram in
fig.~\ref{fig:adjTriangles} or equivalently by the integrals of the form
\begin{align}\begin{split}\label{eq:adjTrianglesGen}
T(\nu_1,\nu_2&,\nu_3,\nu_4,\nu_5,\nu_6,\nu_7;D;p_1^2,p_2^2)\\
&=\left(\frac{e^{\gamma_E\epsilon}}{i\pi^{D/2}}\right)^2
\int d^Dk\int d^Dl
\frac{[(k+p_2)^2]^{-\nu_6}[(l+p_1)^2]^{-\nu_7}}
{[k^2]^{\nu_1}[(k-p_1)^2]^{\nu_2}[l^2]^{\nu_3}[(l-p_2)^2]^{\nu_4}[(k+l)^2]^{\nu_5}},
\end{split}\end{align}
with the usual conditions: $\nu_i$ integer with $\nu_6,\nu_7\leq0$ and $D=n-2\epsilon$ with $n$ even.

We will first consider the case with $p_3^2=0$ and then the
case with $p_2^2=p_3^2=0$. Our motivation to consider this example is twofold:
these diagrams have a richer diagrammatic structure than previous ones, and they
evaluate to a new type of hypergeometric function we have not yet encountered in
this paper, namely the $_3F_2$ hypergeometric function.

Let us start with the case with $p_3^2=0$. The basis for the space of integrals
defined by eq.~\eqref{eq:adjTrianglesGen} has dimension six, with a single
master integral with five propagators which we choose to be
\begin{align}\begin{split}
T(p_1^2,p_2^2)=\epsilon^4(p_1^2-p_2^2)\,T(1,1,1,1,1,0,0;4-2\epsilon;p_1^2,p_2^2)\,.
\end{split}\end{align}
The expressions for this master integral and the six independent cuts
are listed in appendix~\ref{sec:adjTrinSym}. Following the usual procedure
we find the diagrammatic coaction:
\begin{align}
\label{2madjTrianglesCoaction}
\nonumber&\Delta\left[\!
\begin{tikzpicture}[baseline={([yshift=-.5ex]current bounding box.center)}]
\coordinate (G1) at (0,0);
\coordinate (G2) at (-2/3,0);
\coordinate (G3) at (2/3,0);
\coordinate (G4) at (0,1);
\coordinate [below left=1/3 of G2] (H1);
\coordinate [below right=1/3 of G3] (H2);
\coordinate [above=1/3 of G4] (H3);
\coordinate (I1) at (1/3,0);
\coordinate (I2) at (-1/3,0);
\coordinate (I3) at (1/3,1/2);
\coordinate (I4) at (-1/3,1/2);
\coordinate (I5) at (0,1/3);
\coordinate (I11) at (1/3,-8/40);
\coordinate (I12) at (1/3,8/40);
\coordinate (I21) at (-1/3,-8/40);
\coordinate (I22) at (-1/3,8/40);
\coordinate (I31) at (1/3+12/80,1/2+4/40);
\coordinate (I32) at (1/3-12/80,1/2-4/40);
\coordinate (I41) at (-1/3-12/80,1/2+4/40);
\coordinate (I42) at (-1/3+12/80,1/2-4/40);
\coordinate (I51) at (-10/40,1/3);
\coordinate (I52) at (10/40,1/3);
\draw (G1) -- (G2);
\draw (G2) -- (G3);
\draw (G3) -- (G4);
\draw (G4) -- (G1);
\draw (G2) -- (G4);
\draw (G2) [line width=0.75 mm]-- (H1);
\draw (G3) [line width=0.75 mm]-- (H2);
\draw (G4) -- (H3);
\node at (H1) [below left=0,scale=0.7] {$\small{p_1}$};
\node at (H2) [below right=0,scale=0.7] {$\small{p_2}$};
\node at (H3) [above=0,scale=0.7] {$\small{p_3}$};
\node at (-1/3,0) [below=0,scale=0.7] {$\small{1}$};
\node at (0,0.35) [left=0,scale=0.7] {$\small{5}$};
\node at (-1/3,1/2) [above left=0,scale=0.7] {$\small{2}$};
\node at (1/3,0) [below=0,scale=0.7] {$\small{3}$};
\node at (1/3,1/2) [above right=0,scale=0.7] {$\small{4}$};
\end{tikzpicture}\!\right]=\!\!
\begin{tikzpicture}[baseline={([yshift=-.5ex]current bounding box.center)}]
\coordinate (G1) at (0,0);
\coordinate (G2) at (-2/3,0);
\coordinate (G3) at (2/3,0);
\coordinate (G4) at (0,1);
\coordinate [below left=1/3 of G2] (H1);
\coordinate [below right=1/3 of G3] (H2);
\coordinate [above=1/3 of G4] (H3);
\coordinate (I1) at (1/3,0);
\coordinate (I2) at (-1/3,0);
\coordinate (I3) at (1/3,1/2);
\coordinate (I4) at (-1/3,1/2);
\coordinate (I5) at (0,1/3);
\coordinate (I11) at (1/3,-8/40);
\coordinate (I12) at (1/3,8/40);
\coordinate (I21) at (-1/3,-8/40);
\coordinate (I22) at (-1/3,8/40);
\coordinate (I31) at (1/3+12/80,1/2+4/40);
\coordinate (I32) at (1/3-12/80,1/2-4/40);
\coordinate (I41) at (-1/3-12/80,1/2+4/40);
\coordinate (I42) at (-1/3+12/80,1/2-4/40);
\coordinate (I51) at (-10/40,1/3);
\coordinate (I52) at (10/40,1/3);
\draw (G1) -- (G2);
\draw (G2) -- (G3);
\draw (G3) -- (G4);
\draw (G4) -- (G1);
\draw (G2) -- (G4);
\draw (G2) [line width=0.75 mm]-- (H1);
\draw (G3) [line width=0.75 mm]-- (H2);
\draw (G4) -- (H3);
\node at (H1) [below left=0,scale=0.7] {$\small{p_1}$};
\node at (H2) [below right=0,scale=0.7] {$\small{p_2}$};
\node at (H3) [above=0,scale=0.7] {$\small{p_3}$};
\node at (-1/3,0) [below=0,scale=0.7] {$\small{1}$};
\node at (0,0.35) [left=0,scale=0.7] {$\small{5}$};
\node at (-1/3,1/2) [above left=0,scale=0.7] {$\small{2}$};
\node at (1/3,0) [below=0,scale=0.7] {$\small{3}$};
\node at (1/3,1/2) [above right=0,scale=0.7] {$\small{4}$};
\end{tikzpicture}\!\!\otimes\!\!
\begin{tikzpicture}[baseline={([yshift=-.5ex]current bounding box.center)}]
\coordinate (G1) at (0,0);
\coordinate (G2) at (-2/3,0);
\coordinate (G3) at (2/3,0);
\coordinate (G4) at (0,1);
\coordinate [below left=1/3 of G2] (H1);
\coordinate [below right=1/3 of G3] (H2);
\coordinate [above=1/3 of G4] (H3);
\coordinate (I1) at (1/3,0);
\coordinate (I2) at (-1/3,0);
\coordinate (I3) at (1/3,1/2);
\coordinate (I4) at (-1/3,1/2);
\coordinate (I5) at (0,1/3);
\coordinate (I11) at (1/3,-8/40);
\coordinate (I12) at (1/3,8/40);
\coordinate (I21) at (-1/3,-8/40);
\coordinate (I22) at (-1/3,8/40);
\coordinate (I31) at (1/3+12/80,1/2+4/40);
\coordinate (I32) at (1/3-12/80,1/2-4/40);
\coordinate (I41) at (-1/3-12/80,1/2+4/40);
\coordinate (I42) at (-1/3+12/80,1/2-4/40);
\coordinate (I51) at (-10/40,1/3);
\coordinate (I52) at (10/40,1/3);
\draw (G1) -- (G2);
\draw (G2) -- (G3);
\draw (G3) -- (G4);
\draw (G4) -- (G1);
\draw (G2) -- (G4);
\draw (G2) [line width=0.75 mm]-- (H1);
\draw (G3) [line width=0.75 mm]-- (H2);
\draw (G4) -- (H3);
\node at (H1) [below left=0,scale=0.7] {$\small{p_1}$};
\node at (H2) [below right=0,scale=0.7] {$\small{p_2}$};
\node at (H3) [above=0,scale=0.7] {$\small{p_3}$};
\node at (-1/6,0) [below=0,scale=0.7] {$\small{1}$};
\node at (0.1,0.55) [left=0,scale=0.7] {$\small{5}$};
\node at (-1/4,0.6) [above left=0,scale=0.7] {$\small{2}$};
\node at (1/6,0) [below=0,scale=0.7] {$\small{3}$};
\node at (1/4,0.6) [above right=0,scale=0.7] {$\small{4}$};
\draw (I11) [dashed,color=red,line width=0.5 mm] -- (I12);
\draw (I21) [dashed,color=red,line width=0.5 mm] -- (I22);
\draw (I31) [dashed,color=red,line width=0.5 mm] -- (I32);
\draw (I41) [dashed,color=red,line width=0.5 mm] -- (I42);
\draw (I51) [dashed,color=red,line width=0.5 mm] -- (I52);
\end{tikzpicture}\!\!\!\!+
\begin{tikzpicture}[baseline={([yshift=-.5ex]current bounding box.center)}]
\coordinate (G1) at (0,0);
\coordinate (G2) at (1,0.57735026919);
\coordinate (G3) at (1,-0.57735026919);
\coordinate (H1) at (-1/3,0);
\coordinate [above right = 1/3 of G2](H2);
\coordinate [below right = 1/3 of G3](H3);
\coordinate (I1) at (1/2,0.28867513459);
\coordinate [above left=1/4 and 1/8 of I1](I11);
\coordinate [below right = 1/4 and 1/8 of I1](I12);
\coordinate (I2) at (1/2,-0.28867513459);
\coordinate [below left = 1/4 and 1/8 of I2](I21);
\coordinate [above right = 1/4 and 1/8 of I2](I22);
\coordinate (I3) at (1,0);
\coordinate (I31) at (3/4,0);
\coordinate (I32) at (5/4,0);
\coordinate [above left=1/4 of G2] (J1);
\draw (G1) -- (G2);
\draw (G1) -- (G3);
\draw (G2) to[out=-110,in=110] (G3);
\draw (G2) to[out=-70,in=70] (G3);
\draw (G1) [line width=0.75 mm] -- (H1);
\draw (G2) -- (H2);
\draw (G3) [line width=0.75 mm] -- (H3);
\node at (H1) [left=0,scale=0.7] {\small$p_2$};
\node at (H2) [above right=0,scale=0.7] {\small$p_3$};
\node at (H3) [below right=0,scale=0.7] {\small$p_1$};
\node at (1/2,0.57735026919/2) [above left=0,scale=0.7] {\small$4$};
\node at (1/2,-0.57735026919/2) [below left=0,scale=0.7] {\small$3$};
\node at (0.7,0) [scale=0.7] {\small$5$};
\node at (1.3,0) [scale=0.7] {\small$2$};
\end{tikzpicture}\otimes\!\!
\begin{tikzpicture}[baseline={([yshift=-.5ex]current bounding box.center)}]
\coordinate (G1) at (0,0);
\coordinate (G2) at (-2/3,0);
\coordinate (G3) at (2/3,0);
\coordinate (G4) at (0,1);
\coordinate [below left=1/3 of G2] (H1);
\coordinate [below right=1/3 of G3] (H2);
\coordinate [above=1/3 of G4] (H3);
\coordinate (I1) at (1/3,0);
\coordinate (I2) at (-1/3,0);
\coordinate (I3) at (1/3,1/2);
\coordinate (I4) at (-1/3,1/2);
\coordinate (I5) at (0,1/3);
\coordinate (I11) at (1/3,-8/40);
\coordinate (I12) at (1/3,8/40);
\coordinate (I21) at (-1/3,-8/40);
\coordinate (I22) at (-1/3,8/40);
\coordinate (I31) at (1/3+12/80,1/2+4/40);
\coordinate (I32) at (1/3-12/80,1/2-4/40);
\coordinate (I41) at (-1/3-12/80,1/2+4/40);
\coordinate (I42) at (-1/3+12/80,1/2-4/40);
\coordinate (I51) at (-10/40,1/3);
\coordinate (I52) at (10/40,1/3);
\draw (G1) -- (G2);
\draw (G2) -- (G3);
\draw (G3) -- (G4);
\draw (G4) -- (G1);
\draw (G2) -- (G4);
\draw (G2) [line width=0.75 mm]-- (H1);
\draw (G3) [line width=0.75 mm]-- (H2);
\draw (G4) -- (H3);
\node at (H1) [below left=0,scale=0.7] {$\small{p_1}$};
\node at (H2) [below right=0,scale=0.7] {$\small{p_2}$};
\node at (H3) [above=0,scale=0.7] {$\small{p_3}$};
\node at (-1/6,0) [below=0,scale=0.7] {$\small{1}$};
\node at (0.1,0.55) [left=0,scale=0.7] {$\small{5}$};
\node at (-1/4,0.6) [above left=0,scale=0.7] {$\small{2}$};
\node at (1/6,0) [below=0,scale=0.7] {$\small{3}$};
\node at (1/4,0.6) [above right=0,scale=0.7] {$\small{4}$};
\draw (I11) [dashed,color=red,line width=0.5 mm] -- (I12);
\draw (I31) [dashed,color=red,line width=0.5 mm] -- (I32);
\draw (I41) [dashed,color=red,line width=0.5 mm] -- (I42);
\draw (I51) [dashed,color=red,line width=0.5 mm] -- (I52);
\end{tikzpicture}\\
&\qquad
+\begin{tikzpicture}[baseline={([yshift=-.5ex]current bounding box.center)}]
\coordinate (G1) at (0,0);
\coordinate (G2) at (1,0.57735026919);
\coordinate (G3) at (1,-0.57735026919);
\coordinate (H1) at (-1/3,0);
\coordinate [above right = 1/3 of G2](H2);
\coordinate [below right = 1/3 of G3](H3);
\coordinate (I1) at (1/2,0.28867513459);
\coordinate [above left=1/4 and 1/8 of I1](I11);
\coordinate [below right = 1/4 and 1/8 of I1](I12);
\coordinate (I2) at (1/2,-0.28867513459);
\coordinate [below left = 1/4 and 1/8 of I2](I21);
\coordinate [above right = 1/4 and 1/8 of I2](I22);
\coordinate (I3) at (1,0);
\coordinate (I31) at (3/4,0);
\coordinate (I32) at (5/4,0);
\coordinate [above left=1/4 of G2] (J1);
\draw (G1) -- (G2);
\draw (G1) -- (G3);
\draw (G2) to[out=-110,in=110] (G3);
\draw (G2) to[out=-70,in=70] (G3);
\draw (G1) [line width=0.75 mm] -- (H1);
\draw (G2) -- (H2);
\draw (G3) [line width=0.75 mm] -- (H3);
\node at (H1) [left=0,scale=0.7] {\small$p_1$};
\node at (H2) [above right=0,scale=0.7] {\small$p_3$};
\node at (H3) [below right=0,scale=0.7] {\small$p_2$};
\node at (1/2,0.57735026919/2) [above left=0,scale=0.7] {\small$2$};
\node at (1/2,-0.57735026919/2) [below left=0,scale=0.7] {\small$1$};
\node at (0.7,0) [scale=0.7] {\small$5$};
\node at (1.3,0) [scale=0.7] {\small$4$};
\end{tikzpicture}\otimes\!\!
\begin{tikzpicture}[baseline={([yshift=-.5ex]current bounding box.center)}]
\coordinate (G1) at (0,0);
\coordinate (G2) at (-2/3,0);
\coordinate (G3) at (2/3,0);
\coordinate (G4) at (0,1);
\coordinate [below left=1/3 of G2] (H1);
\coordinate [below right=1/3 of G3] (H2);
\coordinate [above=1/3 of G4] (H3);
\coordinate (I1) at (1/3,0);
\coordinate (I2) at (-1/3,0);
\coordinate (I3) at (1/3,1/2);
\coordinate (I4) at (-1/3,1/2);
\coordinate (I5) at (0,1/3);
\coordinate (I11) at (1/3,-8/40);
\coordinate (I12) at (1/3,8/40);
\coordinate (I21) at (-1/3,-8/40);
\coordinate (I22) at (-1/3,8/40);
\coordinate (I31) at (1/3+12/80,1/2+4/40);
\coordinate (I32) at (1/3-12/80,1/2-4/40);
\coordinate (I41) at (-1/3-12/80,1/2+4/40);
\coordinate (I42) at (-1/3+12/80,1/2-4/40);
\coordinate (I51) at (-10/40,1/3);
\coordinate (I52) at (10/40,1/3);
\draw (G1) -- (G2);
\draw (G2) -- (G3);
\draw (G3) -- (G4);
\draw (G4) -- (G1);
\draw (G2) -- (G4);
\draw (G2) [line width=0.75 mm]-- (H1);
\draw (G3) [line width=0.75 mm]-- (H2);
\draw (G4) -- (H3);
\node at (H1) [below left=0,scale=0.7] {$\small{p_1}$};
\node at (H2) [below right=0,scale=0.7] {$\small{p_2}$};
\node at (H3) [above=0,scale=0.7] {$\small{p_3}$};
\node at (-1/6,0) [below=0,scale=0.7] {$\small{1}$};
\node at (0.1,0.55) [left=0,scale=0.7] {$\small{5}$};
\node at (-1/4,0.6) [above left=0,scale=0.7] {$\small{2}$};
\node at (1/6,0) [below=0,scale=0.7] {$\small{3}$};
\node at (1/4,0.6) [above right=0,scale=0.7] {$\small{4}$};
\draw (I21) [dashed,color=red,line width=0.5 mm] -- (I22);
\draw (I31) [dashed,color=red,line width=0.5 mm] -- (I32);
\draw (I41) [dashed,color=red,line width=0.5 mm] -- (I42);
\draw (I51) [dashed,color=red,line width=0.5 mm] -- (I52);
\end{tikzpicture}\!\!
+\begin{tikzpicture}[baseline={([yshift=-.5ex]current bounding box.center)}]
\coordinate (G1) at (0,0);
\coordinate (G2) at (1,0);
\coordinate (G3) at (2,0);
\coordinate (H1) at (-1/3,0);
\coordinate (H2) at (1+4/3,0);
\coordinate (H3) at (1,1/3);
\coordinate (H4) at (1,-1/3);
\coordinate (I1) at (1/2,1/2);
\coordinate (I2) at (1/2,1/8);
\coordinate (I3) at (1/2,-1/2);
\coordinate (I4) at (1/2,-1/8);
\coordinate (J1) at (1,1/4);
\coordinate (J2) at (1,-1/4);
\coordinate (K1) at (1/2,-1/8);
\coordinate (K2) at (3/2,-1/8);
\draw (G1) to[out=80,in=100] (G2);
\draw (G1) to[out=-80,in=-100] (G2);
\draw (G2) to[out=80,in=100] (G3);
\draw (G2) to[out=-80,in=-100] (G3);
\draw (G1) [line width=0.75 mm] -- (H1);
\draw (G3) [line width=0.75 mm]-- (H2);
\draw (G2) -- (H3);
\node at (K1) [above=3.7 mm of K1,scale=0.7] {\small$2$};
\node at (K1) [below=1.3 mm of K1,scale=0.7] {\small$1$};
\node at (K2) [above=3.7 mm of K2,scale=0.7] {\small$4$};
\node at (K2) [below=1.3 mm of K2,scale=0.7] {\small$3$};
\node at (-1/3,0) [left=0,scale=0.7] {\small$p_1$};
\node at (7/3,0) [right=0,scale=0.7] {\small$p_2$};
\node at (1,1/3) [above=0,scale=0.7] {\small$p_3$};
\node at (H4) [below=0 mm of H4] {\vphantom{\small$2$}};
\end{tikzpicture}\otimes\!\!
\begin{tikzpicture}[baseline={([yshift=-.5ex]current bounding box.center)}]
\coordinate (G1) at (0,0);
\coordinate (G2) at (-2/3,0);
\coordinate (G3) at (2/3,0);
\coordinate (G4) at (0,1);
\coordinate [below left=1/3 of G2] (H1);
\coordinate [below right=1/3 of G3] (H2);
\coordinate [above=1/3 of G4] (H3);
\coordinate (I1) at (1/3,0);
\coordinate (I2) at (-1/3,0);
\coordinate (I3) at (1/3,1/2);
\coordinate (I4) at (-1/3,1/2);
\coordinate (I5) at (0,1/3);
\coordinate (I11) at (1/3,-8/40);
\coordinate (I12) at (1/3,8/40);
\coordinate (I21) at (-1/3,-8/40);
\coordinate (I22) at (-1/3,8/40);
\coordinate (I31) at (1/3+12/80,1/2+4/40);
\coordinate (I32) at (1/3-12/80,1/2-4/40);
\coordinate (I41) at (-1/3-12/80,1/2+4/40);
\coordinate (I42) at (-1/3+12/80,1/2-4/40);
\coordinate (I51) at (-10/40,1/3);
\coordinate (I52) at (10/40,1/3);
\draw (G1) -- (G2);
\draw (G2) -- (G3);
\draw (G3) -- (G4);
\draw (G4) -- (G1);
\draw (G2) -- (G4);
\draw (G2) [line width=0.75 mm]-- (H1);
\draw (G3) [line width=0.75 mm]-- (H2);
\draw (G4) -- (H3);
\node at (H1) [below left=0,scale=0.7] {$\small{p_1}$};
\node at (H2) [below right=0,scale=0.7] {$\small{p_2}$};
\node at (H3) [above=0,scale=0.7] {$\small{p_3}$};
\node at (-1/6,0) [below=0,scale=0.7] {$\small{1}$};
\node at (0.1,0.55) [left=0,scale=0.7] {$\small{5}$};
\node at (-1/4,0.6) [above left=0,scale=0.7] {$\small{2}$};
\node at (1/6,0) [below=0,scale=0.7] {$\small{3}$};
\node at (1/4,0.6) [above right=0,scale=0.7] {$\small{4}$};
\draw (I11) [dashed,color=red,line width=0.5 mm] -- (I12);
\draw (I21) [dashed,color=red,line width=0.5 mm] -- (I22);
\draw (I31) [dashed,color=red,line width=0.5 mm] -- (I32);
\draw (I41) [dashed,color=red,line width=0.5 mm] -- (I42);
\end{tikzpicture}\\
\nonumber&
\qquad+\begin{tikzpicture}[baseline={([yshift=-.5ex]current bounding box.center)}]
\coordinate (G1) at (0,0);
\coordinate (G2) at (1,0);
\coordinate (H1) at (-1/3,0);
\coordinate (H2) at (4/3,0);
\coordinate (I1) at (1/2,1/2);
\coordinate (I2) at (1/2,1/8);
\coordinate (I3) at (1/2,-1/2);
\coordinate (I4) at (1/2,-1/8);
\coordinate (J1) at (1,1/4);
\coordinate (J2) at (1,-1/4);
\coordinate (K1) at (1/2,-1/8);
\draw (G1) -- (G2);
\draw (G1) to[out=80,in=100] (G2);
\draw (G1) to[out=-80,in=-100] (G2);
\draw (G1) [line width=0.75 mm] -- (H1);
\draw (G2) [line width=0.75 mm]-- (H2);
\node at (K1) [above=0.7 mm of K1,scale=0.7] {\small$5$};
\node at (K1) [above=3.7 mm of K1,scale=0.7] {\small$2$};
\node at (K1) [below=1.3 mm of K1,scale=0.7] {\small$3$};
\node at (H1) [left=0,scale=0.7] {\small$p_1$};
\node at (H2) [right=0,scale=0.7] {\small$p_1$};
\end{tikzpicture}\otimes
\begin{tikzpicture}[baseline={([yshift=-.5ex]current bounding box.center)}]
\coordinate (G1) at (0,0);
\coordinate (G2) at (-2/3,0);
\coordinate (G3) at (2/3,0);
\coordinate (G4) at (0,1);
\coordinate [below left=1/3 of G2] (H1);
\coordinate [below right=1/3 of G3] (H2);
\coordinate [above=1/3 of G4] (H3);
\coordinate (I1) at (1/3,0);
\coordinate (I2) at (-1/3,0);
\coordinate (I3) at (1/3,1/2);
\coordinate (I4) at (-1/3,1/2);
\coordinate (I5) at (0,1/3);
\coordinate (I11) at (1/3,-8/40);
\coordinate (I12) at (1/3,8/40);
\coordinate (I21) at (-1/3,-8/40);
\coordinate (I22) at (-1/3,8/40);
\coordinate (I31) at (1/3+12/80,1/2+4/40);
\coordinate (I32) at (1/3-12/80,1/2-4/40);
\coordinate (I41) at (-1/3-12/80,1/2+4/40);
\coordinate (I42) at (-1/3+12/80,1/2-4/40);
\coordinate (I51) at (-10/40,1/3);
\coordinate (I52) at (10/40,1/3);
\draw (G1) -- (G2);
\draw (G2) -- (G3);
\draw (G3) -- (G4);
\draw (G4) -- (G1);
\draw (G2) -- (G4);
\draw (G2) [line width=0.75 mm]-- (H1);
\draw (G3) [line width=0.75 mm]-- (H2);
\draw (G4) -- (H3);
\node at (H1) [below left=0,scale=0.7] {$\small{p_1}$};
\node at (H2) [below right=0,scale=0.7] {$\small{p_2}$};
\node at (H3) [above=0,scale=0.7] {$\small{p_3}$};
\node at (-1/6,0) [below=0,scale=0.7] {$\small{1}$};
\node at (0.1,0.55) [left=0,scale=0.7] {$\small{5}$};
\node at (-1/4,0.6) [above left=0,scale=0.7] {$\small{2}$};
\node at (1/6,0) [below=0,scale=0.7] {$\small{3}$};
\node at (1/4,0.6) [above right=0,scale=0.7] {$\small{4}$};
\draw (I11) [dashed,color=red,line width=0.5 mm] -- (I12);
\draw (I41) [dashed,color=red,line width=0.5 mm] -- (I42);
\draw (I51) [dashed,color=red,line width=0.5 mm] -- (I52);
\end{tikzpicture}+
\begin{tikzpicture}[baseline={([yshift=-.5ex]current bounding box.center)}]
\coordinate (G1) at (0,0);
\coordinate (G2) at (1,0);
\coordinate (H1) at (-1/3,0);
\coordinate (H2) at (4/3,0);
\coordinate (I1) at (1/2,1/2);
\coordinate (I2) at (1/2,1/8);
\coordinate (I3) at (1/2,-1/2);
\coordinate (I4) at (1/2,-1/8);
\coordinate (J1) at (1,1/4);
\coordinate (J2) at (1,-1/4);
\coordinate (K1) at (1/2,-1/8);
\draw (G1) -- (G2);
\draw (G1) to[out=80,in=100] (G2);
\draw (G1) to[out=-80,in=-100] (G2);
\draw (G1) [line width=0.75 mm] -- (H1);
\draw (G2) [line width=0.75 mm]-- (H2);
\node at (K1) [above=0.7 mm of K1,scale=0.7] {\small$5$};
\node at (K1) [above=3.7 mm of K1,scale=0.7] {\small$4$};
\node at (K1) [below=1.3 mm of K1,scale=0.7] {\small$1$};
\node at (H1) [left=0,scale=0.7] {\small$p_2$};
\node at (H2) [right=0,scale=0.7] {\small$p_2$};
\end{tikzpicture}\otimes
\begin{tikzpicture}[baseline={([yshift=-.5ex]current bounding box.center)}]
\coordinate (G1) at (0,0);
\coordinate (G2) at (-2/3,0);
\coordinate (G3) at (2/3,0);
\coordinate (G4) at (0,1);
\coordinate [below left=1/3 of G2] (H1);
\coordinate [below right=1/3 of G3] (H2);
\coordinate [above=1/3 of G4] (H3);
\coordinate (I1) at (1/3,0);
\coordinate (I2) at (-1/3,0);
\coordinate (I3) at (1/3,1/2);
\coordinate (I4) at (-1/3,1/2);
\coordinate (I5) at (0,1/3);
\coordinate (I11) at (1/3,-8/40);
\coordinate (I12) at (1/3,8/40);
\coordinate (I21) at (-1/3,-8/40);
\coordinate (I22) at (-1/3,8/40);
\coordinate (I31) at (1/3+12/80,1/2+4/40);
\coordinate (I32) at (1/3-12/80,1/2-4/40);
\coordinate (I41) at (-1/3-12/80,1/2+4/40);
\coordinate (I42) at (-1/3+12/80,1/2-4/40);
\coordinate (I51) at (-10/40,1/3);
\coordinate (I52) at (10/40,1/3);
\draw (G1) -- (G2);
\draw (G2) -- (G3);
\draw (G3) -- (G4);
\draw (G4) -- (G1);
\draw (G2) -- (G4);
\draw (G2) [line width=0.75 mm]-- (H1);
\draw (G3) [line width=0.75 mm]-- (H2);
\draw (G4) -- (H3);
\node at (H1) [below left=0,scale=0.7] {$\small{p_1}$};
\node at (H2) [below right=0,scale=0.7] {$\small{p_2}$};
\node at (H3) [above=0,scale=0.7] {$\small{p_3}$};
\node at (-1/6,0) [below=0,scale=0.7] {$\small{1}$};
\node at (0.1,0.55) [left=0,scale=0.7] {$\small{5}$};
\node at (-1/4,0.6) [above left=0,scale=0.7] {$\small{2}$};
\node at (1/6,0) [below=0,scale=0.7] {$\small{3}$};
\node at (1/4,0.6) [above right=0,scale=0.7] {$\small{4}$};
\draw (I21) [dashed,color=red,line width=0.5 mm] -- (I22);
\draw (I31) [dashed,color=red,line width=0.5 mm] -- (I32);
\draw (I51) [dashed,color=red,line width=0.5 mm] -- (I52);
\end{tikzpicture}\,,
\end{align}
which satisfies all properties of section~\ref{sec:propDiagCoac}
and provides further nontrivial evidence for the existence of 
a diagrammatic coaction beyond one loop. In this coaction there is a single
two-loop diagram that we have not yet discussed, namely the product
of two one-loop bubbles. As discussed at the end of section~\ref{sec:doubTad}, its coaction is fully determined by
the coaction on one-loop integrals \cite{Abreu:2017enx,Abreu:2017mtm}.

As was done for the double-edged triangle, we can obtain the diagrammatic
coaction for the case where $p_2^2=p_3^2=0$ by taking the $p_2^2\to 0$ limit of
the coaction in eq.~(\ref{2madjTrianglesCoaction}). 
To this end one may compute an asymptotic expansion of the hypergeometric functions for the uncut integrals and the various 
cuts in~eq.~(\ref{2madjTrianglesCoaction}) at vanishing $p_2^2$. 
Because the massless limit exposes new infrared divergences, this asymptotic expansion
must be carefully computed\footnote{We provide a pedagogical example in appendix~\ref{sec:adjTrinSym}.} keeping $\epsilon<0$. 
Specifically, considering the second entries in~eq.~(\ref{2madjTrianglesCoaction}) for $p_2^2\to 0$ we find,
using the expressions in 
appendix \ref{sec:adjTrinSym}, that the maximal cut,  $\mathcal{C}_{2,3,4,5}$ and $\mathcal{C}_{1,2,3,4}$ all vanish for $\epsilon<0$, while amongst the remaining first entries the $p_2^2$ sunset vanishes. This leaves just two terms in the coaction, which takes the form: 
\begin{align}
\label{1madjTrianglesCoaction}
\Delta\left[\begin{tikzpicture}[baseline={([yshift=-.5ex]current bounding box.center)}]
\coordinate (G1) at (0,0);
\coordinate (G2) at (-2/3,0);
\coordinate (G3) at (2/3,0);
\coordinate (G4) at (0,1);
\coordinate [below left=1/3 of G2] (H1);
\coordinate [below right=1/3 of G3] (H2);
\coordinate [above=1/3 of G4] (H3);
\coordinate (I1) at (1/3,0);
\coordinate (I2) at (-1/3,0);
\coordinate (I3) at (1/3,1/2);
\coordinate (I4) at (-1/3,1/2);
\coordinate (I5) at (0,1/3);
\coordinate (I11) at (1/3,-8/40);
\coordinate (I12) at (1/3,8/40);
\coordinate (I21) at (-1/3,-8/40);
\coordinate (I22) at (-1/3,8/40);
\coordinate (I31) at (1/3+12/80,1/2+4/40);
\coordinate (I32) at (1/3-12/80,1/2-4/40);
\coordinate (I41) at (-1/3-12/80,1/2+4/40);
\coordinate (I42) at (-1/3+12/80,1/2-4/40);
\coordinate (I51) at (-10/40,1/3);
\coordinate (I52) at (10/40,1/3);
\draw (G1) -- (G2);
\draw (G2) -- (G3);
\draw (G3) -- (G4);
\draw (G4) -- (G1);
\draw (G2) -- (G4);
\draw (G2) [line width=0.75 mm]-- (H1);
\draw (G3) -- (H2);
\draw (G4) -- (H3);
\node at (H1) [below left=0,scale=0.7] {$\small{p_1}$};
\node at (H2) [below right=0,scale=0.7] {$\small{p_2}$};
\node at (H3) [above=0,scale=0.7] {$\small{p_3}$};
\node at (-1/3,0) [below=0,scale=0.7] {$\small{1}$};
\node at (0,0.35) [left=0,scale=0.7] {$\small{5}$};
\node at (-1/3,1/2) [above left=0,scale=0.7] {$\small{2}$};
\node at (1/3,0) [below=0,scale=0.7] {$\small{3}$};
\node at (1/3,1/2) [above right=0,scale=0.7] {$\small{4}$};
\end{tikzpicture}\right]
=\begin{tikzpicture}[baseline={([yshift=-.5ex]current bounding box.center)}]
\coordinate (G1) at (0,0);
\coordinate (G2) at (1,0.57735026919);
\coordinate (G3) at (1,-0.57735026919);
\coordinate (H1) at (-1/3,0);
\coordinate [above right = 1/3 of G2](H2);
\coordinate [below right = 1/3 of G3](H3);
\coordinate (I1) at (1/2,0.28867513459);
\coordinate [above left=1/4 and 1/8 of I1](I11);
\coordinate [below right = 1/4 and 1/8 of I1](I12);
\coordinate (I2) at (1/2,-0.28867513459);
\coordinate [below left = 1/4 and 1/8 of I2](I21);
\coordinate [above right = 1/4 and 1/8 of I2](I22);
\coordinate (I3) at (1,0);
\coordinate (I31) at (3/4,0);
\coordinate (I32) at (5/4,0);
\coordinate [above left=1/4 of G2] (J1);
\draw (G1) -- (G2);
\draw (G1) -- (G3);
\draw (G2) to[out=-110,in=110] (G3);
\draw (G2) to[out=-70,in=70] (G3);
\draw (G1) [line width=0.75 mm] -- (H1);
\draw (G2) -- (H2);
\draw (G3) -- (H3);
\node at (H1) [left=0,scale=0.7] {\small$p_1$};
\node at (H2) [above right=0,scale=0.7] {\small$p_3$};
\node at (H3) [below right=0,scale=0.7] {\small$p_2$};
\node at (1/2,0.57735026919/2) [above left=0,scale=0.7] {\small$2$};
\node at (1/2,-0.57735026919/2) [below left=0,scale=0.7] {\small$1$};
\node at (0.7,0) [scale=0.7] {\small$5$};
\node at (1.3,0) [scale=0.7] {\small$4$};
\end{tikzpicture}\otimes\!\!
\begin{tikzpicture}[baseline={([yshift=-.5ex]current bounding box.center)}]
\coordinate (G1) at (0,0);
\coordinate (G2) at (-2/3,0);
\coordinate (G3) at (2/3,0);
\coordinate (G4) at (0,1);
\coordinate [below left=1/3 of G2] (H1);
\coordinate [below right=1/3 of G3] (H2);
\coordinate [above=1/3 of G4] (H3);
\coordinate (I1) at (1/3,0);
\coordinate (I2) at (-1/3,0);
\coordinate (I3) at (1/3,1/2);
\coordinate (I4) at (-1/3,1/2);
\coordinate (I5) at (0,1/3);
\coordinate (I11) at (1/3,-8/40);
\coordinate (I12) at (1/3,8/40);
\coordinate (I21) at (-1/3,-8/40);
\coordinate (I22) at (-1/3,8/40);
\coordinate (I31) at (1/3+12/80,1/2+4/40);
\coordinate (I32) at (1/3-12/80,1/2-4/40);
\coordinate (I41) at (-1/3-12/80,1/2+4/40);
\coordinate (I42) at (-1/3+12/80,1/2-4/40);
\coordinate (I51) at (-10/40,1/3);
\coordinate (I52) at (10/40,1/3);
\draw (G1) -- (G2);
\draw (G2) -- (G3);
\draw (G3) -- (G4);
\draw (G4) -- (G1);
\draw (G2) -- (G4);
\draw (G2) [line width=0.75 mm]-- (H1);
\draw (G3) -- (H2);
\draw (G4) -- (H3);
\node at (H1) [below left=0,scale=0.7] {$\small{p_1}$};
\node at (H2) [below right=0,scale=0.7] {$\small{p_2}$};
\node at (H3) [above=0,scale=0.7] {$\small{p_3}$};
\node at (-1/6,0) [below=0,scale=0.7] {$\small{1}$};
\node at (0.1,0.55) [left=0,scale=0.7] {$\small{5}$};
\node at (-1/4,0.6) [above left=0,scale=0.7] {$\small{2}$};
\node at (1/6,0) [below=0,scale=0.7] {$\small{3}$};
\node at (1/4,0.6) [above right=0,scale=0.7] {$\small{4}$};
\draw (I21) [dashed,color=red,line width=0.5 mm] -- (I22);
\draw (I31) [dashed,color=red,line width=0.5 mm] -- (I32);
\draw (I41) [dashed,color=red,line width=0.5 mm] -- (I42);
\draw (I51) [dashed,color=red,line width=0.5 mm] -- (I52);
\end{tikzpicture}\!\!\!\!\!\!\!+
\begin{tikzpicture}[baseline={([yshift=-.5ex]current bounding box.center)}]
\coordinate (G1) at (0,0);
\coordinate (G2) at (1,0);
\coordinate (H1) at (-1/3,0);
\coordinate (H2) at (4/3,0);
\coordinate (I1) at (1/2,1/2);
\coordinate (I2) at (1/2,1/8);
\coordinate (I3) at (1/2,-1/2);
\coordinate (I4) at (1/2,-1/8);
\coordinate (J1) at (1,1/4);
\coordinate (J2) at (1,-1/4);
\coordinate (K1) at (1/2,-1/8);
\draw (G1) -- (G2);
\draw (G1) to[out=80,in=100] (G2);
\draw (G1) to[out=-80,in=-100] (G2);
\draw (G1) [line width=0.75 mm] -- (H1);
\draw (G2) [line width=0.75 mm]-- (H2);
\node at (K1) [above=0.7 mm of K1,scale=0.7] {\small$5$};
\node at (K1) [above=3.7 mm of K1,scale=0.7] {\small$2$};
\node at (K1) [below=1.3 mm of K1,scale=0.7] {\small$3$};
\node at (H1) [left=0,scale=0.7] {\small$p_1$};
\node at (H2) [right=0,scale=0.7] {\small$p_1$};
\end{tikzpicture}\otimes\!\!\!\!\!
\begin{tikzpicture}[baseline={([yshift=-.5ex]current bounding box.center)}]
\coordinate (G1) at (0,0);
\coordinate (G2) at (-2/3,0);
\coordinate (G3) at (2/3,0);
\coordinate (G4) at (0,1);
\coordinate [below left=1/3 of G2] (H1);
\coordinate [below right=1/3 of G3] (H2);
\coordinate [above=1/3 of G4] (H3);
\coordinate (I1) at (1/3,0);
\coordinate (I2) at (-1/3,0);
\coordinate (I3) at (1/3,1/2);
\coordinate (I4) at (-1/3,1/2);
\coordinate (I5) at (0,1/3);
\coordinate (I11) at (1/3,-8/40);
\coordinate (I12) at (1/3,8/40);
\coordinate (I21) at (-1/3,-8/40);
\coordinate (I22) at (-1/3,8/40);
\coordinate (I31) at (1/3+12/80,1/2+4/40);
\coordinate (I32) at (1/3-12/80,1/2-4/40);
\coordinate (I41) at (-1/3-12/80,1/2+4/40);
\coordinate (I42) at (-1/3+12/80,1/2-4/40);
\coordinate (I51) at (-10/40,1/3);
\coordinate (I52) at (10/40,1/3);
\draw (G1) -- (G2);
\draw (G2) -- (G3);
\draw (G3) -- (G4);
\draw (G4) -- (G1);
\draw (G2) -- (G4);
\draw (G2) [line width=0.75 mm]-- (H1);
\draw (G3) -- (H2);
\draw (G4) -- (H3);
\node at (H1) [below left=0,scale=0.7] {$\small{p_1}$};
\node at (H2) [below right=0,scale=0.7] {$\small{p_2}$};
\node at (H3) [above=0,scale=0.7] {$\small{p_3}$};
\node at (-1/6,0) [below=0,scale=0.7] {$\small{1}$};
\node at (0.1,0.55) [left=0,scale=0.7] {$\small{5}$};
\node at (-1/4,0.6) [above left=0,scale=0.7] {$\small{2}$};
\node at (1/6,0) [below=0,scale=0.7] {$\small{3}$};
\node at (1/4,0.6) [above right=0,scale=0.7] {$\small{4}$};
\draw (I11) [dashed,color=red,line width=0.5 mm] -- (I12);
\draw (I41) [dashed,color=red,line width=0.5 mm] -- (I42);
\draw (I51) [dashed,color=red,line width=0.5 mm] -- (I52);
\end{tikzpicture}\!\!\!\!\,.
\end{align}
This result agrees with the global coaction obtained directly using the
expressions in appendix~\ref{sec:adjTrinAsy}.
We also note that, consistently with the fact that the maximal cut vanishes for $p_2^2\to 0$,
the integral $T(p_1^2,0)$ itself  reduces to simpler integrals:
\begin{equation}
\label{Tp10}
	T(p_1^2,0)=P(0,0,p_1^2)-\frac{1}{4}S(p_1^2)\,,
\end{equation}
with 
$P(0,0,p_1^2)$ as defined in section~\ref{sec:doubleEdged} and 
$S(p_1^2)$ as given in section~\ref{sec:sunsetMassless}.  
Eq.~(\ref{Tp10}), along with the coactions of eqs.~(\ref{eq:massless_sunset}) and~(\ref{eq:diagCoaction1mSy}),
provides an alternative way to verify the results in~\ref{sec:adjTrinAsy}.

\subsection{Diagonal box}

\begin{figure}[htb]
		\centering
		\resizebox{4cm}{!}{
		\begin{tikzpicture}
		\coordinate (G1) at (0,0);
		\coordinate (G2) at (1,0);
		\coordinate (G3) at (1,1);
		\coordinate (G4) at (0,1);
		\coordinate [below left=1/3 of G1] (H1);
		\coordinate [below right=1/3 of G2] (H2);
		\coordinate [above right=1/3 of G3] (H3);
		\coordinate [above left=1/3 of G4] (H4);
		\coordinate (I1) at (1/2,0);
		\coordinate (I2) at (1,1/2);
		\coordinate (I3) at (1/2,1);
		\coordinate (I4) at (0,1/2);
		\coordinate (I5) at (1/2,1/2);
		\coordinate [below=1/4 of I1] (I11);
		\coordinate [above=1/4 of I1] (I12);
		\coordinate [left=1/4 of I2] (I21);
		\coordinate [right=1/4 of I2] (I22);
		\coordinate [below=1/4 of I3] (I31);
		\coordinate [above=1/4 of I3] (I32);
		\coordinate [left=1/4 of I4] (I41);
		\coordinate [right=1/4 of I4] (I42);
		\coordinate [below left=1/4 of I5] (I51);
		\coordinate [above right=1/4 of I5] (I52);
		\draw (G1) -- (G2);
		\draw (G2) -- (G3);
		\draw (G3) -- (G4);
		\draw (G4) -- (G1);
		\draw (G2) -- (G4);
		\draw (G1) -- (H1);
		\draw (G2) -- (H2);
		\draw (G3) -- (H3);
		\draw (G4) -- (H4);
		\node at (H1) [below left=0,scale=0.7] {$\small{p_1}$};
		\node at (H2) [below right=0,scale=0.7] {$\small{p_2}$};
		\node at (H3) [above right=0,scale=0.7] {$\small{p_3}$};
		\node at (H4) [above left=0,scale=0.7] {$\small{p_4}$};
		\node at (1/2,0) [below=0,scale=0.7] {$\small{1}$};
		\node at (1,1/2) [right=0,scale=0.7] {$\small{2}$};
		\node at (1/2,1) [above=0,scale=0.7] {$\small{3}$};
		\node at (0,1/2) [left=0,scale=0.7] {$\small{4}$};
		\node at (1/2+0.07,1/2+0.07) [below left=0,scale=0.7] {$\small{5}$};
		\end{tikzpicture}
		}
		\caption{Diagonal box. All propagators and external legs are massless.}
		\label{DBfig}
\end{figure}
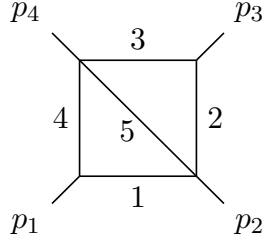

As a final example we consider the four-point two-loop diagram of 
fig.~\ref{DBfig}, which we call the `diagonal box'. 
The space of
integrals defined by this diagram is given by
\begin{align}\begin{split}
&B(\nu_1,\nu_2,\nu_3,\nu_4,\nu_5,\nu_6,\nu_7,\nu_8,\nu_9;D;s,t)\\
=&\left(\frac{e^{\gamma_E\epsilon}}{i\pi^{D/2}}\right)^2
\int d^Dk\,d^Dl\,
\frac{[(l+p_1)^2]^{-\nu_6}[(l+p_2)^2]^{-\nu_7}[(l+p_3)^2]^{-\nu_8}[(k+p_3)^2]^{-\nu_9}}
{[k^2]^{\nu_1}[(k+p_2+l)^2]^{\nu_2}[(k+p_2+p_3+l)^2]^{\nu_3}[(k-p_1)^2]^{\nu_4}[l^2]^{\nu_5}},
\end{split}\end{align}
with integer $\nu_i$ and $\nu_6,\ldots,\nu_9\leq0$ and $D=n-2\eps$ with $n$ even. 
This space is generated by three master integrals out of which a single master integral 
features all five propagators. We choose it to be
\begin{align}\begin{split}\label{eq:basisDiagBox}
	&B(s,t)=\epsilon^4(s+t)
	B(1,1,1,1,1,0,0,0,0;4-2\epsilon;s,t)\,,
\end{split}\end{align}
where $s=(p_1+p_2)^2$ and $t=(p_2+p_3)^2$. The remaining two master integrals
are the sunset integrals with external legs of mass $s$ and $t$.
The five-propagator integral $B(s,t)$ evaluates to Gauss hypergeometric 
functions~\cite{Anastasiou:1999bn}:
\begin{align}
B(s,t)=&-e^{2\gamma_E\epsilon}\frac{\epsilon(s+t)}{2(1-2\epsilon)}
\frac{\Gamma^3(1-\epsilon)\Gamma(1+2\epsilon)}{\Gamma(1-3\epsilon)}
\left(\frac{t^{-2\epsilon}}{s}\,{}_2F_1\left(1-2\epsilon,1-2\epsilon;2-2\epsilon;1+\frac{t}{s}\right)\right.\nonumber\\
&\left.\qquad+\frac{s^{-2\epsilon}}{t} \,{}_2F_1\left(1-2\epsilon,1-2\epsilon;2-2\epsilon;1+\frac{s}{t}\right)\right)\,,
\end{align}
and as such is simpler than the previous ones we considered, despite this being a four-point
function. There are three independent cuts, and they can be found
in appendix~\ref{sec:diagBox}.

Starting with the global coaction on hypergeometric functions given in eq.~\eqref{eq:coaction2F1}
we then obtain the diagrammatic coaction of the diagonal box
\begin{align}\label{DBCoaction}
\Delta\left[\begin{tikzpicture}[baseline={([yshift=-.5ex]current bounding box.center)}]
\coordinate (G1) at (0,0);
\coordinate (G2) at (1,0);
\coordinate (G3) at (1,1);
\coordinate (G4) at (0,1);
\coordinate [below left=1/3 of G1] (H1);
\coordinate [below right=1/3 of G2] (H2);
\coordinate [above right=1/3 of G3] (H3);
\coordinate [above left=1/3 of G4] (H4);
\coordinate (I1) at (1/2,0);
\coordinate (I2) at (1,1/2);
\coordinate (I3) at (1/2,1);
\coordinate (I4) at (0,1/2);
\coordinate (I5) at (1/2,1/2);
\coordinate [below=1/4 of I1] (I11);
\coordinate [above=1/4 of I1] (I12);
\coordinate [left=1/4 of I2] (I21);
\coordinate [right=1/4 of I2] (I22);
\coordinate [below=1/4 of I3] (I31);
\coordinate [above=1/4 of I3] (I32);
\coordinate [left=1/4 of I4] (I41);
\coordinate [right=1/4 of I4] (I42);
\coordinate [below left=1/4 of I5] (I51);
\coordinate [above right=1/4 of I5] (I52);
\draw (G1) -- (G2);
\draw (G2) -- (G3);
\draw (G3) -- (G4);
\draw (G4) -- (G1);
\draw (G2) -- (G4);
\draw (G1) -- (H1);
\draw (G2) -- (H2);
\draw (G3) -- (H3);
\draw (G4) -- (H4);
\node at (H1) [below left=0,scale=0.7] {$\small{p_1}$};
\node at (H2) [below right=0,scale=0.7] {$\small{p_2}$};
\node at (H3) [above right=0,scale=0.7] {$\small{p_3}$};
\node at (H4) [above left=0,scale=0.7] {$\small{p_4}$};
\node at (1/2,0) [below=0,scale=0.7] {$\small{1}$};
\node at (1,1/2) [right=0,scale=0.7] {$\small{2}$};
\node at (1/2,1) [above=0,scale=0.7] {$\small{3}$};
\node at (0,1/2) [left=0,scale=0.7] {$\small{4}$};
\node at (1/2+0.07,1/2+0.07) [below left=0,scale=0.7] {$\small{5}$};
\end{tikzpicture}\right]
=&\begin{tikzpicture}[baseline={([yshift=-.5ex]current bounding box.center)}]
\coordinate (G1) at (0,0);
\coordinate (G2) at (1,0);
\coordinate (G3) at (1,1);
\coordinate (G4) at (0,1);
\coordinate [below left=1/3 of G1] (H1);
\coordinate [below right=1/3 of G2] (H2);
\coordinate [above right=1/3 of G3] (H3);
\coordinate [above left=1/3 of G4] (H4);
\coordinate (I1) at (1/2,0);
\coordinate (I2) at (1,1/2);
\coordinate (I3) at (1/2,1);
\coordinate (I4) at (0,1/2);
\coordinate (I5) at (1/2,1/2);
\coordinate [below=1/4 of I1] (I11);
\coordinate [above=1/4 of I1] (I12);
\coordinate [left=1/4 of I2] (I21);
\coordinate [right=1/4 of I2] (I22);
\coordinate [below=1/4 of I3] (I31);
\coordinate [above=1/4 of I3] (I32);
\coordinate [left=1/4 of I4] (I41);
\coordinate [right=1/4 of I4] (I42);
\coordinate [below left=1/4 of I5] (I51);
\coordinate [above right=1/4 of I5] (I52);
\draw (G1) -- (G2);
\draw (G2) -- (G3);
\draw (G3) -- (G4);
\draw (G4) -- (G1);
\draw (G2) -- (G4);
\draw (G1) -- (H1);
\draw (G2) -- (H2);
\draw (G3) -- (H3);
\draw (G4) -- (H4);
\node at (H1) [below left=0,scale=0.7] {$\small{p_1}$};
\node at (H2) [below right=0,scale=0.7] {$\small{p_2}$};
\node at (H3) [above right=0,scale=0.7] {$\small{p_3}$};
\node at (H4) [above left=0,scale=0.7] {$\small{p_4}$};
\node at (1/2,0) [below=0,scale=0.7] {$\small{1}$};
\node at (1,1/2) [right=0,scale=0.7] {$\small{2}$};
\node at (1/2,1) [above=0,scale=0.7] {$\small{3}$};
\node at (0,1/2) [left=0,scale=0.7] {$\small{4}$};
\node at (1/2+0.07,1/2+0.07) [below left=0,scale=0.7] {$\small{5}$};
\end{tikzpicture}\otimes\begin{tikzpicture}[baseline={([yshift=-.5ex]current bounding box.center)}]
\coordinate (G1) at (0,0);
\coordinate (G2) at (1,0);
\coordinate (G3) at (1,1);
\coordinate (G4) at (0,1);
\coordinate [below left=1/3 of G1] (H1);
\coordinate [below right=1/3 of G2] (H2);
\coordinate [above right=1/3 of G3] (H3);
\coordinate [above left=1/3 of G4] (H4);
\coordinate (I1) at (1/2,0);
\coordinate (I2) at (1,1/2);
\coordinate (I3) at (1/2,1);
\coordinate (I4) at (0,1/2);
\coordinate (I5) at (1/2,1/2);
\coordinate [below=1/4 of I1] (I11);
\coordinate [above=1/4 of I1] (I12);
\coordinate [left=1/4 of I2] (I21);
\coordinate [right=1/4 of I2] (I22);
\coordinate [below=1/4 of I3] (I31);
\coordinate [above=1/4 of I3] (I32);
\coordinate [left=1/4 of I4] (I41);
\coordinate [right=1/4 of I4] (I42);
\coordinate [below left=1/4 of I5] (I51);
\coordinate [above right=1/4 of I5] (I52);
\draw (G1) -- (G2);
\draw (G2) -- (G3);
\draw (G3) -- (G4);
\draw (G4) -- (G1);
\draw (G2) -- (G4);
\draw (G1) -- (H1);
\draw (G2) -- (H2);
\draw (G3) -- (H3);
\draw (G4) -- (H4);
\draw (I11) [dashed,color=red,line width=0.5 mm] -- (I12);
\draw (I21) [dashed,color=red,line width=0.5 mm] -- (I22);
\draw (I31) [dashed,color=red,line width=0.5 mm] -- (I32);
\draw (I41) [dashed,color=red,line width=0.5 mm] -- (I42);
\draw (I51) [dashed,color=red,line width=0.5 mm] -- (I52);
\node at (H1) [below left=0,scale=0.7] {$\small{p_1}$};
\node at (H2) [below right=0,scale=0.7] {$\small{p_2}$};
\node at (H3) [above right=0,scale=0.7] {$\small{p_3}$};
\node at (H4) [above left=0,scale=0.7] {$\small{p_4}$};
\node at (1/3,0) [below=0,scale=0.7] {$\small{1}$};
\node at (1,1/3) [right=0,scale=0.7] {$\small{2}$};
\node at (2/3,1) [above=0,scale=0.7] {$\small{3}$};
\node at (0,2/3) [left=0,scale=0.7] {$\small{4}$};
\node at (1/2+0.33,1/2-0.15) [below left=0,scale=0.7] {$\small{5}$};
\end{tikzpicture}+\begin{tikzpicture}[baseline={([yshift=-.5ex]current bounding box.center)}]
\coordinate (G1) at (0,0);
\coordinate (G2) at (1,0);
\coordinate (H1) at (-1/3,0);
\coordinate (H2) at (4/3,0);
\coordinate (I1) at (1/2,1/2);
\coordinate (I2) at (1/2,1/8);
\coordinate (I3) at (1/2,-1/2);
\coordinate (I4) at (1/2,-1/8);
\coordinate (J1) at (1,1/4);
\coordinate (J2) at (1,-1/4);
\coordinate (K1) at (1/2,-1/8);
\draw (G1) -- (G2);
\draw (G1) to[out=80,in=100] (G2);
\draw (G1) to[out=-80,in=-100] (G2);
\draw (G1) [line width=0.75 mm] -- (H1);
\draw (G2) [line width=0.75 mm]-- (H2);
\node at (K1) [above=0.7 mm of K1,scale=0.7] {\small$5$};
\node at (K1) [above=3.7 mm of K1,scale=0.7] {\small$3$};
\node at (K1) [below=1.3 mm of K1,scale=0.7] {\small$1$};
\node at (H1) [left=0,scale=0.7] {\small$p_{23}$};
\node at (H2) [right=0,scale=0.7] {\small$p_{23}$};
\end{tikzpicture}\otimes\begin{tikzpicture}[baseline={([yshift=-.5ex]current bounding box.center)}]
\coordinate (G1) at (0,0);
\coordinate (G2) at (1,0);
\coordinate (G3) at (1,1);
\coordinate (G4) at (0,1);
\coordinate [below left=1/3 of G1] (H1);
\coordinate [below right=1/3 of G2] (H2);
\coordinate [above right=1/3 of G3] (H3);
\coordinate [above left=1/3 of G4] (H4);
\coordinate (I1) at (1/2,0);
\coordinate (I2) at (1,1/2);
\coordinate (I3) at (1/2,1);
\coordinate (I4) at (0,1/2);
\coordinate (I5) at (1/2,1/2);
\coordinate [below=1/4 of I1] (I11);
\coordinate [above=1/4 of I1] (I12);
\coordinate [left=1/4 of I2] (I21);
\coordinate [right=1/4 of I2] (I22);
\coordinate [below=1/4 of I3] (I31);
\coordinate [above=1/4 of I3] (I32);
\coordinate [left=1/4 of I4] (I41);
\coordinate [right=1/4 of I4] (I42);
\coordinate [below left=1/4 of I5] (I51);
\coordinate [above right=1/4 of I5] (I52);
\draw (G1) -- (G2);
\draw (G2) -- (G3);
\draw (G3) -- (G4);
\draw (G4) -- (G1);
\draw (G2) -- (G4);
\draw (G1) -- (H1);
\draw (G2) -- (H2);
\draw (G3) -- (H3);
\draw (G4) -- (H4);
\draw (I11) [dashed,color=red,line width=0.5 mm] -- (I12);
\draw (I31) [dashed,color=red,line width=0.5 mm] -- (I32);
\draw (I51) [dashed,color=red,line width=0.5 mm] -- (I52);
\node at (H1) [below left=0,scale=0.7] {$\small{p_1}$};
\node at (H2) [below right=0,scale=0.7] {$\small{p_2}$};
\node at (H3) [above right=0,scale=0.7] {$\small{p_3}$};
\node at (H4) [above left=0,scale=0.7] {$\small{p_4}$};
\node at (1/3,0) [below=0,scale=0.7] {$\small{1}$};
\node at (1,1/3) [right=0,scale=0.7] {$\small{2}$};
\node at (2/3,1) [above=0,scale=0.7] {$\small{3}$};
\node at (0,2/3) [left=0,scale=0.7] {$\small{4}$};
\node at (1/2+0.33,1/2-0.15) [below left=0,scale=0.7] {$\small{5}$};
\end{tikzpicture}\nonumber\\
&+\begin{tikzpicture}[baseline={([yshift=-.5ex]current bounding box.center)}]
\coordinate (G1) at (0,0);
\coordinate (G2) at (1,0);
\coordinate (H1) at (-1/3,0);
\coordinate (H2) at (4/3,0);
\coordinate (I1) at (1/2,1/2);
\coordinate (I2) at (1/2,1/8);
\coordinate (I3) at (1/2,-1/2);
\coordinate (I4) at (1/2,-1/8);
\coordinate (J1) at (1,1/4);
\coordinate (J2) at (1,-1/4);
\coordinate (K1) at (1/2,-1/8);
\draw (G1) -- (G2);
\draw (G1) to[out=80,in=100] (G2);
\draw (G1) to[out=-80,in=-100] (G2);
\draw (G1) [line width=0.75 mm] -- (H1);
\draw (G2) [line width=0.75 mm]-- (H2);
\node at (K1) [above=0.7 mm of K1,scale=0.7] {\small$5$};
\node at (K1) [above=3.7 mm of K1,scale=0.7] {\small$4$};
\node at (K1) [below=1.3 mm of K1,scale=0.7] {\small$2$};
\node at (H1) [left=0,scale=0.7] {\small$p_{12}$};
\node at (H2) [right=0,scale=0.7] {\small$p_{12}$};
\end{tikzpicture}\otimes\begin{tikzpicture}[baseline={([yshift=-.5ex]current bounding box.center)}]
\coordinate (G1) at (0,0);
\coordinate (G2) at (1,0);
\coordinate (G3) at (1,1);
\coordinate (G4) at (0,1);
\coordinate [below left=1/3 of G1] (H1);
\coordinate [below right=1/3 of G2] (H2);
\coordinate [above right=1/3 of G3] (H3);
\coordinate [above left=1/3 of G4] (H4);
\coordinate (I1) at (1/2,0);
\coordinate (I2) at (1,1/2);
\coordinate (I3) at (1/2,1);
\coordinate (I4) at (0,1/2);
\coordinate (I5) at (1/2,1/2);
\coordinate [below=1/4 of I1] (I11);
\coordinate [above=1/4 of I1] (I12);
\coordinate [left=1/4 of I2] (I21);
\coordinate [right=1/4 of I2] (I22);
\coordinate [below=1/4 of I3] (I31);
\coordinate [above=1/4 of I3] (I32);
\coordinate [left=1/4 of I4] (I41);
\coordinate [right=1/4 of I4] (I42);
\coordinate [below left=1/4 of I5] (I51);
\coordinate [above right=1/4 of I5] (I52);
\draw (G1) -- (G2);
\draw (G2) -- (G3);
\draw (G3) -- (G4);
\draw (G4) -- (G1);
\draw (G2) -- (G4);
\draw (G1) -- (H1);
\draw (G2) -- (H2);
\draw (G3) -- (H3);
\draw (G4) -- (H4);
\draw (I21) [dashed,color=red,line width=0.5 mm] -- (I22);
\draw (I41) [dashed,color=red,line width=0.5 mm] -- (I42);
\draw (I51) [dashed,color=red,line width=0.5 mm] -- (I52);
\node at (H1) [below left=0,scale=0.7] {$\small{p_1}$};
\node at (H2) [below right=0,scale=0.7] {$\small{p_2}$};
\node at (H3) [above right=0,scale=0.7] {$\small{p_3}$};
\node at (H4) [above left=0,scale=0.7] {$\small{p_4}$};
\node at (1/3,0) [below=0,scale=0.7] {$\small{1}$};
\node at (1,1/3) [right=0,scale=0.7] {$\small{2}$};
\node at (2/3,1) [above=0,scale=0.7] {$\small{3}$};
\node at (0,2/3) [left=0,scale=0.7] {$\small{4}$};
\node at (1/2+0.33,1/2-0.15) [below left=0,scale=0.7] {$\small{5}$};
\end{tikzpicture},
\end{align}
where $p_{ij}=p_i+p_j$.


\section{Summary and discussion}

\label{conclusions}

In this paper we have taken first steps towards generalising the diagrammatic coaction of refs.~\cite{Abreu:2017enx,Abreu:2017mtm} beyond one loop. The main features of this coaction are as follows: first, if (cut) Feynman graphs are replaced by the 
functions they represent in dimensional regularisation, the diagrammatic coaction maps directly to the global coaction on hypergeometric functions, which in turn agrees with the local coaction acting on MPLs order-by-order in the $\eps$ expansion. 
Both are realisations of the same fundamental coaction on integrals~\cite{Abreu:2017enx} which is based on pairing of differential forms in the left entry with integration contours in the right.
Second, it is possible to safely set to zero both propagator masses and external masses in the diagrammatic coaction, even though individual entries in the coaction may develop singularities. While the existence of a coaction with these properties is highly non-trivial, and it involves several new features compared to the well-studied one-loop case, we have provided a set of examples of two-loop integrals with up to four external legs and a variety of mass configurations for which we explicitly derived such a diagrammatic coaction.

At one loop, the construction of the diagrammatic coaction was guided by a solid mathematical understanding of the homology groups associated to one-loop integrals~\cite{teplitz,Abreu:2017ptx}. In particular, at one loop there is a one-to-one correspondence between independent integration contours and cut integrals. This simple correspondence is lost beyond one loop, where there are several distinct contours corresponding to the same set of propagators put on shell. This property mirrors the fact that, starting from two loops, different master integrals may share the very same set of propagators. A cornerstone of our approach is the realisation that, by iterating a known identity among cut and uncut integrals at one loop, it is possible to define a spanning set of cuts for general $L$-loop integrals, where every loop contains at least one cut propagator. Upon using only these \emph{genuine $L$-loop cuts} to express the right entries, the corresponding left entries---which only feature propagators that are cut on the right---retain all $L$ loops of the original integral considered, and may thus appear in the corresponding basis of master integrals. 
The diagrammatic coaction of any Feynman integral can then identify a natural pairing between integration contours and master integrals. 

In all examples considered in this paper, we could evaluate the cut integrals in terms of the Gauss hypergeometric function and its generalisations, such as the Appell functions. The coaction on such hypergeometric functions was conjectured by us in ref.~\cite{Abreu:2019wzk} (and proven for Lauricella functions in ref.~\cite{brown2019lauricella}). In this way we can uniquely identify the form of the diagrammatic coaction for all the Feynman integrals we have considered. We expect that the same strategy can be applied to other classes of Feynman integrals that can be expressed in terms of these or similar hypergeometric functions. We emphasise that there is no obvious obstacle to generalise our approach to more complex two-loop integrals, and indeed to higher-loop integrals. All that is required a priori is the existence of a global coaction on the relevant type of hypergeometric functions. The fact that the entries of this global coaction can be interpreted in terms of cut graphs is of course highly nontrivial, and it supports our expectation that a diagrammatic coaction for (cut) Feynman graphs exists in general. 
This still awaits to be fully established. 

Let us conclude by providing directions for future research. First, our results are restricted to Feynman integrals whose $\eps$-expansion can be expressed in terms of MPLs, such as the sunset integrals with up to two massive propagators. It would be interesting to investigate the predictions of our diagrammatic coaction for the sunset integral with three massive propagators. The latter cannot be expressed in terms of MPLs, but functions associated with elliptic curves are required, cf., e.g., refs.~\cite{Laporta:2004rb,Bloch:2013tra,Adams:2013nia,Adams:2014vja,Adams:2015gva,Adams:2015ydq,Adams:2017ejb,Broedel:2017siw,Broedel:2018iwv,Bogner:2019lfa}. Second, it would be interesting to investigate how our diagrammatic coaction is related to other coactions involving cuts of Feynman integrals, cf.~refs.~\cite{Kreimer:2020mwn,Kreimer:2021jni}. These coactions, however, are not formulated in the context of dimensional regularisation, and, unlike our diagrammatic coaction, they are not applicable to Feynman integrals involving infrared divergences. Finally, we do not currently have a good understanding of the (co)homology groups associated to multi-loop integrals. Extending some of the results of refs.~\cite{Mastrolia:2018uzb,Caron-Huot:2021xqj} could be a first step in this direction. We leave the investigation of these topics to future work.

\acknowledgments

We thank Aris Ioannou for useful discussions.
This work was supported by the U.K. Royal Society through Grant URF\textbackslash{}R1\textbackslash201607 (SA),  by the 
European Research Council under the European Union's Horizon 2020 research and innovation programme through grants 
647356 (RB) and  637019 (CD), and the STFC Consolidated Grant ``Particle Physics at the Higgs Centre'' (EG, JM).




\appendix


\section{Calculation of Cut Integrals at Two Loops}\label{CutsSection}

One of the major difficulties in obtaining a diagrammatic coaction at two-loops lies in the
calculation of the cut integrals, which serve as a basis for the right entries of the global
coaction tensors. In this appendix we give a brief overview of the techniques we used
to obtain the results used in this paper and illustrate their application in the context of the sunset topology.

Our calculations are based on a loop-by-loop approach. For each loop integration,
we use an explicit parametrisation of the loop momentum to impose the cut conditions,
that is we use the same approach as in refs.~\cite{Abreu:2014cla,Abreu:2015zaa,Abreu:2017ptx}.
More explicitly, for each loop momentum $k$ we write the inverse propagators that depend on $k$
in the form $(k+q_i)^2-m_i^2$ for some $q_0,\ldots,q_{n-1}$, where the $q_i$ are linear
combinations of momenta that are external to the loop under consideration. We use translation
invariance to set $q_0=0$ and write
\begin{align}\begin{split}
q_1=\left(q_1^0,\underline{0}_{D-1}\right),\,\,
 q_2=\left(q_2^0,q_2^1,\underline{0}_{D-2}\right),\,\,\,\,\ldots\,\,\,\,,\,\,
 q_{n-1}=\left(q_{n-1}^0,\ldots,q_{n-1}^{n-2},\underline{0}_{D-n+1}\right)\,,
\end{split}\end{align}
where we introduced the notation $\underline{a}_n = (\underbrace{a,\ldots,a}_{n})$.
We then parametrise $k$ in a straightforward way:
\begin{equation}\label{param}
k=k_0\left(1,\beta\,\text{cos}\,\theta_1,\beta\,\text{cos}\,\theta_2\,
\text{sin}\,\theta_1,\ldots,\beta\,\text{cos}\,\theta_{n-2}\prod_{i=1}^{n-3}
\text{sin}\,\theta_i,\beta\left(\prod_{i=1}^{n-2}\text{sin}\,\theta_i\right)
\underline{1}_{D-n+1}\right)\,,
\end{equation}
where $\underline{1}_{D-n+1}$ ranges over unit vectors in the dimensions transverse to the external momenta,
with the corresponding integration measure
\begin{align}
\label{k_measure}
\int d^Dk=\frac{2\pi^{\frac{D-n+1}{2}}}{\Gamma\left(\frac{D-n+1}{2}\right)}
\int_{-\infty}^{+\infty} dk_0k_0^{D-1}\int_0^\infty d\beta\beta^{D-2}
\prod_{j=1}^{n-2}\int_0^\pi d\theta_j\,\text{sin}^{D-2-j}\theta_j\,,
\end{align}
where the coordinates which the propagators do not depend upon have already been integrated over.
It will sometimes be easier to use an alternative parametrisation in Euclidean space after Wick rotating the loop momentum.  
The Euclidean momentum is parametrised as
\begin{align}\label{kE}
k^E=\vert k^E\vert\left(\text{cos}\,\theta_0,\text{cos}\,\theta_1\,\text{sin}\,\theta_0,
\ldots,\text{cos}\,\theta_{n-2}\prod_{j=0}^{n-3}\text{sin}\,\theta_j,
\left(\prod_{j=0}^{n-2}\text{sin}\,\theta_j\right)\underline{1}_{D-n+1}\right),
\end{align}
such that the square of this momentum now depends on only the single variable $\vert k^E\vert$. 
The measure is: 
\begin{align}\label{Measure_kE}
\int d^Dk= i \int d^Dk^E 
=\frac{i \pi^{\frac{D-n+1}{2}}}{\Gamma\left(\frac{D-n+1}{2}\right)}
\int_{0}^{\infty} d|k^E|^2\, \left(|k^E|^2\right)^{\frac{D-2}{2}}
\prod_{j=0}^{n-2}\int_0^\pi d\theta_j\,\text{sin}^{D-2-j}\theta_j\,.
\end{align}
This parametrisation will be convenient to use in the computation of cut loops where only a 
single propagator is placed on shell.

\subsection{Maximal cuts of the one-mass sunset~\label{sec:1massSunsetMaxCut}}
Let us now show in more detail how this approach allows us to compute the maximal cut
of the one-mass sunset master integral defined in eq.~(\ref{SunsetMastInt1}). 
We start from eq.~(\ref{OneMassSunsetMaxCutwoNormalization}) which we rewrite here 
including the prefactors by which we will normalise it, 
\begin{align}\label{eq:temp1ms}
\begin{split}
	\mathcal{C}_{1,2,3}S^{(1)}(p^2;m^2)
&=-\frac{\epsilon}{4} (2\pi i)^2e^{2\gamma_E\epsilon}  (p^2-m^2) \times\\& \hspace*{50pt}\mathcal{C}_1\int\frac{d^{2-2\epsilon}k}{i\pi^{1-\epsilon}}
	\frac{1}{k^2}\,\,\left(
	\mathcal{C}_{2,3}\int\frac{d^{2-2\epsilon}l}{i\pi^{1-\epsilon}}
	\frac{1}{l^2}\frac{1}{(k+l+p)^2-m^2}\right)\\
&=-\frac{\epsilon}{2}(2\pi i) e^{2\gamma_E\epsilon}
	(p^2-m^2)\frac{\Gamma(1-\epsilon)}{\Gamma(1-2\epsilon)}
	\mathcal{C}_1\int\frac{d^{2-2\epsilon}k}{i\pi^{1-\epsilon}}
	\frac{1}{k^2}
	\frac{\left((k+p)^2\right)^{\epsilon}}
	{\left((k+p)^2-m^2\right)^{1+2\epsilon}}\,.
\end{split}
\end{align}
The normalisation is consistent with the one used in eq.~(\ref{SunsetMastInt1}). We also include
powers of $\pi$ to remove trivial overall factors of $\pi$ that are leftover after imposing the
cut conditions. 
We have chosen a particular routing of momenta, and a specific ordering of the integration over the loop momenta, which define the steps of our loop-by-loop approach. We emphasise that while these choices may in general affect the parametric representation one obtains for the cuts, different choices will lead to the same space of cuts. 
In the second step of eq.~(\ref{eq:temp1ms}), we used the expression for the maximal cut of the one-loop
bubble subloop given in eq.~\eqref{bubPMCutP}.
To impose the remaining cut condition, that is $k^2=0$, we use the parametrisation in 
eq.~\eqref{param}. For this particular case ($n=2$), we can trivially integrate all angles
and the measure becomes
\begin{equation}
\label{measure_k_n=2}
	\int d^{2-2\epsilon}k=\frac{2\pi^{\frac12-\epsilon}}{\Gamma\left(\frac12-\epsilon\right)}
	\int dk_0 \,k_0^{1-2\epsilon}\, \int d\beta \beta^{-2\epsilon}.
\end{equation}
In this parametrisation, $k^2=k_0^2(1-\beta^2)$, and the cut condition is thus imposed
by evaluating the residue at $\beta=1$. Assuming $p^2>0$ and noting that
\begin{equation}
	(k+p)^2=k_0^2(1-\beta^2)+p^2+2\sqrt{p^2}k_0\,,
\end{equation}
the cut condition can be easily imposed and we obtain
\begin{align}
\label{MaxCut1massSunsetResult}
	\mathcal{C}_{1,2,3}S^{(1)}(p^2;m^2)=
	  (p^2-m^2)\epsilon e^{2\gamma_E\epsilon} \frac{2^{-2\epsilon}\Gamma^2(1-\epsilon)}{\Gamma^2(1-2\epsilon)}
	\int dk_0\,k_0^{-1-2\epsilon}
	\frac{\left(p^2+2\sqrt{p^2}k_0\right)^{\epsilon}}
	{\left(p^2-m^2+2\sqrt{p^2}k_0\right)^{1+2\epsilon}}\,.
\end{align}
As explained below eq.~\eqref{eq:tempCut123}, we easily recognise this integrand
as that of a Gauss hypergeometric function for which we know how to determine the two independent
integration cycles.

\subsection{Maximal cuts of the two-mass sunset} 
Let us now outline the calculation
of the maximal cuts of the two-mass sunset integral. To demonstrate the methods we will use 
the first of the three master integrals defined in eq.~(\ref{eq:2massBasis}). 
We set up the calculation in exactly the same way as for the one-mass case discussed above. We have
\begin{align}\label{eq:temp2ms}
\begin{split}
	\mathcal{C}_{1,2,3}S^{(1)}(p^2;m_1^2,m_2^2)
&=\frac{\epsilon}{4} (2\pi i)^2  e^{2\gamma_E\epsilon} \sqrt{\lambda(p^2,m_1^2,m_2^2)}\,\, {\cal C}_{\rm\text{max}} \mathcal{S}(p^2;m_1^2,m_2^2)\,,
\end{split}
\end{align}
with
\begin{align}\label{eq:temp2ms_int}
\begin{split}
{\cal C}_{\rm\text{max}} \mathcal{S}(p^2;m_1^2,m_2^2) \equiv \mathcal{C}_1\int\frac{d^{2-2\epsilon}k}{i\pi^{1-\epsilon}}
	\frac{1}{k^2-m_1^2}\,\,\left(
	\mathcal{C}_{2,3}\int\frac{d^{2-2\epsilon}l}{i\pi^{1-\epsilon}}
	\frac{1}{l^2}\frac{1}{(k+l+p)^2-m_2^2}\right)\,,
\end{split}
\end{align}
where in eq.~(\ref{eq:temp2ms}) we normalise the cut integral consistently with the 
one-mass sunset of eq.~(\ref{eq:temp1ms}). 
For brevity we  suppress the arguments of ${\cal C}_{\rm\text{max}} \mathcal{S}$ in what follows.
Inserting the result for the maximal cut of the one-mass bubble into eq.~(\ref{eq:temp2ms_int}) we get
\begin{align}\label{2massSunsetCmaxDef}
\begin{split}
	{\cal C}_{\rm\text{max}} \mathcal{S}=
	\frac{2}{2\pi i}\,\frac{\Gamma(1-\epsilon)}{\Gamma(1-2\epsilon)}\, \mathcal{C}_1\int\frac{d^{2-2\epsilon}k}{i\pi^{1-\epsilon}}
	\frac{1}{k^2-m_1^2}
	\frac{\left((k+p)^2\right)^{\epsilon}}
	{\left((k+p)^2-m_2^2\right)^{1+2\epsilon}}\,.
\end{split}\end{align}
We now explain how  eq.~(\ref{2massSunsetCmaxDef}) can be evaluated by imposing the cut condition $k^2=m_1^2$ 
using the two alternative parametrisations of the momentum $k$, the Minkowski one in eq.~(\ref{param}) and Wick-rotated one in eq.~(\ref{kE}), both leading to the final results summarised in eqs.~(\ref{MaxCut1_2mass_sunsetS1}) through~(\ref{MaxCut3_2mass_sunsetS1}).

\paragraph{Minkowski-space momentum parametrisation.} We first consider the Minkowski-space parametrisation (\ref{param}) used in section~\ref{sec:1massSunsetMaxCut} with the measure given in eq.~(\ref{measure_k_n=2}).
The advantages of this method are that it is suitable for considering massless limits, and it provides a clear physical interpretation for the cut contours in terms of the values taken by real-momentum 
components---here this will be the energy flowing through the cut propagator.
Equation~(\ref{2massSunsetCmaxDef}) becomes:
\begin{align}\label{2massSunsetCmaxStarting}
\begin{split}
	{\cal C}_{\rm\text{max}} \mathcal{S}=&
	\frac{2}{2\pi i}\,\frac{\Gamma(1-\epsilon)}{\Gamma(1-2\epsilon)}\,\frac{1}{i\pi^{1-\epsilon}}  
\frac{2\pi^{\frac12-\epsilon}}{\Gamma\left(\frac12-\epsilon\right)}
	\int dk_0 \,k_0^{1-2\epsilon}\, \\&
{\text{Res}}_{\beta}
	\frac{ \beta^{-2\epsilon}}{k_0^2(1-\beta^2)-m_1^2}
	\frac{\left(k_0^2(1-\beta^2) +p^2+2\sqrt{p^2}k_0\right)^{\epsilon}}
	{\left(k_0^2(1-\beta^2) +p^2+2\sqrt{p^2}k_0-m_2^2\right)^{1+2\epsilon}}\,,
\end{split}\end{align}
where the cut condition ${\cal C}_1$ in eq.~(\ref{2massSunsetCmaxDef}) amounts to taking a residue, localizing the integration over $\beta$. Given that $\beta$ is a priori positive, the residue is taken at $\beta=\sqrt{1-m_1/k_0^2}$, after which the expression simplifies to:
\begin{align}\label{2massSunsetCmaxStarting}
\begin{split}
	{\cal C}_{\rm\text{max}} \mathcal{S}&=
\frac{1}{(2\pi i)^2} \,\frac{2^{2-2\epsilon}\Gamma^2(1-\epsilon)}{\Gamma^2(1-2\epsilon)}
	\int dk_0 
	\left(k_0^2-m_1^2\right)^{-\frac12-\epsilon}
	\frac{\left(m_1^2 +p^2+2\sqrt{p^2}k_0\right)^{\epsilon}}
	{\left(m_1^2 +p^2+2\sqrt{p^2}k_0-m_2^2\right)^{1+2\epsilon}}\,.
\end{split}\end{align}
Naturally, this expression reproduces the one-mass case of eq.~(\ref{MaxCut1massSunsetResult}) upon taking $m_1\to 0$.
In line with the general method to constructing the global coaction~\cite{Abreu:2019wzk}, the contours to be considered are those ranging between the branch points of the integrand in eq.~(\ref{2massSunsetCmaxStarting}). Specifying the integration range amounts to choosing a specific cut within the space of maximal cuts.\footnote{We note in passing that with a suitable shift and rescaling of the energy variable one may recognise eq.~(\ref{2massSunsetCmaxStarting}) as the one-parameter integral representation of an Appell $F_1$ function. We comment on this further following eq.~(\ref{eq:temp2ms_res}) below. Here we proceed to evaluate the integral as an Appell $F_4$ function, as needed for the coaction.} Here we will illustrate the method by computing one such maximal cut, that of eq.~(\ref{MaxCut3_2mass_sunsetS1}).  This cut corresponds to integrating between $k_0=-(m_1^2+p^2)/(2\sqrt{p^2})$ and $k_0=-\sqrt{m_1^2}$, working in the kinematic region where $p^2>m_1^2>0$. To avoid having an additional branch point within this integration domain due to the denominator in eq.~(\ref{2massSunsetCmaxStarting}) we keep $m_2^2<0$ throughout. 

To proceed we use a Mellin-Barnes representation of the denominator in eq.~(\ref{2massSunsetCmaxStarting}), splitting between the $-m_2^2$  and the rest. The $k_0$ integration can then be performed under the Mellin-Barnes integral, yielding ${}_2F_1$ hypergeometric functions: 
\begin{align}\label{2massSunsetCmaxMB}
\begin{split}
	{\cal C}_{\rm\text{max}} \mathcal{S}=\frac{2\Gamma(1-\epsilon) }{\Gamma(1-2\epsilon)\Gamma(1+2\epsilon)}
&\frac{
(-m_2^2)^{-1-2\epsilon}}{(2\pi i)^3} \int_{-i\infty}^{i\infty} dz \Gamma(1+2\epsilon+z)\Gamma(-z)\left(-\frac{m_2^2}{p^2}\right)^{-z}\,\times 
\\ \hspace*{50pt}&\Bigg[\Gamma(\epsilon)  \, \left(\frac{m_1^2}{p^2}\right)^{-\epsilon} {}_2F_1\left(-\epsilon-z,-z;1-\epsilon;\frac{m_1^2}{p^2}\right)
\\& \,\,+\frac{\Gamma(-\epsilon)\Gamma(1+\epsilon+z)}{\Gamma(1-\epsilon+z)}  {}_2F_1\left(\epsilon-z,-z;1+\epsilon;\frac{m_1^2}{p^2}\right)\Bigg]
\end{split}\end{align}
Writing these hypergeometric functions as a power series in $\frac{m_1^2}{p^2}$, and considering $-m_2^2<p^2$ we may close the Mellin-Barnes contour to the left, encircling the sets of poles generated by $\Gamma(1+2\epsilon+z)$ and $\Gamma(1+\epsilon+z)$. 
The resulting double sum directly gives three Appell $F_4$ functions:
\begin{align}\label{2massSunsetCmaxAppellF4}
\begin{split}
	{\cal C}_{\rm\text{max}} \mathcal{S}&=\frac{(p^2)^{-1-2\epsilon}}{(2\pi i)^2} \frac{\Gamma(1+2\epsilon)}{\epsilon}\frac{\sin 2\pi\epsilon }{\sin \pi\epsilon }
 \, \,  \Bigg\{\vphantom{\frac{1}{2}}z_1^{-\epsilon}
F_4\left(1+\epsilon,1+2\epsilon;1-\epsilon,1+\epsilon;z_1,z_2\right)\\
&\hspace*{50pt}+(-z_2)^{-\epsilon} \frac{\sin 2\pi\epsilon }{\sin \pi\epsilon }
F_4\left(1+\epsilon,1+2\epsilon;1+\epsilon,1-\epsilon;z_1,z_2\right)\\
&\hspace*{50pt}-\frac{\Gamma(1+3\epsilon)}{\Gamma^3(1+\epsilon)} \, \frac{\sin 3\pi\epsilon }{\sin \pi\epsilon } 
\frac{\pi\epsilon }{\sin \pi\epsilon }
F_4\left(1+2\epsilon,1+3\epsilon;1+\epsilon,1+\epsilon;z_1,z_2\right)\Bigg\}
\end{split}\end{align}
where we defined $z_i=\frac{m_i^2}{p^2}$. Upon restoring the normalisation in eq.~(\ref{eq:temp2ms}) 
and ignoring the $i\pi$ terms generated in the $\epsilon$ expansion of the factors multiplying 
the Appell $F_4$ functions, one obtains the cut quoted in eq.~(\ref{MaxCut3_2mass_sunsetS1}) 
times an overall factor of $-2$. This factor is required to enforce the duality 
condition that guarantees that eq.~\eqref{eq:coacGenDual} is satisfied.
The other basis elements in the space of maximal cuts can be obtained in a similar way, integrating 
over the energy between pairs of branch points in eq.~(\ref{2massSunsetCmaxStarting}). Alternative 
techniques to set the basis of cuts using differential equations will be discussed in 
appendix~\ref{DiffEqCutsSection}. Before doing that let us briefly examine the same maximal 
cut computation using Wick rotation.

\paragraph{Euclidean-space momentum parametrisation.} 
The second method to impose the cut condition is to use Euclidean momentum parametrisation 
according to eq.~\eqref{kE} following a Wick rotation. As we shall see it is marginally 
simpler to implement, since the cut condition $k^2= - |k_E|^2=m_1^2$ directly fixes the 
magnitude of the Euclidean momentum (however in this way massless limits are not straightforward to take). 
The integration measure from eq.~(\ref{Measure_kE}) is
\begin{equation}
	\int d^{2-2\epsilon}k=
	\frac{i \pi^{\frac12-\epsilon}}{\Gamma\left(\frac12-\epsilon\right)}\int d\abs{k^E}^2
	\left(\abs{k^E}^2\right)^{-\epsilon}\int d\theta\,\sin^{-2\epsilon}\theta\,,
\end{equation}
and, under the cut condition,
\begin{equation}
	(k+p)^2=m_1^2+p^2-2\sqrt{m_1^2p^2}\cos\theta\,.
\end{equation}
The maximal cut is then given by
\begin{align}
\label{eq:temp2ms_res}
\begin{split}
{\cal C}_{\rm\text{max}} \mathcal{S} &=
 \frac{(m_1^2)^{-\epsilon} }{(2\pi i)^2} \,\frac{2^{2-2\epsilon}\Gamma^2(1-\epsilon)}{\Gamma^2(1-2\epsilon)}
\int d\theta\,\frac{\left(\sin \theta\right)^{-2\epsilon}\,\left(m_1^2+p^2-2\sqrt{m_1^2p^2}\,\text{cos}\,\theta\right)^{\epsilon}}
	{\left(m_1^2+p^2-2\sqrt{m_1^2p^2}\,\cos\theta  -m_2^2 \right)^{1+2\epsilon}}\,.
\end{split}\end{align}
Changing variables according to $\cos\theta=2x-1$ we directly obtain:
\begin{align}
\label{eq:temp2ms_res}
\begin{split}
{\cal C}_{\rm\text{max}} \mathcal{S} &=  \frac{(m_1^2)^{-\epsilon}}{(2\pi i)^2} \,\frac{2^{2-4\epsilon}\Gamma^2(1-\epsilon)}{\Gamma^2(1-2\epsilon)}
\int dx\frac{[x(1-x)]^{-\frac{1}{2}-\epsilon}\, [m_1^2+p^2-2\sqrt{m_1^2p^2}(2x-1)]^{\epsilon}}{[m_1^2+p^2-2\sqrt{m_1^2p^2}(2x-1) -m_2^2 ]^{1+2\epsilon}}\,.
\end{split}\end{align}
A couple of comments are due regarding this elegant parametric representation of the maximal cuts. First, we note that it may be directly related to the Minkowski-space integral in eq.~(\ref{2massSunsetCmaxStarting}) by identifying $k_0 = m_1 (1 - 2 x)$. Second,  eq.~(\ref{eq:temp2ms_res}) is readily recognisable as the one-dimensional integral representation of the Appell~$F_1$ function, which 
was used in ref.~\cite{Abreu:2019wzk} to construct the coaction on this class of function. 
From this it immediately follows that there are \emph{three} independent cut contours.  
However, we already know that it is also possible to express the maximal cuts in terms of Appell $F_4$ functions, as we have just shown in eq.~(\ref{2massSunsetCmaxAppellF4}). 
Writing the maximal cuts in terms of Appell $F_4$ functions is indeed the natural space of function to express the right entries in the coaction, given that the uncut master integrals are themselves Appell $F_4$ functions (see appendix~\ref{sec:appTwoMassExp}) and, as shown in ref.~\cite{Abreu:2019wzk}, all entries in the coaction of  Appell $F_4$ are expressible using the same type of functions.  
The fact that, despite this, the maximal cuts are also expressible in terms of Appell~$F_1$ functions is of conceptual significance: they form a three-dimensional subspace within the larger, four-dimensional, space of cuts. From the perspective of the coaction this relates to the fact that the maximal cut contours are dual to the master integrands at the top topology, while the additional cut completing the four-dimensional space is dual to the double tadpole integrand, which features only two of the three propagators. 
Known reduction formulae allow one to express the three Appell $F_4$ functions of the form found 
in eq.~(\ref{2massSunsetCmaxAppellF4}), as Appell $F_1$ functions---see for example eqs.~(14-15) 
in ref.~\cite{Kniehl:2011ym} which were shown there to apply to the two-mass bubble integral.
 An additional perspective on how the three-dimensional  Appell~$F_1$  subspace is accommodated within the larger Appell $F_4$ space will be discussed in appendix~\ref{DiffEqCutsSection}.

\subsection{Two-propagator cut of the two-mass sunset}
For the two-mass sunset we must also compute the two-propagator cut that encircles the poles associated
with the two massive propagators. We proceed in the same way as for the maximal cut, defining
\begin{align}\label{eq:temp2ms_two_prop_cut}
\begin{split}
	\mathcal{C}_{1,2}S^{(1)}(p^2;m_1^2,m_2^2)
&=-e^{2\gamma_E\epsilon} \sqrt{\lambda(p^2,m_1^2,m_2^2)}\,\,
\mathcal{C}_1\int\frac{d^{2-2\epsilon}k}{i\pi^{1-\epsilon}}
	\frac{1}{k^2-m_1^2}\,\mathcal{C}_2B_2\,.
\end{split}
\end{align}
The inner integral $\mathcal{C}_2B_2$ is the single-propagator cut of a one-mass
bubble with external mass $(p+k)^2$. It evaluates to~\cite{Abreu:2017ptx,Abreu:2017mtm} 
 \begin{align}
\begin{split}
 \mathcal{C}_2B_2
&=\,
	\mathcal{C}_{2}\int\frac{d^{2-2\epsilon}l}{i\pi^{1-\epsilon}}
	\frac{1}{l^2}\frac{1}{(k+l+p)^2-m_2^2}\\
 &=-\frac{1}{\Gamma(1-\epsilon)}(-m_2^2)^{-\epsilon}\frac{1}{(k+p)^2}\,{}_2F_1\left(1,1+\epsilon;1-\epsilon;\frac{m_2^2}{(k+p)^2}\right)\,.
\end{split}
 \end{align}
The remaining cut integral over $k$ can be computed using the same parametrisation for the 
loop momentum as in the calculation of the maximal cuts. 
We expand the $_2F_1$ function as a series, and then sum the double series into an 
Appell $F_4$ function:
\begin{align}\label{aequals0}
&\mathcal{C}_{1}\int\frac{d^Dk}{i\pi^{D/2}}\frac{1}{k^2-m_1^2}\mathcal{C}_2\int\frac{d^Dl}{i\pi^{D/2}}\frac{1}{(l^2-m_2^2)(k+l+p)^2}\\
\nonumber
&\hspace*{60pt}=\frac{(p^2)^{-1-2\epsilon}}{\Gamma^2(1-\epsilon)}
\, \left(-\frac{m_1^2}{p^2}\right)^{-\epsilon}\left(-\frac{m_2^2}{p^2}\right)^{-\epsilon} \,F_4\left(1,1+\epsilon;1-\epsilon,1-\epsilon;\frac{m_1^2}{p^2},\frac{m_2^2}{p^2}\right).
\end{align}
Modulo $i\pi$ and after including the normalisation factor of eq.~(\ref{eq:temp2ms_two_prop_cut}),
we recover the result quoted in eq.~(\ref{C12_2mass_sunsetS1}).


\section{Differential equations and the basis of cuts \label{DiffEqCutsSection}}

The direct computation of cuts via residues, 
as outlined in section~\ref{sec:first-examples} and appendix~\ref{CutsSection}, is a good strategy 
to derive an integral representation for a particular type of cut.
An alternative perspective can however be provided by analysing the differential equations
satisfied by the cut integrals. Indeed, a lot is known about the solution space of
differential equations of hypergeometric type, and we can leverage this knowledge to
construct completed bases of integrals where a fixed set of propagators is cut.

As explained in section~\ref{sec:coactionEntries}, the same
set of linear first-order differential equations obeyed by the basis of master integrals is also
satisfied by any of its cuts. This has a very important consequence: 
from the point of view of the representation of these integrals
as hypergeometric functions, it implies that all the cuts must be expressible in terms of the
same type of functions as the uncut integrals. This is crucial for the relation between
the diagrammatic coaction and the global coaction, as we have shown that for the later
the same type of functions appear in the right entries of the coaction tensor \cite{Abreu:2019wzk}.

Let us now discuss two different ways to interpret the differential equations for cut integrals.
The first is the one mentioned in the previous paragraph:
the uncut integral and the various cut integrals satisfy
the same differential equation and are distinguished solely by the choice of 
boundary conditions used to solve these equations.
Imposing cut conditions might lead to some integrals being set to zero, 
which simplifies the solution of the whole system and implies that the non-vanishing 
cut integrals evaluate to simpler functions,
in this case to hypergeometric functions of the same type as the uncut integral but 
where the parameters take degenerate values. From this perspective, the simplicity
of integrals with many cut propagators compared to integrals with fewer or no cut propagators
might not be apparent, as it relies on 
knowing how the relevant hypergeometric functions degenerate with a particular
set of parameters.
The second way to interpret the differential equation for cut integrals
is to consider the subsystem of equations obtained by removing all integrals that are set
to zero by the cut conditions. We then get a smaller system of first-order equations, which 
is easier to solve.
This implies that the solutions can also be written in terms of a simpler
type of hypergeometric functions. 
Broadly speaking, from the first interpretation we obtain complex
hypergeometric functions (of the type that can be used to express the uncut integral) 
with degenerate parameters, while from the second
we obtain simpler hypergeometric functions involving fewer parameters. 
The first perspective
makes the connection with the global coaction straightforward, while the second one is more 
convenient to get compact expressions for the cut integrals. While the two are compatible, this
might be obscured by the fact that the relation involves nontrivial identities between different classes
of hypergeometric functions.

In the remainder of this appendix, we provide two examples illustrating the use of differential 
equations in the calculation of cut integrals. Our first example
concerns the one-mass sunset of section~\ref{sec:first-examples}. As discussed there, in this
example there are only two master integrals which both have the same three propagators, and therefore
the system will not become simpler for any number of cut propagators:
the two maximal cuts evaluate to the same type of functions as the uncut integrals and not
to degenerate versions of them. By considering the differential equations, we 
can nevertheless check that the expressions for the cut integrals
do indeed satisfy the differential equation.
Our second example is the two-mass sunset discussed in section~\ref{sec:sunsetTwoMass}. 
It features four master integrals, three of which are of the top topology, 
and the fourth is the double tadpole with only two out of the three propagators. If we consider
the solutions of the differential equation corresponding to the maximal cuts, the boundary condition
for the double tadpole is zero (as must happen when a cut condition in placed on a propagator that is absent). 
The maximal cut can then also be shown to satisfy a smaller three-by-three system of differential equations. 
As we will see, these two perspectives lead respectively to the two 
hypergeometric representations for the maximal cuts found in appendix~\ref{CutsSection}, in terms of either Appell $F_4$ or Appell $F_1$ functions.

\subsection{Maximal Cuts of the One-Mass Sunset\label{One-mass-sunset_DiffEq}}

The one-mass sunset defined in eq.~\eqref{eq:ssetFamily} depends on 
two variables, $p^2$ and $m^2$. It is however sufficient to consider
the differential equation with respect to the dimensionless variable
$z\equiv p^2/m^2$, since the dependence on $m^2$ can then be trivially
restored by dimensional analysis. Through standard techniques,
we find that $S^{(1)}$ and $S^{(2)}$, given in eqs.~\eqref{SunsetMastInt1}
 and \eqref{SunsetMastInt2} respectively, obey the following system of first-order differential equations: 
 \begin{align}\label{SunsetDiffEqs}
 \frac{d}{dz}\left(\begin{array}{c}S^{(1)}\\S^{(2)}\end{array}\right)=\frac{\epsilon}{2z}\left(\begin{array}{cc}\frac{(3+5z)}{1-z}&-3\\1&-1\end{array}\right)\left(\begin{array}{c}S^{(1)}\\S^{(2)}\end{array}\right)\,,
 \end{align}
where the derivative with respect to $z$ is taken for constant $m^2$.
The cuts must satisfy the same system of equations:
 \begin{align}\label{SunsetCutDiffEqs}
 \frac{d}{dz}\left(\begin{array}{c}\mathcal{C}^{(i)}_{1,2,3}S^{(1)}\\
 \mathcal{C}^{(i)}_{1,2,3}S^{(2)}\end{array}\right)=
 \frac{\epsilon}{2z}\left(\begin{array}{cc}\frac{(3+5z)}{1-z}&-3\\1&-1\end{array}\right)
 \left(\begin{array}{c}\mathcal{C}^{(i)}_{1,2,3}S^{(1)}\\
 \mathcal{C}^{(i)}_{1,2,3}S^{(2)}\end{array}\right),
 \end{align}
for $i=1,2$. We recall that in section \ref{sec:first-examples}
we distinguished the cases $i=1$ and $i=2$ by a different choice of integration
contour, whereas here they would be distinguished by different choices of boundary conditions.
One may indeed verify that our explicit expressions for 
$\mathcal{C}^{(1)}_{1,2,3}$, given in eqs.~(\ref{SS_C1S1}) 
and (\ref{SS_C1S2}), satisfy the differential equation in 
eq.~(\ref{SunsetCutDiffEqs}); 
the same is true for $\mathcal{C}^{(2)}_{1,2,3}$, 
given in eqs.~(\ref{SS_C2S1}) and (\ref{SS_C2S2}). 

Instead of considering a system of first-order differential equations,
we can consider an equivalent second-order differential equation, which is
a more standard way to establish the connection with hypergeometric
functions~\cite{handbook}.
By differentiating eq.~(\ref{SunsetDiffEqs}) a second time and expressing 
the result using the operator $\theta=z\frac{d}{dz}$ 
we obtain
 \begin{align}\label{1mSunsetSecondOrderDiffEq2}
 \left[z(\theta+1+2\epsilon)(\theta+1+\epsilon)-\theta(\theta-\epsilon)\right]\frac{1}{1-z}S^{(1)}&=0\,.
 \end{align}
Any of the cuts of $S^{(1)}$ should obey this same equation
and indeed we find
 \begin{align}\label{1mSunsetSecondOrderDiffEq}
 \left[z(\theta+1+2\epsilon)(\theta+1+\epsilon)-\theta(\theta-\epsilon)\right]\frac{1}{1-z}\mathcal{C}_{1,2,3}S^{(1)}=0\,.
 \end{align}
This second-order equation is directly recognisable as the differential 
equation of a ${}_2F_1$ function~\cite{handbook}.
It has a two-dimensional solution space spanned by the results of eqs.~(\ref{SS_C1S1}) and~(\ref{SS_C2S1}).
In particular, it then follows that the uncut integral should be expressible in terms of the maximal
cuts, and this was already established in eq.~\eqref{eq:unctuAsCut}. In fact,
any cut of~$S^{(1)}$ must be expressible in terms of these two functions as they all satisfy the same 
second-order differential equation. In this respect the one-mass sunset example is rather special. 
Generically, non-maximal cuts are also required to span the full space of the cuts, 
in direct correspondence with the fact that there are master integrals with fewer propagators.

\subsection{Cuts of the two-mass sunset\label{2masssunset}}

Let us now consider the differential equations obeyed by the cuts of the two-mass sunset integral. 
Starting from its differential equation,
we will  demonstrate that the cuts presented in section~\ref{sec:appTwoMassExp}
span the entire space of cuts of these integrals. 
As a consequence, this set of cuts is sufficient to express the second entries of the 
coaction of the uncut master integrals.
Furthermore, we will demonstrate the relation between the three-dimensional subspace spanned by the maximal cuts, which may be expressed in terms of Appell $F_1$ integrals, and the full four-dimensional space of cuts which also includes the two-propagator cut and contains the uncut integral
in its span.

To derive the differential equations we consider the three master integrals $S^{(1)}$, $S^{(2)}$ 
and $S^{(3)}$ defined in eq.~(\ref{eq:2massBasis}) and given explicitly in 
section~\ref{sec:appTwoMassExp}, along with the double tadpole master integral $J$ of eq.~(\ref{eq:tadsq}),
as functions of the variables $z_1=\frac{m_1^2}{p^2}$ and $z_2=\frac{m_2^2}{p^2}$, 
regarding $p^2$ as constant.
We arrange these four functions in a vector $\vec S$,
\begin{equation}
	\vec S=\left(\begin{array}{c}S^{(1)}\\S^{(2)}\\S^{(3)}\\J\end{array}\right)\,,
\end{equation}
which satisfies the differential equations
\begin{align}\label{FirstOrderJi}
\theta\vec S=\epsilon \,\mathcal{A}\,\vec S,
\qquad \qquad
\phi\vec S=\epsilon\, \mathcal{B}\,\vec S,
\end{align}
where $\theta\equiv z_1\frac{\partial}{\partial z_1}$, $\phi\equiv z_2\frac{\partial}{\partial z_2}$,
and the matrices $\mathcal{A}$ and $\mathcal{B}$ are given by
\begin{align}\begin{split}
\label{calAB}
\mathcal{A}&\,=\left(\begin{array}{cccc}\frac{1-5z_1^2+z_2^2+4z_1-2z_2+4z_1z_2}{\lambda(1,z_1,z_2)}
&\frac{2(z_2-1)}{\sqrt{\lambda(1,z_1,z_2)}}
&\frac{z_2-1-3z_1}{\sqrt{\lambda(1,z_1,z_2)}}
&\frac{z_2-1-z_1}{\sqrt{\lambda(1,z_1,z_2)}}
\\
\frac{1-z_1-z_2}{\sqrt{\lambda(1,z_1,z_2)}}
&-2
&-1
&-1
\\
\frac{2z_1}{\sqrt{\lambda(1,z_1,z_2)}}
&0
&0
&1
\\
0
&0
&0
&-1\end{array}\right)\,,\\
\mathcal{B}&\,=\left(\begin{array}{cccc}
\frac{1+z_1^2-5z_2^2-2z_1+4z_2+4z_1z_2}{\lambda(1,z_1,z_2)}
&\frac{z_1-1-3z_2}{\sqrt{\lambda(1,z_1,z_2)}}
&\frac{2(z_1-1)}{\sqrt{\lambda(1,z_1,z_2)}}
&\frac{z_1-1-z_2}{\sqrt{\lambda(1,z_1,z_2)}}
\\
\frac{2z_2}{\sqrt{\lambda(1,z_1,z_2)}}
&0
&0
&1
\\
\frac{1-z_1-z_2}{\sqrt{\lambda(1,z_1,z_2)}}
&
-1
&
-2
&-1
\\0
&0
&0
&-1
\end{array}\right).
\end{split}\end{align}

All the cuts satisfy the same system of first-order differential equations. In particular,
the two-propagator cut satisfies
\begin{equation}\label{FirstOrderC12Ji}
	\theta\,\mathcal{C}_{1,2}\vec S=\epsilon \,\mathcal{A}\,\mathcal{C}_{1,2}\vec  S\,,
	\qquad \qquad
	\phi\,\mathcal{C}_{1,2}\vec S=\epsilon\, \mathcal{B}\,\mathcal{C}_{1,2}\vec S\,,
\end{equation}
and the maximal cuts satisfy
\begin{equation}
\label{maxcut2masssunset}
\bsp
	\theta\,\mathcal{C}^{(i)}_{1,2,3}\vec S=\epsilon \,\mathcal{A}\,\mathcal{C}^{(i)}_{1,2,3}\vec  S\,,
	\qquad \qquad
	\phi\,\mathcal{C}^{(i)}_{1,2,3}\vec S=\epsilon\, \mathcal{B}\,\mathcal{C}^{(i)}_{1,2,3}\vec S\,,
\esp\end{equation}
with $i=1,2,3$. For the maximal cuts, we note that the double tadpole $J$ trivially vanishes upon taking a residue on propagator 3, which is absent there, so we have:
\begin{equation}
\label{maxcutwithJeq0}
	\mathcal{C}^{(i)}_{1,2,3}\vec S=
	\left(\begin{array}{c}\mathcal{C}^{(i)}_{1,2,3}S^{(1)}\\
	\mathcal{C}^{(i)}_{1,2,3}S^{(2)}\\
	\mathcal{C}^{(i)}_{1,2,3}S^{(3)}\\
	0\end{array}\right)\,.
\end{equation}
Note that $J$ vanishing under the maximal cut is consistent with the fact 
that the first three entries in the last row of the matrices $\mathcal{A}$ 
and $\mathcal{B}$ are zero. The latter is a consequence of the fact that the
differential equations for Feynman integrals have a natural hierarchical 
structure, where the differential equation of an integral with a given set 
of propagators can always be written in a form that does not involve 
integrals with more propagators.
It then follows from eqs.~(\ref{maxcut2masssunset}) 
and~(\ref{maxcutwithJeq0})  that the maximal cuts of the two-mass sunset 
satisfy the simpler system of equations:
\begin{align}\label{FirstOrderC123Ji}
\theta\left(\begin{array}{c}\mathcal{C}^{(i)}_{1,2,3}S^{(1)}\\
\mathcal{C}^{(i)}_{1,2,3}S^{(2)}\\\mathcal{C}^{(i)}_{1,2,3}S^{(3)}\end{array}\right)
=\epsilon \tilde{\mathcal{A}}\left(\begin{array}{c}\mathcal{C}^{(i)}_{1,2,3}S^{(1)}\\
\mathcal{C}^{(i)}_{1,2,3}S^{(2)}\\\mathcal{C}^{(i)}_{1,2,3}S^{(3)}\end{array}\right)\,,\qquad\,\,
\phi\left(\begin{array}{c}\mathcal{C}^{(i)}_{1,2,3}S^{(1)}\\\mathcal{C}^{(i)}_{1,2,3}S^{(2)}\\
\mathcal{C}^{(i)}_{1,2,3}S^{(3)}\end{array}\right)=\epsilon \tilde{\mathcal{B}}
\left(\begin{array}{c}\mathcal{C}^{(i)}_{1,2,3}S^{(1)}\\\mathcal{C}^{(i)}_{1,2,3}S^{(2)}\\
\mathcal{C}^{(i)}_{1,2,3}S^{(3)}\end{array}\right),
\end{align}
with $\tilde{\mathcal{A}}$ and $\tilde{\mathcal{B}}$ respectively given by the 
upper-left three-by-three sub-matrix of ${\mathcal{A}}$ and ${\mathcal{B}}$ of eq.~(\ref{calAB}):
\begin{align}
\tilde{\mathcal{A}}\,=\,\left(\begin{array}{ccc}{\cal A}_{1,1}&{\cal A}_{1,2}&{\cal A}_{1,3}\\
{\cal A}_{2,1}&{\cal A}_{2,2}&{\cal A}_{2,3}\\{\cal A}_{3,1}&A_{3,2}&A_{3,3}\end{array}\right)\,,
\qquad\quad 
\tilde{\mathcal{B}}\,=\,\left(\begin{array}{ccc}{\cal B}_{1,1}&{\cal B}_{1,2}&{\cal B}_{1,3}\\
{\cal B}_{2,1}&{\cal B}_{2,2}&{\cal B}_{2,3}\\
{\cal B}_{3,1}&{\cal B}_{3,2}&{\cal B}_{3,3}\end{array}\right)\,.
\end{align}

From these first-order systems we can derive a set of second-order equations for the first master $S^{(1)}$ and its cuts. These all satisfy
\begin{align}\label{SecondOrderJ1}
\mathcal{D}^{(1)}_1\frac{1}{\sqrt{\lambda(1,z_1,z_2)}}f^{(1)}(z_1,z_2)=&0\,,
\qquad\quad 
\mathcal{D}^{(1)}_2\frac{1}{\sqrt{\lambda(1,z_1,z_2)}}f^{(1)}(z_1,z_2)=0\,,
\end{align}
where $f^{(1)}(z_1,z_2)$ can be $S^{(1)}$ or \emph{any} of its cuts (including all maximal cuts $\mathcal{C}_{1,2,3}^{(i)}S^{(1)}$ and the 
two-propagator cut $\mathcal{C}_{1,2}S^{(1)}$)
and the differential operators are
\begin{align}\begin{split}
\mathcal{D}^{(1)}_1=\,&(1-z_1-z_2)\theta^2-2z_1\theta\phi-[(2+5\epsilon)z_1-\epsilon(1-z_2)]\theta
\\&
-2(1+2\epsilon)z_1\phi-(1+2\epsilon)(1+3\epsilon)z_1\,,\\
\mathcal{D}^{(1)}_2=\,&(1-z_1-z_2)\phi^2-2z_2\theta\phi-[(2+5\epsilon)z_2-\epsilon(1-z_1)]\phi\\
&-2(1+2\epsilon)z_2\theta-(1+2\epsilon)(1+3\epsilon)z_2\,.
\end{split}\end{align}
To derive these second-order equations we start from the expressions for $\theta S^{(1)}$ and 
$\phi S^{(1)}$ from eq.~(\ref{FirstOrderJi}), along with the expression for $\theta\phi S^{(1)}$, and 
solve them for $S^{(2)}$, $S^{(3)}$ and~$J$. 
This allows the objects $\theta^2S^{(1)}$ and $\phi^2S^{(1)}$
to be written solely in terms of $S^{(1)}$, $\theta S^{(1)}$, $\phi S^{(1)}$ and $\theta\phi S^{(1)}$. 
For the two-propagator cut we start from eq.~(\ref{FirstOrderC12Ji}) and repeat the very same procedure.
For the maximal cuts we similarly start from eq.~(\ref{maxcut2masssunset}). 
However, given eq.~(\ref{maxcutwithJeq0}) we are only required to 
eliminate the maximal cuts of $S^{(2)}$ and $S^{(3)}$, but not of~$J$, and so there is an extra 
independent relation:
\begin{align}\label{D13eq0}
\mathcal{D}^{(1)}_3\frac{1}{\sqrt{\lambda(1,z_1,z_2)}}\mathcal{C}_{1,2,3}^{(i)}S^{(1)}=0,
\end{align}
with
\begin{align}
&\mathcal{D}^{(1)}_3 =\theta\phi-\frac{1+2\epsilon}{\lambda(1,z_1,z_2)}\left[z_2(1+3z_1-z_2)\theta+z_1(1-z_1+3z_2)\phi+2(1+3\epsilon)z_1z_2\right].
\end{align}

Solving the equations \eqref{SecondOrderJ1}, 
we find the general solution
\begin{align}\label{GenSolf}
\begin{split}
f^{(1)}(z_1,z_2)=
\,\sqrt{\lambda\left(1,z_1,z_2\right)}\bigg[&A\,F_4(1+2\epsilon,1+3\epsilon;1+\epsilon,1+\epsilon;z_1,z_2)
\\&+B\,z_1^{-\epsilon} F_4(1+2\epsilon,1+\epsilon;1-\epsilon,1+\epsilon;z_1,z_2)
\\&+C\,z_2^{-\epsilon}F_4(1+2\epsilon,1+\epsilon;1+\epsilon,1-\epsilon;z_1,z_2)
\\&+D\,z_1^{-\epsilon}z_2^{-\epsilon}F_4(1+\epsilon,1;1-\epsilon,1-\epsilon;z_1,z_2)\bigg]\,.
\end{split}
\end{align}
where the coefficients $A$ through $D$ depend on $\epsilon$, but not on the kinematic variables.
These four independent solutions correspond~\cite{Matthew:2020dpb} to the four independent contours 
emerging from the integral representation of the $F_4$ function of ref.~\cite{goto2014monodromy}, and $S^{(1)}$ and any of its cuts can be written
in terms of them, each corresponding to different values of the coefficients
$A$ through $D$. This may be verified with the explicit expressions in section~\ref{sec:appTwoMassExp}.

The general solution in eq.~(\ref{GenSolf}) was chosen to have a specific property, namely that
only the first three terms obey eq.~\eqref{D13eq0}.
This may be demonstrated by employing eqs.~(14-15) in ref.~\cite{Kniehl:2011ym} to re-express these solutions using $F_1$ functions, and then using
 \begin{align}\label{F1ThirdDiffEq}
 \left(\theta\phi-\frac{\beta^\prime y}{x-y}\theta+\frac{\beta x}{x-y}\phi\right)F_1(\alpha;\beta,\beta^\prime;\gamma;x,y)=0\,,
 \end{align}
 which is satisfied by the generic $F_1$ function~\cite{Mullen}. The extra differential equation (\ref{D13eq0}) obeyed by the maximal cuts can thus be interpreted as arising from the relation (\ref{F1ThirdDiffEq}) which sets apart a three dimensional subspace within the four-dimensional space of solutions of~eqs.~\eqref{SecondOrderJ1}.
This explains why the maximal cuts can both be written as Appell~$F_4$ and as Appell $F_1$ functions, spanning a three-dimensional subspace of the former. It follows that the maximal cuts of $S^{(1)}$ can be written as a linear combination of the first three functions in eq.~(\ref{GenSolf}) and this is indeed what one finds using the method of appendix~\ref{CutsSection}, with the three cuts summarised in  eqs.~(\ref{MaxCut1_2mass_sunsetS1}) through (\ref{MaxCut3_2mass_sunsetS1}).

The same reasoning can be applied to the integrals corresponding to the masters $S^{(2)}$ and $S^{(3)}$.
The differential equations have a more complex form than for the case of $S^{(1)}$. 
We examine only $S^{(2)}$, as the case of $S^{(3)}$ will follow from swapping 
$z_1$ and $z_2$. We define the three operators:
\begin{align}
\nonumber\mathcal{D}^{(2)}_1=\,&(1-z_1-z_2)\theta^2-2z_1\theta\phi-[5\epsilon z_1-\epsilon(1-z_2)]\theta-(1+4\epsilon)z_1\phi-6\epsilon^2z_1\,,\\
\mathcal{D}^{(2)}_2=\,&(1-z_1-z_2)\phi^2-2z_2\theta\phi-[5\epsilon z_2-(\epsilon-1)(1-z_1)]\phi-4\epsilon z_2\theta-6\epsilon^2z_2\,,\\
\nonumber\mathcal{D}^{(2)}_3=\,&\theta\phi-\frac{2\epsilon z_2(1+3z_1-z_2)\theta+[1-z_1+z_2+\epsilon(2-2z_1+6z_2)]z_1\phi+12\epsilon^2z_1z_2}{\lambda(1,z_1,z_2)}\,.
\end{align}
It then follows from eqs.~(\ref{FirstOrderJi}), (\ref{FirstOrderC12Ji}) and (\ref{FirstOrderC123Ji}) that $S^{(2)}$ and all of its cuts satisfy
\begin{align}\begin{split}\label{J2DiffEqs}
\left[\mathcal{D}^{(2)}_1+\frac{z_1\lambda(1,z_1,z_2)}{-z_1(1-z_1+z_2)+\epsilon(-1+z_1^2-z_2^2+2z_2)}
\mathcal{D}^{(2)}_3\right]f^{(2)}(z_1,z_2)=&\,0\,,\\
\left[\mathcal{D}^{(2)}_2+\frac{(1-z_1)\lambda(1,z_1,z_2)}{-z_1(1-z_1+z_2)
+\epsilon(-1+z_1^2-z_2^2+2z_2)}\mathcal{D}^{(2)}_3\right]f^{(2)}(z_1,z_2)=&\,0\,,
\end{split}\end{align}
with the maximal cut obeying the additional constraint
\begin{equation}\label{D23eq0}
\mathcal{D}^{(2)}_3\mathcal{C}_{1,2,3}S^{(2)}=0\,.
\end{equation}
Similar to the case of $S^{(1)}$, we find that the general solution
to eqs.~\eqref{J2DiffEqs} is
\begin{align}
\label{diff_eq_solution_second_master}
\begin{split}
f^{(2)}(z_1,z_2)=&\,A\,F_4(2\epsilon,3\epsilon;1+\epsilon,\epsilon;z_1,z_2)\\
&+B\,z_1^{-\epsilon}F_4(\epsilon,2\epsilon;1-\epsilon,\epsilon;z_1,z_2)\\
&+C\,z_2^{1-\epsilon}F_4(1+\epsilon,1+2\epsilon;1+\epsilon,2-\epsilon;z_1,z_2)\\
&+D\,z_1^{-\epsilon}z_2^{-\epsilon}\left[1+\frac{2\epsilon}{1-\epsilon}
z_2 \,F_4(1,1+\epsilon;1-\epsilon,2-\epsilon;z_1,z_2)\right]\,,
\end{split}
\end{align}
where only the first three terms satisfy eq.~\eqref{D23eq0},
and are thus sufficient to span the three-dimensional space of the maximal
cuts.


\section{Expressions for Master Integrals and Cuts}\label{Expressions}

In this appendix we collect expressions for the Feynman integrals and their cuts
that we used as examples for the diagrammatic coaction.

\subsection{Sunsets}

\subsubsection{Massless Sunset}
For completeness, we reproduce the results given in section \ref{sec:sunsetMassless}.
The uncut integral is
\begin{align}\begin{split}
S(p^2)=(-p^2)^{-2\epsilon}e^{2\gamma_E\epsilon}
\frac{\Gamma^3(1-\epsilon)\Gamma(1+2\epsilon)}{\Gamma(1-3\epsilon)}\,,
\end{split}\end{align}
and the cut integral is
\begin{align}\begin{split}
\mathcal{C}_{1,2,3}S(p^2)=(p^2)^{-2\epsilon}e^{2\gamma_E\epsilon}
\frac{\Gamma^3(1-\epsilon)\Gamma(1+2\epsilon)}{\Gamma(1-3\epsilon)}\,.
\end{split}\end{align}

\subsubsection{One-mass Sunset}

The uncut integrals considered in section \ref{sec:first-examples} are
\begin{align}\label{SS1m1}\begin{split}
S^{(1)}(p^2,m^2)=&(m^2)^{-2\epsilon}\left(1-\frac{p^2}{m^2}\right)e^{2\gamma_E\epsilon}
\Gamma(1+2\epsilon)\Gamma(1-\epsilon)\Gamma(1+\epsilon)\\
&\,{}_2F_1\left(1+2\epsilon,1+\epsilon;1-\epsilon;\frac{p^2}{m^2}\right)\,,
\end{split}\end{align}
\begin{align}\label{SS1m2}
\begin{split}
 S^{(2)}(p^2,m^2)=\,(m^2)^{-2\epsilon}e^{2\gamma_E\epsilon}\Gamma(1+2\epsilon)
\Gamma(1-\epsilon)\Gamma(1+\epsilon)
\,{}_2F_1\left(2\epsilon,\epsilon;1-\epsilon;\frac{p^2}{m^2}\right)\,.
\end{split}\end{align}

The associated cuts are:
\begin{align}\label{SS_C1S1}\begin{split}
\mathcal{C}^{(1)}_{1,2,3}S^{(1)}=\,&e^{2\gamma_E\epsilon}\frac{\Gamma^2(1-\epsilon)}{\Gamma(1-4\epsilon)}(p^2)^{2\epsilon}(p^2-m^2)^{-4\epsilon}{}_2F_1\left(-2\epsilon,-\epsilon;-4\epsilon;1-\frac{m^2}{p^2}\right)\,,
\end{split}\end{align}
\begin{align}\label{SS_C2S1}
\begin{split}
\mathcal{C}^{(2)}_{1,2,3}S^{(1)}=\,&e^{2\gamma_E\epsilon}
	\frac{\Gamma(1+\epsilon)\Gamma(1-\epsilon)}
	{\Gamma(1-2\epsilon)}(p^2-m^2)^{-2\epsilon}
	{}_2F_1\left(-2\epsilon,1+2\epsilon;1-\epsilon;\frac{p^2}{p^2-m^2}\right)\\
&-e^{2\gamma_E\epsilon}\frac{\Gamma^2(1-\epsilon)}{\Gamma(1-4\epsilon)}(p^2)^{2\epsilon}(p^2-m^2)^{-4\epsilon}{}_2F_1\left(-2\epsilon,-\epsilon;-4\epsilon;1-\frac{m^2}{p^2}\right),
\end{split}
\end{align}
\begin{align}\label{SS_C1S2}\begin{split}
\mathcal{C}^{(1)}_{1,2,3}S^{(2)}=\,&e^{2\gamma_E\epsilon}\frac{\epsilon\Gamma^2(1-\epsilon)}{2\,\Gamma(2-4\epsilon)}\frac{(p^2-m^2)^{1-4\epsilon}}{(p^2)^{1-2\epsilon}}{}_2F_1\left(1-2\epsilon,1-\epsilon;2-4\epsilon;1-\frac{m^2}{p^2}\right),
\end{split}\end{align}
\begin{align}\label{SS_C2S2}
\begin{split}
\mathcal{C}^{(2)}_{1,2,3}S^{(2)}
=\,&e^{2\gamma_E\epsilon}\frac{\Gamma(1+\epsilon)\Gamma(1-\epsilon)}{\Gamma(1-2\epsilon)}(p^2-m^2)^{-2\epsilon}{}_2F_1\left(1-2\epsilon,2\epsilon;1-\epsilon;\frac{p^2}{p^2-m^2}\right)\\
-&e^{2\gamma_E\epsilon}\frac{\epsilon\Gamma^2(1-\epsilon)}{2\,\Gamma(2-4\epsilon)}
\frac{(p^2-m^2)^{1-4\epsilon}}{(p^2)^{1-2\epsilon}}{}_2F_1\left(1-2\epsilon,1-\epsilon;2-4\epsilon;1-\frac{m^2}{p^2}\right).
\end{split}\end{align}

\subsubsection{Two-mass Sunset}\label{sec:appTwoMassExp}

The uncut integrals defined in eq.~\eqref{eq:2massBasis} are given by~\cite{Jegerlehner:2003py,Matthew:2020dpb} 
\begin{align}\label{STwoMasses}
&S^{(1)}(p^2,m_1^2,m_2^2)=-\epsilon^2\sqrt{\lambda\left(p^2,m_1,m_2^2\right)}\,
S(1,1,1,0,0;2-2\epsilon;p^2,m_1^2,m_2^2)\nonumber\\
&=e^{2\gamma_E\epsilon}\sqrt{\lambda\left(p^2,m_1,m_2^2\right)}(m_1^2)^{-1-2\epsilon}
\left\{-\Gamma^2(1+\epsilon)\left(\frac{m_2^2}{m_1^2}\right)^{-\epsilon}F_4\left(1+\epsilon,1;1-\epsilon,1-\epsilon,\frac{p^2}{m_1^2},\frac{m_2^2}{m_1^2}\right)\right.\nonumber\\
&\left.+\Gamma(1+2\epsilon)\Gamma(1-\epsilon)\Gamma(1+\epsilon)F_4\left(1+2\epsilon,1+\epsilon;1-\epsilon,1+\epsilon;\frac{p^2}{m_1^2},\frac{m_2^2}{m_1^2}\right)\right\}\,,
\end{align}
\begin{align}\label{STwoMasses2}
&S^{(2)}(p^2,m_1^2,m_2^2)=\epsilon^2S(1,1,1,-1,0;2-2\epsilon;p^2,m_1^2,m_2^2)\nonumber\\
&=e^{2\gamma_E\epsilon}\Gamma(1+\epsilon)^2\left(m_1^2\right)^{-\epsilon}\left(m_2^2\right)^{-\epsilon}\left\{-1-\frac{2\epsilon}{1-\epsilon}\frac{m_2^2}{m_1^2}F_4\left(1+\epsilon,1;1-\epsilon,2-\epsilon;\frac{p^2}{m_1^2},\frac{m_2^2}{m_1^2}\right)\right.\nonumber\\
&+\left.\left(\frac{m_2^2}{m_1^2}\right)^\epsilon\frac{\Gamma(1+2\epsilon)\Gamma(1-\epsilon)}{\Gamma(1+\epsilon)}F_4\left(2\epsilon,\epsilon;1-\epsilon,\epsilon;\frac{p^2}{m_1^2},\frac{m_2^2}{m_1^2}\right)\right\}\,,
\end{align}
\begin{align}\label{STwoMasses3}
&S^{(3)}(p^2,m_1^2,m_2^2)=\epsilon^2S(1,1,1,0,-1;2-2\epsilon;p^2,m_1^2,m_2^2)\nonumber\\
&=e^{2\gamma_E\epsilon}\Gamma(1+\epsilon)^2\left(m_1^2\right)^{-\epsilon}\left(m_2^2\right)^{-\epsilon}
\left\{-1+2F_4\left(\epsilon,1;1-\epsilon,1-\epsilon;\frac{p^2}{m_1^2},
\frac{m_2^2}{m_1^2}\right)\right.\nonumber\\
&-\left.\left(\frac{m_2^2}{m_1^2}\right)^\epsilon\frac{\Gamma(1+2\epsilon)\Gamma(1-\epsilon)}{\Gamma(1+\epsilon)}F_4\left(2\epsilon,1+\epsilon;1-\epsilon,1+\epsilon;\frac{p^2}{m_1^2},\frac{m_2^2}{m_1^2}\right)\right\}\,.
\end{align}

In order to apply the coaction it is convenient to first obtain symmetric representations of these master integrals in terms of the arguments $z_1=\frac{m_1^2}{p^2}$ and $z_2=\frac{m_2^2}{p^2}$. This is the basis of functions in which the cuts are computed in appendices \ref{CutsSection} and \ref{DiffEqCutsSection}, and it is therefore our basis-of-choice for applying the global coaction.

To this end one applies the following analytic continuation relation~\cite{MSM_1925__3__1_0}:
\begin{align}
\label{AnalContRelations}
\begin{split}
F_4\left(\alpha, \beta; \gamma, \tilde{\gamma}; U,  V\right) =&\,\left(-U\right)^{-\alpha}
 \frac{\Gamma(\beta-\alpha)\Gamma(\gamma)}{ \Gamma(\beta)\Gamma(\gamma-\alpha) }
 F_4\left(\alpha, 1 + \alpha - \gamma; 1 + \alpha - \beta, \tilde{\gamma}; \frac{1}{U}, \frac{V}{U}\right) 
 \\&\,+\,\left(-U\right)^{-\beta}
 \frac{\Gamma(\alpha-\beta)\Gamma(\gamma)}{ \Gamma(\alpha)\Gamma(\gamma-\beta) }
 F_4\left(\beta, 1 + \beta- \gamma; 1 + \beta - \alpha, \tilde{\gamma}; \frac{1}{U}, \frac{V}{U}\right) \,.
\end{split}
\end{align}
Considering the first master integral, this transformation converts each of the two Appell $F_4$ functions in eq.~(\ref{STwoMasses}), depending on $\frac{p^2}{m_1^2}$ and $\frac{m_2^2}{m_1^2}$, into two Appell $F_4$ functions depending on $z_1$ and $z_2$. The four functions obtained this way correspond directly to the ones emerging as independent solutions to the differential equations, given in eq.~(\ref{GenSolf}). Upon applying eq.~(\ref{AnalContRelations}) to (\ref{STwoMasses}) one thus readily determines the four coefficients $A$, $B$, $C$ and $D$ in eq.~(\ref{GenSolf}), in terms of $\epsilon$-dependent Gamma functions along with phases associated with the analytic continuation. The very same procedure applies to the second master integral in eq.~(\ref{STwoMasses2}), yielding the result in terms of the four functions appearing in eq.~(\ref{diff_eq_solution_second_master}). Finally, the third master integral may be obtained from the second 
using the $z_1\leftrightarrow z_2$ symmetry. We refer the reader to ref.~\cite{Matthew:2020dpb} for the explicit results.

For the cuts, we set $z_1=\frac{m_1^2}{p^2}$ and $z_2=\frac{m_2^2}{p^2}$ and find the following results. For the cuts of $S^{(1)}$:
\begin{align}\label{MaxCut1_2mass_sunsetS1}\begin{split}
\mathcal{C}_{1,2,3}^{(1)}S^{(1)}(p^2,m_1^2,m_2^2)=&\,\,e^{2\gamma_E\epsilon}
\sqrt{\lambda\left(1,z_1,z_2\right)}(p^2)^{-2\epsilon} \Gamma(1+2\epsilon)\\
\times&\left\{\vphantom{\frac{1}{2}}\,- \, z_1^{-\epsilon}
F_4\left(1+\epsilon,1+2\epsilon;1-\epsilon,1+\epsilon;z_1,z_2\right)\right.\\
-&z_2^{-\epsilon}
F_4\left(1+\epsilon,1+2\epsilon;1+\epsilon,1-\epsilon;z_1,z_2\right)\\
+&3\left.\frac{\Gamma(1+3\epsilon)}{\Gamma^3(1+\epsilon)}
F_4\left(1+2\epsilon,1+3\epsilon;1+\epsilon,1+\epsilon;z_1,z_2\right)\right\}\,.
\end{split}\end{align}
\begin{align}\label{MaxCut2_2mass_sunsetS1}\begin{split}
\mathcal{C}_{1,2,3}^{(2)}S^{(1)}(p^2,m_1^2,m_2^2)=&\,\,
e^{2\gamma_E\epsilon}\sqrt{\lambda\left(1,z_1,z_2\right)}(p^2)^{-2\epsilon} \Gamma(1+2\epsilon)\\
\times&\left\{\vphantom{\frac{1}{2}}
2z_1^{-\epsilon}
F_4\left(1+\epsilon,1+2\epsilon;1-\epsilon,1+\epsilon;z_1,z_2\right)\right.\\
+&z_2^{-\epsilon}
F_4\left(1+\epsilon,1+2\epsilon;1+\epsilon,1-\epsilon;z_1,z_2\right)\\
-&3\left.\frac{\Gamma(1+3\epsilon)}{\Gamma^3(1+\epsilon)}
F_4\left(1+2\epsilon,1+3\epsilon;1+\epsilon,1+\epsilon;z_1,z_2\right)\vphantom{\frac{1}{2}}\right\}\,.
\end{split}\end{align}
\begin{align}\label{MaxCut3_2mass_sunsetS1}
\begin{split}
\mathcal{C}_{1,2,3}^{(3)}S^{(1)}(p^2,m_1^2,m_2^2)=&\,\,
e^{2\gamma_E\epsilon}\sqrt{\lambda\left(1,z_1,z_2\right)}(p^2)^{-2\epsilon} \Gamma(1+2\epsilon)\\
\times&\left\{\vphantom{\frac{1}{2}}z_1^{-\epsilon}
F_4\left(1+\epsilon,1+2\epsilon;1-\epsilon,1+\epsilon;z_1,z_2\right)\right.\\
+&2z_2^{-\epsilon}
F_4\left(1+\epsilon,1+2\epsilon;1+\epsilon,1-\epsilon;z_1,z_2\right)\\
-&3\left.\frac{\Gamma(1+3\epsilon)}{\Gamma^3(1+\epsilon)}
F_4\left(1+2\epsilon,1+3\epsilon;1+\epsilon,1+\epsilon;z_1,z_2\right)\right\}\,.
\end{split}\end{align}
\begin{align}\label{C12_2mass_sunsetS1}
\begin{split}
\mathcal{C}_{1,2}S^{(1)}(p^2,m_1^2,m_2^2)=
&\,\,-e^{2\gamma_E\epsilon}\sqrt{\lambda(1,z_1,z_2)}(p^2)^{-2\epsilon}
\Gamma^2(1+\epsilon)z_1^{-\epsilon}z_2^{-\epsilon}\\
\times&F_4\left(1,1+\epsilon;1-\epsilon,1-\epsilon;z_1,z_2\right)\,.
\end{split}\end{align}

For the cuts of $S^{(2)}$:
\begin{align}\begin{split}
\mathcal{C}_{1,2,3}^{(1)}S^{(2)}(p^2,m_1^2,m_2^2)=&\,\,
-e^{2\gamma_E\epsilon}(p^2)^{-2\epsilon}\Gamma(1+2\epsilon)
\left\{\vphantom{\frac{1}{2}}
z_1^{-\epsilon}
F_4\left(\epsilon,2\epsilon;1-\epsilon,\epsilon;z_1,z_2\right)\right.\\
+&\frac{2\epsilon}{1-\epsilon}z_2^{1-\epsilon}
F_4\left(1+\epsilon,1+2\epsilon;1+\epsilon,2-\epsilon;z_1,z_2\right)\\
+&\left.\frac{\Gamma(1+3\epsilon)}{\Gamma^3(1+\epsilon)}
F_4\left(2\epsilon,3\epsilon;1+\epsilon,\epsilon;z_1,z_2\right)\right\}\,.
\end{split}\end{align}
\begin{align}\begin{split}
\mathcal{C}_{1,2,3}^{(2)}S^{(2)}(p^2,m_1^2,m_2^2)=&\,\,
e^{2\gamma_E\epsilon}(p^2)^{-2\epsilon}\Gamma(1+2\epsilon)
\left\{\vphantom{\frac{1}{2}}2z_1^{-\epsilon}
F_4\left(\epsilon,2\epsilon;1-\epsilon,\epsilon;z_1,z_2\right)\right.\\
+&\frac{2\epsilon}{1-\epsilon}z_2^{1-\epsilon}
F_4\left(1+\epsilon,1+2\epsilon;1+\epsilon,2-\epsilon;z_1,z_2\right)\\
+&\left.\frac{\Gamma(1+3\epsilon)}{\Gamma^3(1+\epsilon)}
F_4\left(2\epsilon,3\epsilon;1+\epsilon,\epsilon;z_1,z_2\right)\right\}\,.
\end{split}\end{align}
\begin{align}\begin{split}
\mathcal{C}_{1,2,3}^{(3)}S^{(2)}(p^2,m_1^2,m_2^2)=&\,\,
e^{2\gamma_E\epsilon}(p^2)^{-2\epsilon}\Gamma(1+2\epsilon)
\left\{\vphantom{\frac{1}{2}}z_1^{-\epsilon}
F_4\left(\epsilon,2\epsilon;1-\epsilon,\epsilon;z_1,z_2\right)\right.\\
+&\frac{4\epsilon}{1-\epsilon}z_2^{1-\epsilon}
F_4\left(1+\epsilon,1+2\epsilon;1+\epsilon,2-\epsilon;z_1,z_2\right)\\
+&\left.\frac{\Gamma(1+3\epsilon)}{\Gamma^3(1+\epsilon)}
F_4\left(2\epsilon,3\epsilon;1+\epsilon,\epsilon;z_1,z_2\right)\right\}\,.
\end{split}\end{align}
\begin{align}\begin{split}
\mathcal{C}_{1,2}S^{(2)}(p^2,m_1^2,m_2^2)=&\,\,
-e^{2\gamma_E\epsilon}\Gamma^2(1+\epsilon)(p^2)^{-2\epsilon}z_1^{-\epsilon}z_2^{-\epsilon}\\
\times&\left\{1+\frac{2\epsilon}{1-\epsilon}z_2
F_4\left(1,1+\epsilon;1-\epsilon,2-\epsilon;z_1,z_2\right)\right\}\,.
\end{split}\end{align}

For the cuts of $S^{(3)}$:
\begin{align}\begin{split}
\mathcal{C}_{1,2,3}^{(1)}S^{(3)}(p^2,m_1^2,m_2^2)=\,\,&
-e^{2\gamma_E\epsilon}(p^2)^{-2\epsilon}\Gamma(1+2\epsilon)\left\{
z_2^{-\epsilon}F_4(\epsilon,2\epsilon;\epsilon,1-\epsilon;z_1,z_2)\right.\\
+&\frac{2\epsilon}{1-\epsilon}z_1^{1-\epsilon}
F_4(1+\epsilon,1+2\epsilon;2-\epsilon,1+\epsilon;z_1,z_2)\\
+&\left.\frac{\Gamma(1+3\epsilon)}{\Gamma^3(1+\epsilon)}
F_4(2\epsilon,3\epsilon;\epsilon,1+\epsilon;z_1,z_2)\right\}\,.
\end{split}\end{align}
\begin{align}\begin{split}
\mathcal{C}_{1,2,3}^{(2)}S^{(3)}(p^2,m_1^2,m_2^2)=\,\,&
e^{2\gamma_E\epsilon}(p^2)^{-2\epsilon}\Gamma(1+2\epsilon)\left\{
z_2^{-\epsilon}F_4(\epsilon,2\epsilon;\epsilon,1-\epsilon;z_1,z_2)\right.\\
+&\frac{4\epsilon}{1-\epsilon}z_1^{1-\epsilon}
F_4(1+\epsilon,1+2\epsilon;2-\epsilon,1+\epsilon;z_1,z_2)\\
+&\left.\frac{\Gamma(1+3\epsilon)}{\Gamma^3(1+\epsilon)}
F_4(2\epsilon,3\epsilon;\epsilon,1+\epsilon;z_1,z_2)\right\}\,.
\end{split}\end{align}
\begin{align}\begin{split}
\mathcal{C}_{1,2,3}^{(3)}S^{(3)}(p^2,m_1^2,m_2^2)=\,\,&
e^{2\gamma_E\epsilon}(p^2)^{-2\epsilon}\Gamma(1+2\epsilon)\left\{
2z_2^{-\epsilon}F_4(\epsilon,2\epsilon;\epsilon,1-\epsilon;z_1,z_2)\right.\\
+&\frac{2\epsilon}{1-\epsilon}z_1^{1-\epsilon}
F_4(1+\epsilon,1+2\epsilon;2-\epsilon,1+\epsilon;z_1,z_2)\\
+&\left.\frac{\Gamma(1+3\epsilon)}{\Gamma^3(1+\epsilon)}
F_4(2\epsilon,3\epsilon;\epsilon,1+\epsilon;z_1,z_2)\right\}\,.
\end{split}\end{align}
\begin{align}\begin{split}
\mathcal{C}_{1,2}S^{(3)}(z_1,z_2)=&\,\,
-e^{2\gamma_E\epsilon}\Gamma^2(1+\epsilon)(p^2)^{-2\epsilon}z_1^{-\epsilon}z_2^{-\epsilon}\\
\times&\left\{1+\frac{2\epsilon}{1-\epsilon}z_1 
F_4\left(1,1+\epsilon;2-\epsilon,1-\epsilon;z_1,z_2\right)\right\}\,.
\end{split}\end{align}

\subsection{Double-Edged Triangles}

\subsubsection{Symmetric One Scale}\label{sec:doubleEdged1sSy}

\begin{align}\begin{split}
P(0,0,p_3^2)=&-\,p_3^2\,\epsilon^3P(2,1,1,1,0,0,0;4-2\epsilon;0,0,p_3^2)\\
=&\,e^{2\gamma_E\epsilon}
\frac{\Gamma(1+2\epsilon)\Gamma(1-2\epsilon)\Gamma(1+\epsilon)\Gamma^2(1-\epsilon)}
{4\,\Gamma(1-3\epsilon)}(-p_3^2)^{-2\epsilon}\,.
\end{split}\end{align}
\begin{align}\begin{split}
\mathcal{C}_{1,2,3,4}P(0,0,p_3^2)=&\,e^{2\gamma_E\epsilon}
\frac{\Gamma(1-\epsilon)}
{\Gamma(1-3\epsilon)}(p_3^2)^{-2\epsilon}\,.
\end{split}\end{align}

\subsubsection{Asymmetric One Scale}\label{sec:doubleEdged1sAsy}

\begin{align}\begin{split}
P(p_1^2,0,0)=&-p_1^2\,\epsilon^3P(2,1,1,1,0,0,0;4-2\epsilon;p_1^2,0,0)\\
=&-\,e^{2\gamma_E\epsilon}\frac{\Gamma^3(1-\epsilon)\Gamma(1+2\epsilon)}{2\,\Gamma(1-3\epsilon)}
(-p_1^2)^{-2\epsilon}\,.
\end{split}\end{align}
\begin{align}\begin{split}
\mathcal{C}_{1,2,4}P(p_1^2,0,0)=&\,
e^{2\gamma_E\epsilon}\frac{\Gamma^3(1-\epsilon)\Gamma(1+2\epsilon)}{2\,\Gamma(1-3\epsilon)}
(p_1^2)^{-2\epsilon}\,.
\end{split}\end{align}

\subsubsection{Symmetric Two Scale}\label{sec:doubleEdged2sSy}

\begin{align}\begin{split}
P(p_1^2,p_2^2,0)=&-(p_1^2-p_2^2)\epsilon^3 P(2,1,1,1,0,0,0;4-2\epsilon;p_1^2,p_2^2,0)\\
=&\,e^{2\gamma_E\epsilon}\frac{\Gamma^3(1-\epsilon)\Gamma(1+2\epsilon)}{2\,\Gamma(1-3\epsilon)}
\left((-p_2^2)^{-2\epsilon}-(-p_1^2)^{-2\epsilon}\right)\,.
\end{split}\end{align}
\begin{align}\begin{split}
\mathcal{C}_{1,2,4}P(p_1^2,p_2^2,0)=&\,
-e^{2\gamma_E\epsilon}\frac{\Gamma^3(1-\epsilon)\Gamma(1+2\epsilon)}
{2\,\Gamma(1-3\epsilon)}
(p_1^2)^{-2\epsilon}\,,\\
\mathcal{C}_{1,2,3}P(p_1^2,p_2^2,0)=&\,e^{2\gamma_E\epsilon}
\frac{\Gamma^3(1-\epsilon)\Gamma(1+2\epsilon)}{2\,\Gamma(1-3\epsilon)}(p_2^2)^{-2\epsilon}\,.
\end{split}\end{align}

\subsubsection{Asymmetric Two Scale}\label{sec:doubleEdged2sAsy}

\begin{align}\begin{split}
P(p_1^2,0,p_3^2)=&\,-(p^2_3-p_1^2)\epsilon^3 P(2,1,1,1,0,0,0;4-2\epsilon;p_1^2,0,p_3^2)\\
=&\,e^{2\gamma_E\epsilon}\left(p_3^2-p_1^2\right)(-p_3^2)^{-1-2\epsilon}
\frac{\epsilon\,\Gamma^3(1-\epsilon)\Gamma(1+\epsilon)\Gamma(1+2\epsilon)}
{2\,\Gamma(1-3\epsilon)\Gamma(2+\epsilon)}\\
&{}_2F_1\left(1+\epsilon,1+2\epsilon;2+\epsilon;1-\frac{p_1^2}{p_3^2}\right)\,.
\end{split}\end{align}
\begin{align}\begin{split}
\mathcal{C}_{1,2,3,4}P(p_1^2,0,p_3^2)=&e^{2\gamma_E\epsilon}\frac{\Gamma(1-\epsilon)}{\Gamma(1-3\epsilon)}
\left(p_1^2-p_3^2\right)^{-\epsilon}\left(p_3^2\right)^{-\epsilon}\,,\\
\mathcal{C}_{1,2,4}P(p_1^2,0,p_3^2)=&P(p_1^2,0,p_3^2)\,.
\end{split}\end{align}

\subsubsection{Three Scales}\label{sec:doubleEdged3s}

We do not list the cuts $\mathcal{C}_{1,2,4}$ as these can be deduced from 
$\mathcal{C}_{1,2,3}$ by symmetry. We set $z_1=\frac{p_1^2}{p_3^2}$ and $z_2=\frac{p_2^2}{p_3^2}$.

\begin{align}\begin{split}
&P^{(1)}(p_1^2,p_2^2,p_3^2)=-\epsilon^3\sqrt{\lambda\left(p_1^2,p_2^2,p_3^2\right)}
P(2,1,1,1,0,0,0;4-2\epsilon;p_1^2,p_2^2,p_3^2)\\
&=- e^{2\gamma_E\epsilon}\frac{\sqrt{\lambda\left(p_1^2,p_2^2,p_3^2\right)}}{4}(-p_3^2)^{-2\epsilon}\\
&\times\bigg\{\frac{\Gamma^3(1-\epsilon)\Gamma(1+2\epsilon)}{\Gamma(1-3\epsilon)}z_1^{-2\epsilon}
F_4(1,1-\epsilon;1-2\epsilon,1+2\epsilon;z_1,z_2)\\
&+\frac{\Gamma^3(1-\epsilon)\Gamma(1+2\epsilon)}{\Gamma(1-3\epsilon)}z_2^{-2\epsilon}
F_4(1,1-\epsilon;1+2\epsilon,1-2\epsilon;z_1,z_2)\\
&-\frac{\Gamma(1-2\epsilon)\Gamma^2(1-\epsilon)\Gamma(1+\epsilon)\Gamma(1+2\epsilon)}
{\Gamma(1-3\epsilon)}F_4(1+\epsilon,1+2\epsilon;1+2\epsilon,1+2\epsilon;z_1,z_2)\\
&-\Gamma^2(1-\epsilon)\Gamma^2(1+2\epsilon)z_1^{-2\epsilon}z_2^{-2\epsilon}
F_4(1-3\epsilon,1-2\epsilon;1-2\epsilon,1-2\epsilon;z_1,z_2)\bigg\}
\end{split}\end{align}

\begin{align}\begin{split}
&P^{(2)}(p_1^2,p_2^2,p_3^2)=-\epsilon^2p_3^2P(1,1,1,1,0,0,-1;2-2\epsilon;p_1^2,p_2^2,p_3^2)\\
&=e^{2\gamma_E\epsilon}\frac{(-p_3^2)^{-2\epsilon}}{2}\bigg\{
-\frac{3\,\Gamma^3(1-\epsilon)\Gamma(1+2\epsilon)}{\Gamma(1-3\epsilon)}z_1^{-2\epsilon}F_4(1,-\epsilon;1-2\epsilon,1+2\epsilon;z_1,z_2)\\
&-\frac{3\,\Gamma^3(1-\epsilon)\Gamma(1+2\epsilon)}{\Gamma(1-3\epsilon)}z_2^{-2\epsilon}F_4(1,-\epsilon;1+2\epsilon,1-2\epsilon;z_1,z_2)\\
&-\frac{3\,\Gamma(1-2\epsilon)\Gamma^2(1-\epsilon)\Gamma(1+\epsilon)\Gamma(1+2\epsilon)}{\Gamma(1-3\epsilon)}F_4(\epsilon,1+2\epsilon;1+2\epsilon,1+2\epsilon;z_1,z_2)\\
&+\,\Gamma^2(1+2\epsilon)\Gamma^2(1-\epsilon)z_1^{-2\epsilon}z_2^{-2\epsilon}F_4(-3\epsilon,1-2\epsilon;1-2\epsilon,1-2\epsilon;z_1,z_2)\vphantom{\frac{1}{2}}\bigg\}
\end{split}\end{align}

\begin{align}\begin{split}
\mathcal{C}_{1,2,3,4}^{(1)}P^{(1)}(p_1^2,p_2^2,p_3^2)
&=
e^{2\gamma_E\epsilon} \epsilon\,
\sqrt{\lambda\left(1,z_1,z_2\right)}(p_3^2)^{-2\epsilon}
\frac{\Gamma(1+2\epsilon)\Gamma^2(1-\epsilon)}{8\,\Gamma(2-2\epsilon)}\\
&\times\left(\frac{1}{2}\left[z_1+z_2-1+\sqrt{\lambda(1,z_1,z_2)}\right]\right)^{-1-\epsilon}\\
&\times{}_2F_1\left(1-\epsilon,1+\epsilon;2-2\epsilon;\frac{2\sqrt{\lambda(1,z_1,z_2)}}{z_1+z_2-1+\sqrt{\lambda(1,z_1,z_2)}}\right) \\
&+ e^{2\gamma_E\epsilon}(p_3^2)^{-2\epsilon}
\frac{\Gamma(1+2\epsilon)\Gamma^2(1-\epsilon)}{\Gamma(1-2\epsilon)}
\left[\sqrt{\lambda(1,z_1,z_2)}\right]^{\epsilon}\\
&
\times
\left(\frac{1}{2}
\left[z_1+z_2-1+\sqrt{\lambda(1,z_1,z_2)}\right]\right)^{-2\epsilon}\\
&
\times
{}_2F_1\left(1-\epsilon,\epsilon;1-2\epsilon;
\frac{z_1+z_2-1+\sqrt{\lambda(1,z_1,z_2)}}{2\sqrt{\lambda(1,z_1,z_2)}}\right)
\end{split}\end{align}

\begin{align}\begin{split}
\mathcal{C}_{1,2,3,4}^{(2)}P^{(1)}(p_1^2,p_2^2,p_3^2)
&=
e^{2\gamma_E\epsilon} \epsilon\,
\sqrt{\lambda\left(1,z_1,z_2\right)}(p_3^2)^{-2\epsilon}
\frac{\Gamma(1+2\epsilon)\Gamma^2(1-\epsilon)}{4\,\Gamma(2-2\epsilon)}\\
&\times\left(\frac{1}{2}\left[z_1+z_2-1+\sqrt{\lambda(1,z_1,z_2)}\right]\right)^{-1-\epsilon}\\
&\times{}_2F_1\left(1-\epsilon,1+\epsilon;2-2\epsilon;\frac{2\sqrt{\lambda(1,z_1,z_2)}}{z_1+z_2-1+\sqrt{\lambda(1,z_1,z_2)}}\right) 
\end{split}\end{align}

\begin{align}\begin{split}
\mathcal{C}_{1,2,3}P^{(1)}&(p_1^2,p_2^2,p_3^2) 
=  e^{2\gamma_E\epsilon}\epsilon^2\frac{\Gamma^2(1-\epsilon)}{\Gamma(1-2\epsilon)}\sqrt{\lambda(1,z_1,z_2)}(p_2^2)^{-2\epsilon}\\
&\times\biggg\{\frac{\Gamma(-1+2\epsilon)}{2}\left(\frac{1}{2}\left[z_1-z_2-1+\sqrt{\lambda(1,z_1,z_2)}\right]\right)^{-2+2\epsilon}\\
&\times F_2\left(2-2\epsilon,1-\epsilon;1+\epsilon;2-2\epsilon;2-2\epsilon;\vphantom{\frac{\sqrt{(1)}}{\sqrt{(1)}}}\right.\\
&\hspace{2cm}\left.\frac{2\sqrt{\lambda(1,z_1,z_2)}}{z_1-z_2-1+\sqrt{\lambda(1,z_1,z_2)}},-\frac{2}{z_1-z_2-1+\sqrt{\lambda(1,z_1,z_2)}}\right)\\
&-\frac{\Gamma(1-2\epsilon)\Gamma(-\epsilon)}{\Gamma(1-3\epsilon)\Gamma(2-2\epsilon)}\frac{2}{z_1-z_2-1+\sqrt{\lambda(1,z_1,z_2)}}(p_3^2)^{-2\epsilon}\\
&\times F_2\left(1,1-\epsilon;3\epsilon;2-2\epsilon;2\epsilon;\vphantom{\frac{\sqrt{(1)}}{\sqrt{(1)}}}\right.\\
&\left.\hspace{2cm}\frac{2\sqrt{\lambda(1,z_1,z_2)}}{z_1-z_2-1+\sqrt{\lambda(1,z_1,z_2)}},-\frac{2}{z_1-z_2-1+\sqrt{\lambda(1,z_1,z_2)}}\right)\biggg\}
\end{split}\end{align}

\begin{align}\begin{split}
\mathcal{C}_{1,2,3,4}^{(1)}P^{(2)}(p_1^2,p_2^2,p_3^2)
&= 2 e^{2\gamma_E\epsilon}\Gamma^2(1+2\epsilon)\Gamma^2(1-\epsilon)(p_3^2)^{-2\epsilon}\\
&\times\left(\frac{1}{2}\left[z_1+z_2-1+\sqrt{\lambda(1,z_1,z_2)}\right]\right)^{-\epsilon}\\
&\times{}_2F_1\left(-\epsilon,\epsilon;-2\epsilon;\frac{2\sqrt{\lambda(1,z_1,z_2)}}{z_1+z_2-1+\sqrt{\lambda(1,z_1,z_2)}}\right) \\
&- 2 e^{2\gamma_E\epsilon}(-p_3^2)^{-2\epsilon}
\frac{\Gamma(1+2\epsilon)\Gamma^2(1-\eps)}{\Gamma(1-2\epsilon)} \\
&\times
\left[\sqrt{\lambda(1,z_1,z_2)}\right]^{\epsilon}
\left(\frac{1}{2}\left[z_1+z_2-1+\sqrt{\lambda(1,z_1,z_2)}\right]\right)^{-2\epsilon}\\
&\times{}_2F_1\left(-\epsilon,1+\epsilon;1-2\epsilon;
\frac{z_1+z_2-1+\sqrt{\lambda(1,z_1,z_2)}}{2\sqrt{\lambda(1,z_1,z_2)}}\right) 
\end{split}\end{align}

\begin{align}\begin{split}
\mathcal{C}_{1,2,3,4}^{(2)}P^{(2)}(p_1^2,p_2^2,p_3^2)
&= e^{2\gamma_E\epsilon}\Gamma^2(1+2\epsilon)\Gamma^2(1-\epsilon)(p_3^2)^{-2\epsilon}\\
&\times\left(\frac{1}{2}\left[z_1+z_2-1+\sqrt{\lambda(1,z_1,z_2)}\right]\right)^{-\epsilon}\\
&\times{}_2F_1\left(-\epsilon,\epsilon;-2\epsilon;\frac{2\sqrt{\lambda(1,z_1,z_2)}}{z_1+z_2-1+\sqrt{\lambda(1,z_1,z_2)}}\right) 
\end{split}\end{align}

\begin{align}\begin{split}
\mathcal{C}_{1,2,3}P^{(2)}&(p_1^2,p_2^2,p_3^2) 
 =e^{2\gamma_E\epsilon}\epsilon\frac{\Gamma^2(1-\epsilon)\Gamma(-\epsilon)}{\Gamma(1-2\epsilon)} (p_2^2)^{-2\epsilon}\\
&\times\biggg\{\frac{\Gamma(1+2\epsilon)}{\Gamma(1-\epsilon)}\left(\frac{1}{2}\left[z_1-z_2-1+\sqrt{\lambda(1,z_1,z_2)}\right]\right)^{2\epsilon}\\
&\times F_2\left(-2\epsilon,-\epsilon;\epsilon;-2\epsilon;-2\epsilon;\vphantom{\frac{\sqrt{(1)}}{\sqrt{(1)}}}\right.\\
&\left.\hspace{2cm}\frac{2\sqrt{\lambda(1,z_1,z_2)}}{z_1-z_2-1+\sqrt{\lambda(1,z_1,z_2)}},-\frac{2}{z_1-z_2-1+\sqrt{\lambda(1,z_1,z_2)}}\right)\\
&+\frac{\Gamma(-1-2\epsilon)}{\Gamma(-3\epsilon)\Gamma(-2\eps)}\frac{2}{z_1-z_2-1+\sqrt{\lambda(1,z_1,z_2)}}\\
&\times F_2\left(1,-\epsilon;1+3\epsilon;-2\epsilon;2+2\epsilon;\vphantom{\frac{\sqrt{(1)}}{\sqrt{(1)}}}\right.\\
&\left.\hspace{2cm}\frac{2\sqrt{\lambda(1,z_1,z_2)}}{z_1-z_2-1+\sqrt{\lambda(1,z_1,z_2)}},-\frac{2}{z_1-z_2-1+\sqrt{\lambda(1,z_1,z_2)}}\right)\biggg\}
\end{split}\end{align}

\subsection{Adjacent Triangles}

The adjacent triangles are defined in eq.~\eqref{eq:adjTrianglesGen}.
We will consider the case with two massive external legs (see fig.~\ref{fig:adjTriangles})
and the case with a single massive external leg (i.e., we set $p_2^2=0$ in the diagram of 
fig.~\ref{fig:adjTriangles}), which are known to all orders in $\epsilon$
in~\cite{Gehrmann:1999as}.

\subsubsection{Asymmetric One Scale}\label{sec:adjTrinAsy}

\begin{align}\label{TOneScale}\begin{split}
T(p^2)=&p^2\epsilon^4T(1,1,1,1,1;4-2\epsilon;p^2,0)\\
=&(-p^2)^{-2\ep}e^{2\gamma_E\epsilon}
\frac{\Gamma^2(1-\epsilon)\Gamma(1+2\epsilon)}{4\,\Gamma(1-3\epsilon)}[
\Gamma(1-2\epsilon)\Gamma(1+\epsilon)-\Gamma(1-\epsilon)]\\
\mathcal{C}_{1,2,4,5}T=&(-p^2)^{-2\ep}e^{2\gamma_E\epsilon}
\frac{\Gamma(1-\epsilon)}{\Gamma(1-3\epsilon)}\\
\mathcal{C}_{2,3,5}T=&-(-p^2)^{-2\ep}e^{2\gamma_E\epsilon}
\frac{\Gamma^3(1-\epsilon)\Gamma(1+2\epsilon)}{4\,\Gamma(1-3\epsilon)}
\end{split}\end{align}

\subsubsection{Symmetric Two Scales}\label{sec:adjTrinSym}

We do not list the cuts $\mathcal{C}_{1,4,5}$ and $\mathcal{C}_{2,3,4,5}$ as they
can be deduced by symmetry from the ones given below.

\begin{align}\label{TTwoScales}
&T(p_1^2,p_2^2)=\epsilon^4(p_1^2-p_2^2)\,T(1,1,1,1,1;4-2\epsilon;p_1^2,p_2^2)\nonumber\\
&=e^{2\gamma_E\epsilon}\epsilon\left\{
\frac{\Gamma^2(1+\epsilon)\Gamma^4(1-\epsilon)}{\Gamma(1-2\epsilon)\Gamma(2-2\epsilon)}
\frac{p_1^2-p_2^2}{p_2^2}(-p_1^2)^{-2\epsilon}
{}_2F_1\left(1-\epsilon,1-2\epsilon;2-2\epsilon;1-\frac{p_1^2}{p_2^2}\right)\right.\nonumber\\
&-\frac{\Gamma(1+2\epsilon)\Gamma^3(1-\epsilon)}{2(1-2\epsilon)\Gamma(1-3\epsilon)}
\left[\frac{p_1^2-p_2^2}{p_1^2}(-p_2^2)^{-2\epsilon}
{}_3F_2\left(1-\epsilon,1,1-2\epsilon;1+\epsilon,2-2\epsilon;1-\frac{p_2^2}{p_1^2}\right)\right.\nonumber\\
&\left.\left.+\frac{p_1^2-p_2^2}{p_2^2}(-p_1^2)^{-2\epsilon}
{}_3F_2\left(1-\epsilon,1,1-2\epsilon;1+\epsilon,2-2\epsilon;1-\frac{p_1^2}{p_2^2}\right)
\vphantom{\frac{1}{2}}\right]\right\}\,,
\end{align}

\begin{align}
\label{AjTr2mCut235}
\begin{split}
\mathcal{C}_{2,3,5}T(p_1^2,p_2^2)=&-\,e^{2\gamma_E\epsilon} 
\frac{\epsilon\,\Gamma^3(1-\epsilon)}{2\,\Gamma(1-3\epsilon)\Gamma(2-2\epsilon)}
\frac{p_1^2-p_2^2}{p_2^2}(p_1^2)^{-2\epsilon}\\
&\,{}_3F_2\left(1-2\epsilon,1,1-\epsilon;1+\epsilon,2-2\epsilon;1-\frac{p_1^2}{p_2^2}\right)\,,
\end{split}\end{align}

\begin{align}
\label{AjTr2mCut1245}
\begin{split}
\mathcal{C}_{1,2,4,5}T(p_1^2,p_2^2)=\,&-e^{2\gamma_E\epsilon}
\frac{2\epsilon\,\Gamma(1-\epsilon)}{\Gamma(2-3\epsilon)}\left(1-\frac{p_2^2}{p_1^2}\right)^{1-\epsilon}
(p_2^2)^{-2\epsilon}\\
&\,{}_2F_1\left(1-2\epsilon,1-3\epsilon;2-3\epsilon;1-\frac{p_2^2}{p_1^2}\right)\,,
\end{split}\end{align}

\begin{align}
\label{AjTr2mCut1234}
\begin{split}
\mathcal{C}_{1,2,3,4}T(p_1^2,p_2^2)=\,&e^{2\gamma_E\epsilon}\frac{\epsilon\,\Gamma^2(1-\epsilon)}
{\Gamma(1-2\epsilon)\Gamma(2-2\epsilon)}\frac{p_1^2-p_2^2}{p_2^2}
(p_1^2)^{-2\epsilon}\\
&\,{}_2F_1\left(1-\epsilon,1-2\epsilon;2-2\epsilon;1-\frac{p_1^2}{p_2^2}\right)\,,
\end{split}\end{align}

\begin{align}
\label{AjTr2mCut12345}
\begin{split}
\mathcal{C}_{1,2,3,4,5}T(p_1^2,p_2^2)=&e^{2\gamma_E\epsilon}\Gamma(1+2\epsilon)
(p_1^2)^{-2\epsilon}(p_2^2)^{-2\epsilon}\left(p_1^2-p_2^2\right)^{2\epsilon}\,.
\end{split}\end{align}

As discussed in section~\ref{sec:adjacentTriangles}, the diagrammatic coaction of the one-mass integral $T(p_1^2)$ in eq.~(\ref{TOneScale}) 
can be obtained by taking the $p_2^2\to 0$ limit of $T(p_1^2,p_2^2)$ in~eq.~(\ref{TTwoScales}). 
The corresponding diagrammatic representations are given in eqs.~(\ref{2madjTrianglesCoaction}) and (\ref{1madjTrianglesCoaction}), respectively. 
As stressed in the main text, the $p_2^2\to 0$ limit must be carefully taken keeping $\epsilon<0$, as it can (and does) generate new infrared divergences.
 Let us illustrate this in the simple but illustrative example of $\mathcal{C}_{1,2,4,5}T(p_1^2,p_2^2)$ in eq.~(\ref{AjTr2mCut1245}). 
The hypergeometric function appearing there is
\begin{equation*}
{}_2F_1\left(1-2\epsilon,1-3\epsilon;2-3\epsilon;1-\frac{p_2^2}{p_1^2}\right) = (1-3\epsilon) \int_0^1 du  u^{-3\epsilon} 
	\left[1-\left(1-\frac{p_2^2}{p_1^2}\right)u\right]^{2\epsilon-1} \,,
\end{equation*}
where we used Euler's integral representation in eq.~(\ref{eq:2f1Simp}) to expose the branch point at \hbox{$p_2^2=0$}.
We note that  for $\epsilon<0$ the integral converges on the real line for $p_2^2/p_1^2>0$, but diverges for $p_2^2/p_1^2<0$.  Naively setting $p_2^2=0$ would yield a wrong (vanishing) result for the cut in~eq.~(\ref{AjTr2mCut1245}).
The correct procedure is to consider a representation that is a priori valid for~\hbox{$\epsilon<0$}, where we can approach the $p_2^2/p_1^2\to 0$ limit from any direction. Such a representation can be readily derived using known transformations bringing the argument of the hypergoemetric function to be ${p_2^2}/{p_1^2}$ before considering the limit. In our particular example we may use:
 \begin{align*}
{}_2F_1\left(1-2\epsilon,1-3\epsilon;2-3\epsilon;1-\frac{p_2^2}{p_1^2}\right)& =
\left(1 - \frac{p_2^2}{p_1^2}\right)^{3 \epsilon-1} 
\frac{\Gamma(2 - 3 \epsilon) \Gamma(2 \epsilon)}{\Gamma(1 - \epsilon) }\\
&-  \left(\frac{p_2^2}{p_1^2}\right)^{2 \epsilon} 
\frac{(1 - 3 \epsilon) }{2\epsilon}
{}_2F_1\left(1, 1 - \epsilon;1 + 2 \epsilon; \frac{p_2^2}{p_1^2}\right)\,.
\end{align*}
Given $\epsilon<0$, the first term is finite for vanishing $p_2^2$, while the second is divergent in this limit, precisely cancelling the factor $(p_2^2)^{-2\epsilon}$ in eq.~(\ref{AjTr2mCut1245}) and yielding a finite non-vanishing result for the cut $\mathcal{C}_{1,2,4,5}T(p_1^2,0)$. 
Furthermore, one can confirm that the result coincides with the one presented in eq.~(\ref{TOneScale}). The limits of all other hypergeometric
functions appearing in the uncut and cut expressions of the two-mass case can be computed following similar steps.

\subsection{Diagonal Box}\label{sec:diagBox}
The diagonal box was obtained in ref.~\cite{Anastasiou:1999bn} as a linear
combination of Gauss hypergeometric functions:
\begin{align}\begin{split}
B(s,t)=&\epsilon^4(s+t) B(1,1,1,1,1;4-2\epsilon;s,t)\\
=&-e^{2\gamma_E\epsilon}\frac{\epsilon(s+t)}{2(1-2\epsilon)}
\frac{\Gamma^3(1-\epsilon)\Gamma(1+2\epsilon)}{\Gamma(1-3\epsilon)}
\left[\frac{t^{-2\epsilon}}{s}\,{}_2F_1\left(1-2\epsilon,1-2\epsilon;2-2\epsilon;1+\frac{t}{s}\right)
\right.\\
&\left.+\frac{s^{-2\epsilon}}{t} \,{}_2F_1\left(1-2\epsilon,1-2\epsilon;2-2\epsilon;1+\frac{s}{t}\right)\right],
\end{split}\end{align}

A basis of cut integrals is given by:
\begin{align}\begin{split}
\mathcal{C}_{1,3,5}B(s,t)=\,&-e^{2\gamma_E\epsilon}
\frac{\Gamma^3(1-\epsilon)\Gamma(1+2\epsilon)}{2(1-2\epsilon)\Gamma(1-3\epsilon)}
\left(1+\frac{s}{t}\right) s^{-2\epsilon}
{}_2F_1\left(1-2\epsilon,1-2\epsilon;2-2\epsilon;1+\frac{s}{t}\right),
\end{split}\end{align}
\begin{align}\begin{split}
\mathcal{C}_{2,4,5}B(s,t)=\,&-e^{2\gamma_E\epsilon}
\frac{\Gamma^3(1-\epsilon)\Gamma(1+2\epsilon)}{2(1-2\epsilon)\Gamma(1-3\epsilon)}\left(1+\frac{t}{s}\right) t^{-2\epsilon}
{}_2F_1\left(1-2\epsilon,1-2\epsilon;2-2\epsilon;1+\frac{t}{s}\right),
\end{split}\end{align}
\begin{align}\begin{split}
\mathcal{C}_{1,2,3,4,5}B(s,t)=\,&e^{2\gamma_E\epsilon}\frac{\Gamma^3(1-\epsilon)\Gamma(1+2\epsilon)}{\Gamma(1-3\epsilon)}\frac{(s+t)^{2\epsilon}}{s^{2\epsilon} t^{2\epsilon}}.
\end{split}
\end{align}

\bibliographystyle{JHEP}
\bibliography{bibMain.bib}

\end{document}